# Roadmap on Integrated Quantum Photonics

Galan Moody[1], Volker J. Sorger[2], Daniel. J. Blumenthal[1], Paul W. Juodawlkis[3], William Loh[3], Cheryl Sorace-Agaskar[3], Alex E. Jones[4], Krishna C. Balram[4], Jonathan C. F. Matthews[4], Anthony Laing[4], Marcelo Davanco[5], Lin Chang[1], John E. Bowers[1], Niels Quack[6], Christophe Galland[6], Igor Aharonovich[7], Martin A. Wolff,[8] Carsten Schuck[8], Neil Sinclair[9], Marko Lončar[9], Tin Komljenovic[10], David Weld[1], Shayan Mookherjea[11], Sonia Buckley[5], Marina Radulaski[12], Stephan Reitzenstein[13], Benjamin Pingault[9,14], Bartholomeus Machielse[9], Debsuvra Mukhopadhyay[15], Alexey Akimov[15,16,17], Aleksei Zheltikov[16,17,18], Girish S. Agarwal[15], Kartik Srinivasan[5], Juanjuan Lu[19], Hong X. Tang[19], Wentao Jiang[20], Timothy P. McKenna[20], Amir H. Safavi-Naeini[20], Stephan Steinhauer[21], Ali W. Elshaari[21], Val Zwiller[21], Paul S. Davids[22], Nicholas Martinez[22], Michael Gehl[22], John Chiaverini[3,23], Karan K. Mehta[24], Jacquiline Romero[25,26], Navin B. Lingaraju[27], Andrew M. Weiner[27], Daniel Peace[28], Robert Cernansky[28], Mirko Lobino[28], Eleni Diamanti[29,30], Luis Trigo Vidarte[29,30], and Ryan M. Camacho[31]

1 - University of California, Santa Barbara, USA
2 - George Washington University, USA
3 - MIT Lincoln Laboratory, USA
4 – University of Bristol, UK
5 - National Institute of Standards and Technology, USA
6 - École Polytechnique Dédérale de Lausanne, Switzerland
7 - University of Technology Sydney, Australia
8 - University of Münster, Germany
9 - Harvard University, USA
10 - Nexus Photonics, Goleta, USA
11 - University of California, San Diego, USA
12 - University of California, Davis, USA
13 - Technische Universität Berlin, Germany
14 - QuTech, Delft University of Technology, Delft, The Netherlands
15 - Texas A&M University, USA
16 - PN Lebedev Physical Institute, Moscow, Russia
17 - Russian Quantum Center, Moscow Region, Russia
18 - M.V. Lomonosov Moscow State University, Moscow, Russia
19 - Yale University, USA
20 - Stanford University, USA
21 - KTH Royal Institute of Technology, Sweden
22 - Sandia National Labs, USA
23 - Massachusetts Institute of Technology, USA
24 - ETH Zurich, Switzerland
25 - Australian Research Council Centre of Excellence for Engineered Quantum Systems (EQUS), Australia
26 - University of Queensland, Australia
27 - Purdue University, USA
28 - Griffith University, Australia
29 - Centre National de la Recherche Scientifique, France
30 - Sorbonne University, France
31 - Brigham Young University, USA

## Contents

# 1 - Introduction

Galan Moody[1], Volker J. Sorger[2], and Daniel J. Blumenthal[1]
[1] University of California Santa Barbara, Santa Barbara, CA, USA
[2] George Washington University, Washington D.C., USA

Integrated photonics is at the heart of many classical technologies, from optical communications to biosensors, LIDAR, and data center fiber interconnects. There is strong evidence that these integrated technologies will play a key role in quantum systems as they grow from few-qubit prototypes to tens of thousands of qubits [1]. The underlying laser and optical quantum technologies, with the required functionality and performance, can only be realized through the integration of these components onto quantum photonic integrated circuits (QPICs) with accompanying electronics. In the last decade, remarkable advances in quantum photonic integration and a dramatic reduction in optical losses [2] have enabled benchtop experiments to be scaled down to prototype chips with improvements in efficiency, robustness, and key performance metrics [3,4]. The reduction in size, weight, power, and improvement in stability that will be enabled by QPICs will play a key role in increasing the degree of complexity and scale in quantum demonstrations. As an example, the timeline in Figure 1 illustrates this rapid progression from few-component circuits enabling two-photon quantum interference in 2008 [5] to a decade later with devices combining more than 650 components capable of arbitrary and programmable two-qubit operations [6], enabling advances in foundational quantum mechanics, computing, communications, and metrology [7]. Today, experiments that until recently occupied an optical table, such as Boson sampling [8], diamond color-center emitter arrays [9], and multi-ion quantum logic [10,11], have moved on-chip. Despite these early advancements, the level of QPIC complexity lags that of conventional PICs, which today comprise ~$5\times10^3$ components on chip. Interestingly, this progression mirrors the development of digital electronics that occurred through the 1960s and 1970s, leading to potential quantum integration scaling laws and roadmaps. Today's state-of-the-art integration, with current materials, fabrication, and packaging technologies, gives a snapshot of on-chip complexity currently achievable.

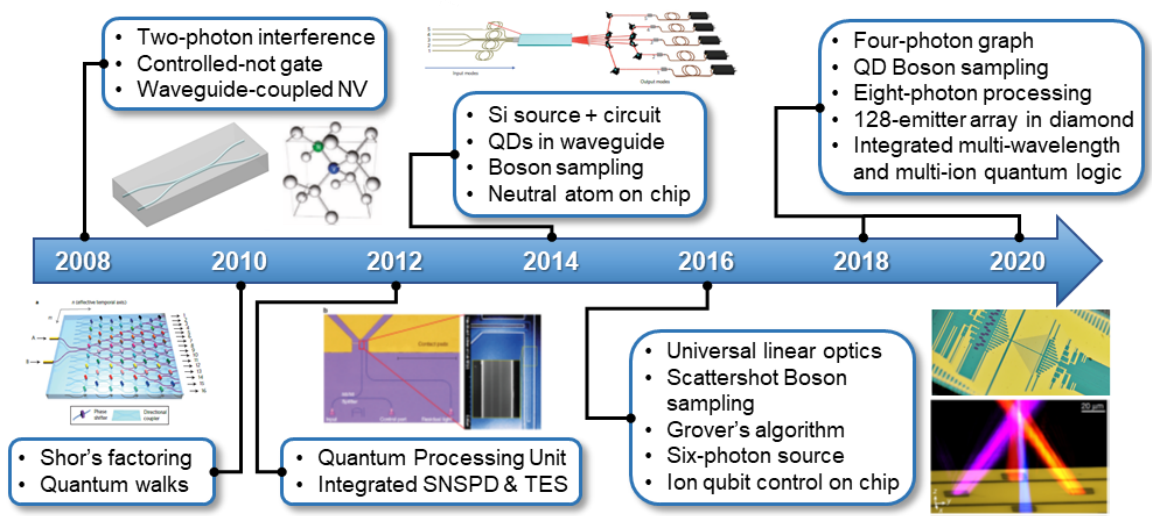

**Figure 1.** Key milestones in integrated quantum photonics in the past decade, beginning with two-photon interference and fundamental quantum gates prior to 2008 to large-scale devices comprising over 650 photonic components and arrays of deterministic and probabilistic quantum light sources on a single chip in 2020.

**Roadmap Organization and Goals**

In this roadmap article, we highlight the status, current and future challenges, and emerging technologies in several key research areas in integrated quantum photonics, which serves as a complementary resource to the recent OIDA Roadmap on Quantum Photonics [12]. With advances in materials, PIC-based platforms, devices and circuits, fabrication and integration processes, packaging, and testing and benchmarking, we can expect a transition from single- and few-function prototypes to large-scale integration of multi-functional and reconfigurable QPICs. These circuits will play a key role in how quantum information is processed, stored, transmitted, and utilized for quantum computing, communications, metrology, and sensing. This roadmap highlights the current progress in the field of integrated quantum photonics, future challenges, and advances in science and technology needed to meet these challenges. Key areas of research and technology addressed include:

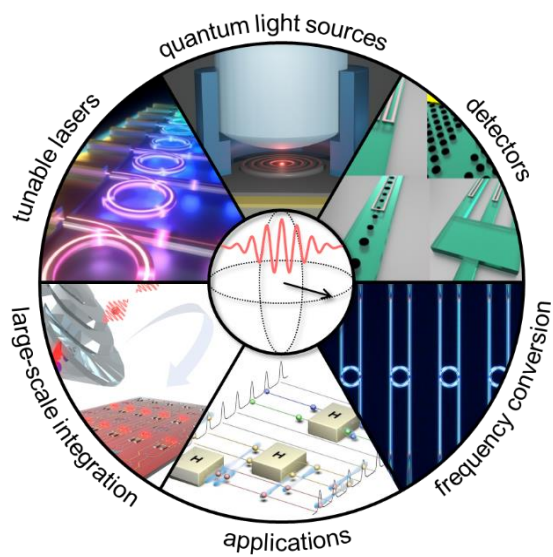

**Figure 2.** The Roadmap on Integrated Quantum Photonics covers topics spanning classical and quantum light sources, detectors, frequency conversion and transduction, photonic material platforms, methods for large-scale integration, and applications in computing, communications, and sensing.

1. **PIC Platforms:** While silicon-based photonics has been the workhorse for QPICs owing to the low waveguide loss and the existing foundry infrastructure, other material platforms have been developed to further extend the capabilities beyond what silicon can offer [13], including III-V semiconductors, lithium niobate, silicon carbide, nitrides, diamond, and tantalum pentoxide. Future advances in quantum photonics will likely require a heterogeneous approach [14] that combines multiple PIC materials to achieve high-level functionality with integrated lasers and amplifiers, passive components, modulators, quantum frequency converters, efficient detectors or chip-to-fiber couplers, and control logic.

2. **Quantum and Classical Light Sources and Qubits:** Quantum light sources generating single photons, entangled-photon pairs, squeezed light, and other non-classical states are fundamental resources for quantum information science. Quantum sources typically fall into two categories: Quantum emitters that can produce photons on-demand, and sources based on $\chi^{(2)}$ or $\chi^{(3)}$ nonlinear optics that produce photons probabilistically. Rapid progress with quantum emitters based on InAs quantum dots, defects in diamond and silicon carbide, emissive centers in silicon, and probabilistic sources in several nonlinear PICs have already been used successfully for numerous applications in communications, computing, and networking. Further technological developments that improve PIC integration must follow to increase the photon generation quality, efficiency, and rates. Likewise, as these sources currently rely on off-chip optical pumping, improvement in their efficiency will facilitate the integration of precision and tunable pump lasers, filters, and control electronics on chip for packaged, turn-key systems. Such heterogeneous integration may demand co-design approaches beyond current single platform design rules.

3. **Quantum Frequency Conversion:** Quantum frequency conversion is essential to establish interconnections between quantum systems operating in different wavelength regimes, for example, to connect a trapped ion quantum processor to a neutral atom ensemble quantum memory or to an optical network featuring microwave-frequency superconducting qubits. Numerous approaches to frequency conversion exist, including optomechanics and nonlinear photonics and optoelectronics, providing exciting prospects for achieving high conversion efficiency spanning ultraviolet to telecommunications wavelengths with low added noise.

4. **Integrated Detectors:** To take full advantage of the low loss and high throughput afforded by integrated photonics, effective schemes for on-chip coupling of high-performance photodetectors with single- and photon-number resolving capabilities are required. Detectors based on superconducting nanowires, which operate at cryogenic temperatures, are the most promising technology to date with near-unity quantum efficiency, low jitter, and low dark count rates. However, the required cooling apparatus sets system limitations, and room-temperature approaches defeating $k_BT$ noise would be a game-changer for QPICs. Single-photon avalanche photodiodes are appealing due to their near-room-temperature operation, but improvements in their performance are required to compete with superconducting detectors. Notably, both technologies can be integrated with various PIC platforms through direct growth, thin film deposition, or chip-to-chip bonding.

5. **Applications:** The potential impact of quantum photonic technologies is vast, ranging from all-optical quantum computing, quantum encryption, and networking to machine learning, sensing, and interfacing with other quantum systems, such as chip-scale ion traps. The key challenges are scaling up the number of integrated components, improving on-chip functionality and performance, and maintaining low excessive noise and loss, while enabling seamless assembly and packaging.

**Concluding Remarks**

In addition to the technological advances, transitioning from proof-of-principle prototypes to packaged and deployable systems used in quantum experiments and applications will require the investment in, and development of, a sustainable quantum photonic ecosystem that brings together interdisciplinary researchers, scientists and engineers, infrastructure and testbeds, and federal, academic, and private partnerships. A key long-term goal of this infrastructure should be to broaden access to quantum integrated technologies and cultivate future generations of a quantum workforce through professional development and mentorship. Other aspects important to the long-term ecosystem include modernized teaching labs and activities that enable hands-on approaches to quantum mechanics and photonics education. Each article in this roadmap provides an overview of a specific research area, highlighting the status, current and future challenges, and advances in the science and technology to meet these challenges, together covering the plurality of fields in this exciting space.

# PHOTONIC CIRCUIT INTEGRATED PLATFORMS
## 2 - Heterogeneous Integrated Photonics for Quantum Information Science and Engineering

Paul W. Juodawlkis, William Loh, Cheryl Sorace-Agaskar
MIT Lincoln Laboratory

**Status**

Integrated photonic technologies will play a central role in advancing the frontiers of quantum information science and in the full realization of quantum processing, sensing, and communication applications. The required scaling (*i.e.*, thousands to millions of quantum processing or sensing elements) and functional complexity of these quantum systems will only be possible through the high density, enhanced performance, and environmental stability afforded by photonic integrated circuits (PICs). Photonic functions that will be needed include light generation (both uncorrelated photon beams and entangled-photons), amplification, modulation, switching, routing and qubit input/output (I/O) interfacing, passive splitting/combining, filtering, frequency translation, and detection (both single photon and linear). Due to the distinct material properties required to realize these functions over the operating wavelengths (*i.e.*, ultraviolet to near-infrared) associated with the various quantum modalities, it is highly unlikely that the required PICs can be made using a homogeneous material platform such as silicon or a single compound-semiconductor material system. Therefore, advances in quantum information science and engineering will require the development of heterogeneous material PICs that are made available to the community through a collaborative ecosystem of design, fabrication, packaging and evaluation resources.

Figure 1 depicts a generic heterogeneous PIC platform that combines the fabrication process control and yield of advanced silicon foundry toolsets with the wide-ranging materials options and flexibility of post-silicon-fab hybrid integration of materials and components. The base platform is comprised of multi-layer dielectric (i.e., silicon nitride (SiN), alumina ($Al_2O_3$)) and silicon waveguides that are used to implement passive (e.g., Fig. 1(a) where a subset of layers were used [1]) and active components, and to create optimized interfaces for heterogeneous integration. Low optical losses have been achieved in both medium-confinement waveguides (< 0.2 dB/cm in SiN from 600-1650 nm, 3 dB/cm in $Al_2O_3$ at 370 nm [2]) and low-confinement waveguides (0.001 dB/cm in SiN from 1300-1650 nm). Optical sources and other non-silicon-based active devices can be added through a combination of material growth and deposition, layer transfer, wafer bonding, and pick-and-place techniques. Exemplar demonstrations of such a heterogeneous PIC platform include high-power (> 300 mW) low-noise lasers incorporating flip-chipped semiconductor optical amplifiers (Fig. 1(d)) [3], surface-bonded optical gain media (Fig. 1(c)) [4], [5], and low-loss surface-bonded thin-film lithium-niobate (TFLN) waveguides (Fig. 1(b)) [6]. While this initial work along with the work of other photonic developers are important first steps, a more significant and coordinated effort is required to develop the platforms needed by the community.

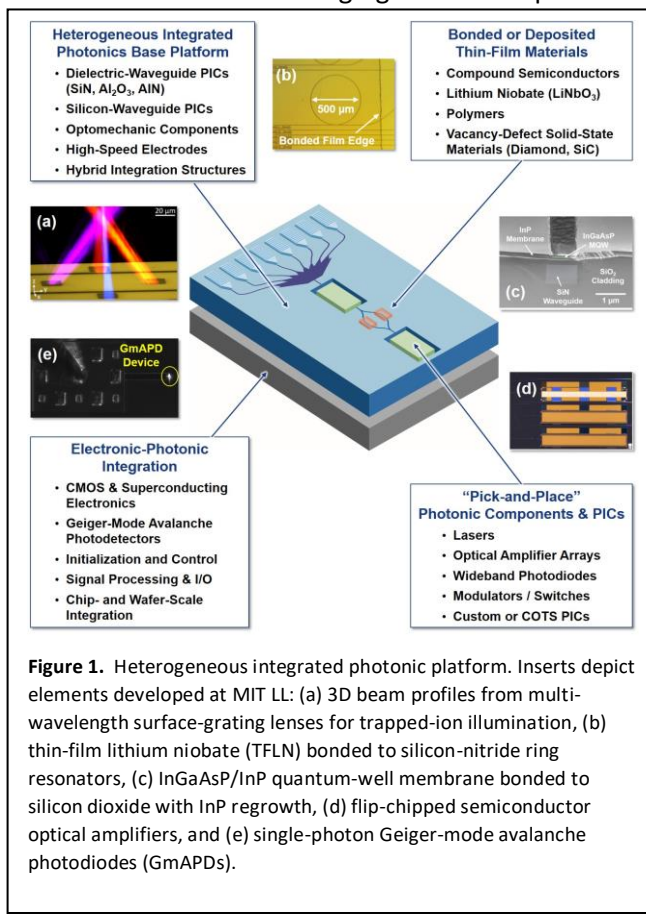

**Figure 1.** Heterogeneous integrated photonic platform. Inserts depict elements developed at MIT LL: (a) 3D beam profiles from multi-wavelength surface-grating lenses for trapped-ion illumination, (b) thin-film lithium niobate (TFLN) bonded to silicon-nitride ring resonators, (c) InGaAsP/InP quantum-well membrane bonded to silicon dioxide with InP regrowth, (d) flip-chipped semiconductor optical amplifiers, and (e) single-photon Geiger-mode avalanche photodiodes (GmAPDs).

**Current and Future Challenges**
The principal challenges to developing heterogeneous PICs for the quantum community are:

- The breadth of the requirements across the different quantum modalities
- The evolution of these requirements as the field transitions from fundamental scientific investigations to scaled engineering prototypes and products
- The variety of photonic materials, devices, and integrated circuits needed to meet the requirements
- The electronic-photonic integration needed to initialize and control the photonic functions, and to process, convert, and transfer the I/O information

We briefly address each of these principal challenges in the following paragraphs.

*Breadth of requirements.* The photonic functions and their associated requirements can be divided into two major classes based on whether the photons serve as the qubits or if they are used to interface to the physical qubit. Quantum systems based on photon qubits require sources of entangled photons, ultralow optical propagation loss, control and maintenance of optical coherence, and high-efficiency detection. For systems in which the photons serve as the I/O to the physical qubit, many of the above requirements can be relaxed. However, other qubit-specific requirements, which will depend on the physics of the photon-to-qubit coupling will be present. The need to operate at many different wavelengths is also a challenge.

*Evolution of requirements.* The requirements of photonics in large-scale quantum systems are still being defined as early demonstrations are performed. For example, consider trapped-ion-based quantum systems for quantum computing and optical atomic clocks. As mentioned by Chiaverini *et al*. in Section 21, MIT LL has integrated SiN waveguides and vertical-grating lenses with ion-trap surface-electrodes to deliver the six wavelengths required to perform qubit operations on a $Sr^+$ ion (Fig. 1(a)) [1]. Significant challenges must be overcome to scale this architecture to the number of trapped ions required for array-based optical atomic clocks and practical quantum computing. These include waveguide loss, size and efficiency of the grating couplers, the lack of high-extinction optical switches and frequency shifters having small enough size to be integrated at each trapped-ion site, crosstalk due to scattered light in high-density PICs, fabrication yield and reproducibility, and the need for integrated optical amplifiers, single-photon sensitive photodetectors (Fig. 1(e)), and electronics. Equally difficult scaling challenges exist for other quantum modalities. One example is networking of superconducting quantum processors via optical fiber interconnects that will require the development of coherent microwave to optical photon conversion. Transition from the lab to the field will also require improvements in environmental stability and packaging.

*Variety of materials, devices, and integrated circuits.* It will be very costly, both in terms of dollars and time, to develop and maintain uncoordinated integrated photonics platforms to address the wide range of photonic functions and operating wavelengths required across the quantum modalities (see Status section above).

*Electronic-photonic integration.* Most quantum photonic functions will require integration with electronics to provide DC power, control, and I/O with the classical world. The level of integration increases with the number and density of qubits to support the required interconnects. It will be critical that this integration be implemented in a way that is both scalable and shields the quantum elements from electromagnetic interference (EMI) generated by the electronics.

**Advances in Science and Technology to Meet Challenges**

To address the significant challenges to providing broadly applicable, scalable, and cost-effective heterogeneous integrated photonics to the quantum community, the following advances will be required:

- A community-defined set of broad-wavelength (~0.35-2.0 µm), multi-waveguide-layer base platforms containing both core photonic devices and structures for implementing a range of heterogeneous photonic integration techniques (*e.g.*, flip-chip bonding, surface bonding, epitaxial growth), and that are fabricated using silicon-foundry-compatible materials and processes well described by open process design kits (PDKs)
- A variety of application-specific materials and devices that can be heterogeneously integrated onto the base platforms using either primary foundry or secondary hybrid-integration facilities
- Three-dimensional (3D) integration with electronics using wafer-bonding techniques to decouple the photonic and electronic fabrication processes and to facilitate enhanced EMI shielding

Table 1 summarizes a number of key integrated photonic functions/devices, the challenges to developing them, and the advances that will be required to meet the challenges. Several of the challenges require material, fabrication and device advances to reduce optical loss, realize efficient optical gain, and enable frequency translation over a wide spectral range. Scaling to large numbers of on-chip gain elements is expected to require improvements in heteroepitaxial material growth and integration with silicon [7] and dielectric waveguides. Comparable material and integration advances will be required to realize high-rate, reproducible sources of single and entangled photons integrated with low-loss waveguides for transmission and optical processing. Integration of nonlinear optical materials such as surface-bonded TFLN for switches and frequency shifters will need to be scaled from chip to wafer level. Silicon-foundry-compatible piezoelectric materials (e.g., AlN) will likely be required in the base platform to implement electro-optomechanical resonators to enable coherent transduction between microwave and optical photons. New materials and device concepts will also be required to realize large arrays of optical switches having high extinction, low excess loss, and low static power dissipation. Transitioning from the lab to the field will require low-cost, environmentally robust subsystems such as ultra-narrow-linewidth lasers based on stimulated Brillouin scattering (SBS) in optical fibers [8], [9] or ultralow-loss on-chip waveguides.

**Concluding Remarks**

The quantum community will benefit from a collaborative and coordinated effort to develop heterogeneous PIC base platforms, materials, devices and circuits, flexible integration processes, and advanced packaging techniques. Heterogeneous integration will be necessary due to the breadth of the photonic requirements across the various quantum modalities, the evolution of these requirements as the field matures, and the wide range of technologies needed to meet these evolving requirements. The critical need for tight integration with electronics for control, processing, and I/O will also eventually require 3D electronic-photonic integration. Enabling rapid progress in the field will require cost-effective heterogeneous solutions that combine the existing silicon-foundry infrastructure with post-foundry integration of application-specific materials and devices at custom hybrid-integration facilities.

Table 1. Summary of some key integrated photonics functions, challenges, and required advances for quantum information science and engineering.

| Integrated Photonic Functions / Devices | Challenges | Required Scientific and Technological Advances |
|---|---|---|
| Low-Loss Waveguides in UV/Visible/Near-IR | • Loss limits transmission and resonator Q<br>• Rayleigh scattering significant at visible/UV λ's | • Develop low-loss materials and fabrication techniques<br>• Use low-confinement (Γ) for reducing scattering loss |
| High-Q Resonators | • Material loss limits achievable Q<br>• Low-Γ resonators consume large area | • Develop high-Γ ultralow-loss platform<br>• Mature couplers to transition between low-Γ and high-Γ |
| Gain Media Covering UV/Visible/Near-IR | • Multiple compound-semiconductor (III-V) material systems needed to cover UV to near-IR λ range | • Develop broadband III-V and solid-state gain media<br>• Use nonlinear frequency conversion to cover λ gaps |
| Compact Lasers with Ultra-Narrow Linewidth | • Achieving required stability with compact cavity<br>• Low power conversion efficiency limits scaling | • Mature on-chip SBS lasers at visible and near-IR λ's<br>• Improve power conversion efficiency of III-V gain media |
| Quantum Sources | • Entangled-photon generation rate too low<br>• Precise location of single-photon sources in PICs | • Mature on-chip SPDC sources having high efficiency<br>• Develop and mature new materials (SiC, AlN, diamond) |
| High-Efficiency Frequency Conversion | • Acousto-optic frequency shifters are too large<br>• Low conversion efficiency between microwave and optical photons | • Develop improved electro-optic materials and devices<br>• Develop optomechanic quantum transducers |
| Compact High-Extinction Optical Switches | • Required extinction (e.g., >70 dB for most optically active qubits) difficult to achieve via interferometric switches | • Develop materials and devices for non-interferometric switches (e.g., acousto-optic, electroabsorption) |
| Low-Static-Power, Low-Loss Optical Switches | • Power of thermo-optic switches limits scaling<br>• Loss of non-thermal switches limits scaling | • Develop improved materials and devices (e.g., polymers, strain-induced $\chi^{(2)}$, chalcogenides) |
| Single-Photon Detectors | • Geiger-mode APDs generate large EMI<br>• Superconducting detectors require low-K temps | • Develop EMI shielding techniques<br>• Optimize heterogeneous PIC platform for low temps |
| Very-Large Scale Integration (VLSI) PICs | • Required number (>$10^6$) of current photonic devices will require area >> stepper reticle area | • Develop sub-wavelength photonic materials & devices<br>• Develop wafer-scale lithography and fab processes |
| Electronic-Photonic Integration | • Difference in photonic & electronic device sizes<br>• Proximity of photonics & electronics induces EMI | • Mature wafer-bonded 3D integration incorporating metallic-electrode EMI shielding layers |

**Acknowledgements**

The authors gratefully acknowledge useful discussions with MIT LL colleagues John Chiaverini, Siddhartha Ghosh, Christopher Heidelberger, Robert McConnell, and Danna Rosenberg. This material is based upon work supported by the Under Secretary of Defense for Research and Engineering (USDR&E), the United States Air Force, and the Defense Advanced Research Projects Agency (DARPA) under Air Force Contract No. FA8702-15-D-0001. Any opinions, findings, conclusions or recommendations expressed in this material are those of the author(s) and do not necessarily reflect the views of the USDR&E, the United States Air Force and DARPA.

## 3 – Scaling Integrated Quantum Photonics Beyond Silicon

Alex E. Jones, Krishna C. Balram, Jonathan C. F. Matthews, Anthony Laing
QET Labs, University of Bristol, UK

**Status**

Silicon photonics has been pivotal for increasing the complexity and component-number of on-chip quantum circuitry and revealing the disruptive potential an engineered integrated approach can take for a quantum technology [1]. Optical circuits in silicon benefit from compatibility with established CMOS processes, and silicon's high component density, operation at telecom wavelengths, and nonlinear properties mean it ticks a lot of boxes on the wish list for an integrated photonics platform. Research also benefits from the maturity of the platform that has to date been largely developed for classical telecommunications. As a result, it has been the material of choice for many early demonstrations of building blocks for photonic quantum computing and has seen rapid growth in the number of generated photons and integrated components over the last 10 years [1-3].

Yet while silicon as a largely homogenous platform for quantum photonics provides a straightforward route to manufacturability, relying on a single material to deliver almost all of the core functions required for quantum information processing incurs severe overheads. In silicon, or indeed in other homogenous integrated quantum photonics platforms, photon generation is spontaneous and there is no direct photon-photon interaction. The lack of deterministic methods to generate, store, and entangle photons is addressed by deploying large amounts of probabilistic measurement and feed-forward operations. The feasibility of these schemes to deliver fault tolerant quantum computing is as yet unproven.

Taking a step back, we can consider the hardware elements of a more sophisticated and intrinsically scalable photonic quantum processor. Ideally, this would comprise deterministic sources of single-photons or photonic entangled states, frequency conversion modules, quantum memories, and controlled light-matter interaction. To support the integration of some, or even all of these capabilities, a heterogeneous platform would benefit from an interfacing material with a broad transparency, such as silicon nitride, lithium niobate, or aluminium nitride, to name a few. However, manufacturing such a heterogenous device is also extremely challenging.

Here we discuss the challenges to further scale silicon quantum photonics as a contemporary manufacturable solution for quantum information processing. We then discuss the necessary long-term endeavour, to mass-integrate arrays of functional quantum components, as well as classical readout and control, into a heterogeneous platform for quantum computing.

**Current and Future Challenges for Silicon**

To maintain its role as a contemporary option for quantum information processing, silicon quantum photonics must continue to make progress against a number of key challenges:

*Photon Loss* - While loss-tolerant approaches to photon quantum information processing tasks exist, reducing propagation losses remains desirable e.g., for integrated delay lines. Tight confinement of light in silicon waveguides increases sensitivity to sidewall roughness, typically leading to significant propagation losses. Tapering from single- to multi-mode waveguides for straight sections can drastically reduce this. A different approach is to interface silicon with a different material such as silicon nitride (SiN), where a lower index contrast means better propagation losses that can be pushed down to <0.1 dB/m [4]. Another important loss mechanism in silicon is two-photon absorption (TPA).

When operating in the near-IR telecom band, this nonlinear process limits the strength of pump fields when generating photons through spontaneous four-wave mixing. Once again, moving to a material with a wider band gap like SiN eliminates this problem. Alternatively, recent work has demonstrated the advantages of operating at longer wavelengths within silicon, to heavily suppress TPA, thus reducing nonlinear propagation losses whilst retaining the full toolbox of silicon photonic components [5]. Although supporting infrastructure for longer wavelength operation is still required. A variety of approaches for reduced loss in/out-coupling to photonic chips exists, but many specialised processes are not currently available in commercial foundries [6].

*Detectors and Cryogenic Operation* - Superconducting nanowire detectors are increasingly being integrated with optical circuitry [7] (also, see Sections 19 and 20 of the roadmap). Associated challenges are pump field rejection from photon generation and co-integration of cryogenic amplifiers and logical electronics for processing detector signals: silicon electronics becoming susceptible to carrier freeze-out that alters e.g., transistor behaviour. Reading out large arrays of superconducting detectors for processing using room temperature electronics may require many coax cables, increasing heat load from cryostats, but optical readout methods could circumvent this. Integrated arrays of high-efficiency room temperature avalanche photodiodes would allow full operation at ambient temperatures, leveraging one of the greatest advantages photons have over other quantum technology platforms.

*Switching* - Fast, low-loss modulation and switching are critical for many quantum computing protocols and multiplexing strategies for making e.g., probabilistic photon generation near-deterministic. Common switching mechanisms in silicon are carrier injection modulation and thermal phase shifting, but these respectively suffer from phase-dependent losses and limited speed. While silicon lacks a natural second-order nonlinearity, it is possible to induce one by applying DC fields that enables fast, cryogenically compatible switching [8]. An alternative approach is to integrate materials capable of supporting fast electro-optic switching such as lithium niobate or barium titanate, and engineer high efficiency coupling to silicon [9,10].

*Interfacing Photonics and Electronics* - The ability to drive and process electrical signals to manipulate and to measure light on a chip motivates better integration of electronics and photonics. Recent work on high-bandwidth detection of squeezed light using a silicon photonic integrated chip wire-bonded directly to a silicon electronics die demonstrated the importance of leveraging the best performing photonic and electronics components, even if they're not available from the same foundry process [11]. By bringing the components close together, the total device capacitance in [11] was curtailed, thus enabling the highest bandwidth of shot-noise limited detector performance yet reported — a tangible advantage to a quantum application brought about by miniaturization and integration. Research into the performance of monolithic electronic-photonic integrated circuits is currently hampered by limited availability of high-performance components across both at commercial foundries and a reduced pool of simultaneous expertise in integrated photonics, high speed, low noise microelectronics and quantum technology.

As quantum processors increase in size and complexity, chip-to-chip connectivity will become critical, not just for exploiting superior component performance in different materials, but also for devices to expand beyond a single wafer. Transmission of quantum information between silicon chips has been demonstrated using optical fibre and sophisticated locking mechanisms for stability [12]. Like approaches for powerful modern computer processors, integrated optical and electronic interposers would provide a more stable, manufacturable and scalable method of quantum and classical communication between different dies.

*Advances in Science and Technology for photonic quantum computing hardware*

A grand ambition for photonic quantum technologies is to bring forth a CMOS-like manufacturable approach for heterogeneous integrated quantum photonic processors. This would enable the promised generations of disruptive quantum technologies, including fault tolerant quantum computing. Some immediate and important advancements required to achieve this can be specified.

*Interfacing Integrated Photonics and Functional Components* - Incompatible with the top-down fabrication methods in today's commercial foundries, heterogeneous integration at scale is a key challenge for modern nanofabrication. While it is relatively straightforward to integrate a few atom-like systems in a cavity, it is not understood how to integrate O(10) solid state atoms in cavities that can then be connected to deliver a real system-level advantage for quantum information processing. We must therefore develop methods to interface foundry fabricated photonic devices, as a photonic interconnecting backbone, with solid state systems, which are typically in the form of nanoscale inclusions such as NV centres in nanodiamonds or membranes such as InAs quantum dots.

*Control and Readout* - While there are multiple electronic methods to control to tune, stabilize, and drive the spin transitions of a single emitter in a cavity, the electrical wiring for tens of emitters becomes problematic. Here, quantum photonics is different from other purely electrical quantum platforms like superconducting and spin qubits, as one needs to keep the metallic wires far enough away from the optical cavities to reduce the excess scattering losses, while still being close enough to drive the spins efficiently. Two chip solutions where the photonic circuitry is implemented in one platform and the RF and control circuitry is implemented on another ASIC provide one way to address this problem. Yet technical challenges, mainly related to efficiency, must be addressed for this approach to work reliably at cryogenic temperatures with low thermal power budgets.

*Characterisation* - New rapid characterisation techniques for emitter identification are required. The current gold standard for single emitter identification at a given site requires measuring the second-order intensity correlation ($g(2)$), that can take up to 10 minutes per site (depending on the photon flux). This method does not scale if one is interested in building O(10) emitter systems, which might require going through 100s of sites to find suitable emitters. Developing novel spectroscopic and algorithmic techniques to speed up this process is also critical if hybrid integrated photonics can be implemented at scale.

**Concluding Remarks**

We expect that silicon photonics will remain an important testbed for further scaling of quantum information processing, if the key challenges can be overcome. However, the advances required to deliver a CMOS-like infrastructure for heterogeneous integrated quantum photonics, and delivery of fault tolerant quantum computing, will require significant and coordinated effort across the photonics community.

**Acknowledgements**

The authors thank Lawrence Rosenfeld, Jacob Bulmer, and Patrick Yard for useful discussions and suggestions. J.C.F.M. acknowledges support from European Research Council starting grant ERC-2018-STG 803665. Support from the Engineering and Physical Sciences Research Council (EPSRC) Hub in Quantum Computing and Simulation (EP/T001062/1) is acknowledged. Fellowship support from EPSRC is acknowledged by A.L. (EP/N003470/1).

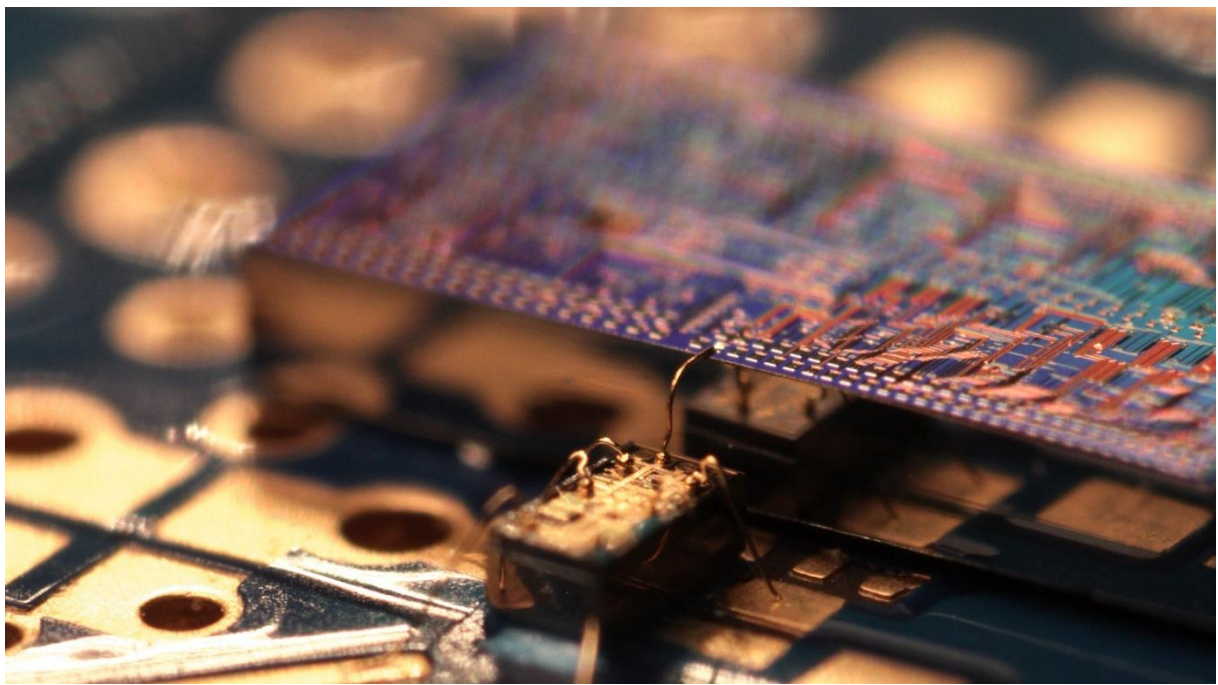

*Figure 1.* An integrated electronics die directly wire-bonded with a silicon photonics integrated circuit reduces total device capacitance, thereby enabling in this instance [11] a high bandwidth homodyne detection for squeezed light detection on-chip. [Photo credit: Joel Tasker, University of Bristol].

## 4 - Heterogeneous Integration Methods with III-V Quantum Dots

Marcelo Davanco, National Institute of Standards and Technology

**Status**

Heterogeneous integration encompasses bonding of two dissimilar materials followed by device fabrication in a single process flow. The approach has been applied in the fabrication of optoelectronic [Bowers2019] and nonlinear photonic [Moody2020] devices on silicon substrates and on-chip waveguides, with gain, absorption and enhanced optical nonlinearities provided by III-V semiconductor materials. While much of the initial work involved bonding at the chip scale, bonding at the wafer scale was later also demonstrated [Liang 2010]. Heterogeneous integration at the chip scale was used by Ben Bakir *et al*. [BenBakir2006] to produce, for the first time, optically pumped lasers based on III-V epitaxial quantum dots on a silicon substrate. Further progress led to demonstration of electrically injected quantum dot lasers on silicon [Tanabe2012]. In 2017, the approach was used at the chip scale to produce $Si_3N_4$-based photonic circuits with integrated single-photon sources based on single InAs quantum dots [Davanco2017], shown in Figs.1(a)-(d). Whereas devices in this work contained multiple, randomly positioned quantum dots, deterministic fabrication of single-photon sources with quantum dots precisely located within GaAs nanowaveguides were demonstrated in 2019 by Schnauber, *et al.* [Schnauber2019], shown in Figs.1(e)-(f). This later work also showed triggered and post-selected indistinguishable single-photon emission into $Si_3N_4$ waveguides, with reported coherence lengths comparable to those obtained from InAs self-assembled quantum dots in GaAs-only devices, under similar optical pumping conditions. The introduction of single quantum dots in quantum silicon photonic circuits has the potential to greatly scale integrated quantum photonic information systems that rely on probabilistic gates, which generally require highly efficient single-photon qubit generation, low-loss waveguide interferometric networks, and highly efficient single-photon detection. On-chip single quantum dots may act as triggered, high-rate sources of indistinguishable photons, and can couple with high efficiency to the silicon photonic circuit through carefully designed GaAs nanophotonic structures. Single quantum dots strongly coupled to on-chip cavities furthermore offer a path towards single-photon nonlinearities, which could enable deterministic quantum gates on chip [Kim2016]. It is worth noting that hybrid integration techniques, in which III-V and silicon-based devices are produced in separate runs and then brought together through methods such as pick-and-place or transfer-printing, have also successfully produced silicon photonic devices with on-chip quantum dot single-photon sources [Elshaari2020].

**Current and Future Challenges**

Deterministic fabrication of on-chip devices with functionality based on single epitaxial quantum dots currently poses a formidable challenge towards scalable integration. Single-dot device functionality requires the creation of an efficient interface between a single quantum dot and a single, efficiently accessible, spatially confined optical mode – e.g., a cavity resonance or waveguide mode. This in turn requires positioning the single dot with sub-wavelength-scale precision at specific locations within the extent of the spatial optical mode, in order to maximize coupling. While site-selective quantum dot growth techniques have allowed such an achievement [Sunner2008], self-assembled quantum dots produced via the Stranski-Krastanow (S-K) growth mode have generally demonstrated superior coherence, a critical characteristic for quantum photonic applications. In contrast with site-controlled dots, self-assembled quantum dots are produced at random positions across the growth wafer surface, and generally present a wide heterogeneity of optical properties such as transition energies, quantum yield and coherence times. Deterministic fabrication of a single S-K dot device, then, requires first that an individual emitter with the desirable optical properties be identified within largely heterogeneous, as-grown ensemble. Subsequently, the photonic geometry that supports the desired, interfacing optical mode must be fabricated around the identified quantum dot. It is worth noting that such challenges exist for any device that derive their functionality from single quantum emitters, in both homogeneous and heterogeneous material platforms, and in hybrid and heterogenous fabrication approaches.

**Advances in Science and Technology to Meet Challenges**

Advances in site-selective growth to allow better control of quantum dot spectral homogeneity and coherence properties would significantly improve integration scalability, by allowing deterministic single quantum dot device fabrication through standard, scalable top-down processing techniques. In the absence of such possibility, scalable integration of single S-K quantum dot devices would greatly benefit from the development of high-throughput single-quantum dot localization and spectroscopy systems. Ideally, such a system would allow fast identification of large number of dots emitting indistinguishable single-photons at desired wavelengths, for integration with nanophotonic geometries. Because S-K quantum dots occur at random locations, however, such an approach would result in randomly located single-dot devices. These devices would then have to be connected to the photonic circuit through *ad hoc* routes, which cannot be produced predictably with optical projection lithography. Hybrid integration techniques can in principle circumvent such issues, since prefabricated quantum dot devices (even if produced at random locations on the origin chip) can be deterministically placed onto pre-determined locations on the destination chip. Scalable integration would, however, require high throughput device transfer techniques, which may be challenging depending particularly on alignment tolerances.

**Concluding Remarks**

In summary, heterogeneous integration has considerable potential for large-scale quantum silicon photonic circuits with functionality based on single III-V semiconductor quantum dots. Although the approach in principle allows completely top-down processing, integration scalability is principally hampered by the necessity to locate and analyse quantum dots at an individual level prior to fabrication. Importantly, though, such non-trivial and time-consuming tasks are generally necessary in any integration platform for devices that employ single quantum emitters.

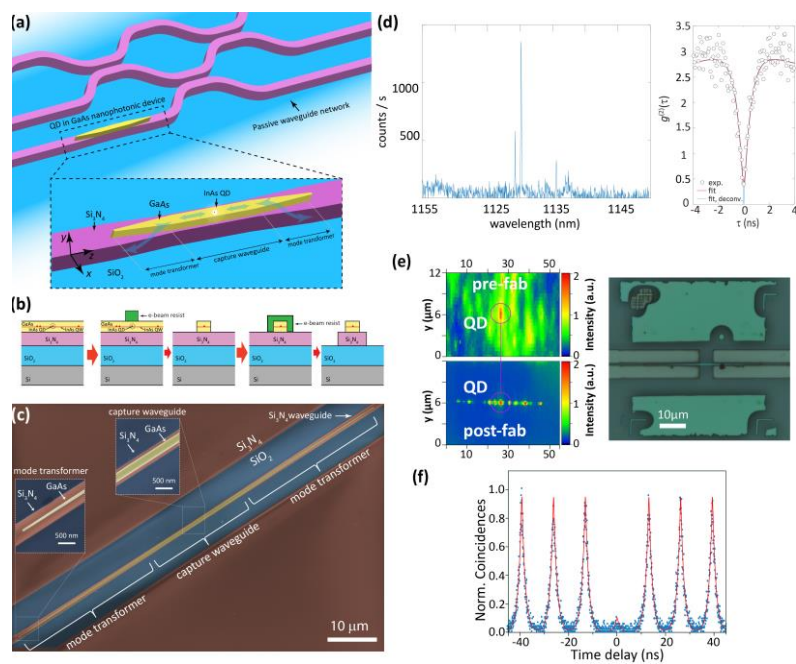

**Figure 1. Heterogenous integration for single quantum dot devices.** (a) Schematic of Si$_3$N$_4$ waveguide-based quantum photonic circuits with directly integrated single-photon sources. As shown in the inset, the sources are GaAs-based nanophotonic geometries (here a straight waveguide with adiabatic mode transformers) that host single quantum dots (QDs), which can be triggered to emit single-photons at a time. (b) Possible process flow for the geometry in (a). (b) False-color scanning electron micrograph of fabricated Si3N4-coupled single-photon source as in (a). (d) Emission spectrum and second-order correlation for a single quantum dot in the device in (c), indicating single-photon emission [Davanco2017]. (e) While in (c) the QDs were randomly located within the GaAs waveguides, in [Schnauber2019] cathodoluminescence (CL) images were used to identify single quantum dots before fabrication (top left). The CL image was used to deterministically produce the waveguide shown on the right-hand side around the located QD with high precision, as evidenced in the post-fab CL map. (f) Second-order correlation of light emission from the QD in (e), where the strong anti-bunching at zero time delay indicated pure, triggered single-photon emission into the Si$_3$N$_4$ waveguide [Schnauber2019].

## 5 - AlGaAsOI Integrated Quantum Photonics

Lin Chang, John E. Bowers, Galan Moody, Department of Electrical and Computer Engineering, University of California, Santa Barbara, CA 93106, USA

**Status**

A primary goal in quantum photonics is the construction of fully integrated and versatile quantum photonic circuits comprising tunable classical and quantum light sources, active and passive components, programmable networks, and detectors [1], [2]. Over the last decade, remarkable progress along this direction has been achieved on the silicon-on-insulator (SOI) platform, which leverages the mature manufacturing infrastructure of the semiconductor industry for high volume and low-cost production. To further extend the capabilities beyond what silicon can offer, a variety of material candidates, such as silicon nitride, lithium niobite ($LiNbO_3$), and III-V semiconductors have been developed and investigated. Among the diverse photonic platforms, (aluminium) gallium arsenide [(Al)GaAs] on insulator has attracted significant interest recently [3], [4] due to its direct bandgap structure for light generation, large bandgap minimizing two-photon absorption at telecom wavelengths, and large $\chi^{(2)}$ and $\chi^{(3)}$ optical nonlinear coefficients, which are orders of magnitude higher than commonly used dielectric photonic media.

Traditionally, III-V photonic platforms suffered from high waveguide loss and low optical confinement because all of the devices were processed on a native III-V substrate [5]. (Al)GaAs-on-insulator [(Al)GaAsOI] overcomes this constraint by heterogeneously integrating an (Al)GaAs film with an oxidized Si substrate by wafer bonding technology (Fig.1a). The high index contrast attained by this approach brings a plethora of new opportunities in nonlinear and quantum applications by enhancing the light intensity and tailoring the waveguide geometry. Another key advance for this platform is the significant reduction in waveguide loss. By combining optimized lithography, etching, and passivation, the propagation loss of AlGaAsOI waveguides (<0.2 dB/cm) is one order of magnitude lower compared to those of previous III-V platforms, which enables optical resonators with a quality factor *Q* beyond 3 million (Fig.1b) [6], on par with many state-of-the-art dielectric material platforms.

Presently, various types of nonlinear devices have been demonstrated on (Al)GaAsOI (Fig.1c-f). Based on the novel direct-phase-matching between TE and TM modes, a record-high normalized second harmonic generation (SHG) efficiency > 40,000% $W^{-1}cm^{-2}$ has been attained in a GaAsOI waveguide (Fig.1d) [7]. For $\chi^{(3)}$-based nonlinear processes, a record-low threshold around 20 µW for Kerr comb generation is achieved in a high-*Q* AlGaAsOI resonator (Fig.1e) [6], where the AlGaAs material bandgap is engineered to avoid two photon absorption at the telecom band. Similar types of waveguides are also used for efficient wavelength conversion in optical signal processing (Fig.1f) [8] with pump power compatible with integrated laser sources.

Such high-efficiency nonlinear processes are essential for multiple quantum purposes. A recent milestone with AlGaAsOI quantum photonics is the demonstration of entangled photon pair generation from a microresonator through spontaneous four-wave-mixing (SFWM) [9]. Combining the high *Q*-factor and the strong Kerr coefficient, a waveguide-integrated source exhibits a pair-generation rate greater than 20 x $10^9$ pairs $sec^{-1}$ $mW^{-2}$ near 1550 nm, with a heralded single photon purity > 99%, entanglement visibility > 97% and coincidence-to-accidental ratio (CAR) > 4300. The brightness of this source (2 x $10^{11}$ pair $sec^{-1}$ $mw^{-2}$ $GHz^{-1}$ bandwidth) is about 1000-fold higher compared to traditional on-chip entangled-pair sources. Such high-efficiency and high-quality quantum light sources in AlGaAsOI points towards exciting prospects for combining state-of-the-art nonlinear components into the quantum regime and can potentially benefit a wide range of system-level applications.

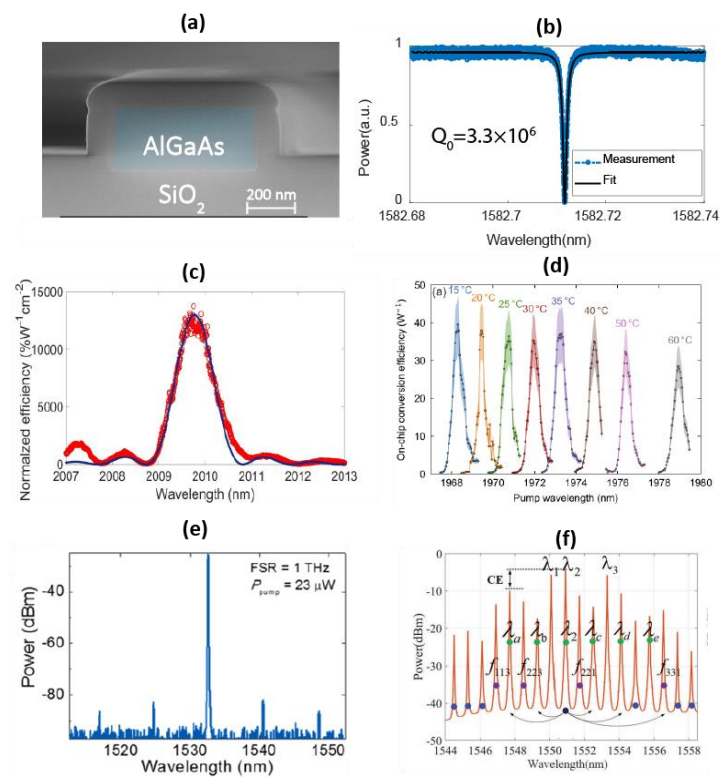

**Figure 1. Demonstrations on (Al)GaAsOI platform.** (a) SEM image of the cross section of an AlGaAsOI waveguide. (b) Transmission spectrum of an AlGaAs resonator. The extracted $Q$ is 3.3 x 10$^6$. (c) The transfer function of SHG normalized efficiency for a GaAsOI waveguide. (d) SHG spectrums under different temperatures by a GaAsOI waveguide, whose normalized efficiency is > 40,000 % W$^{-1}$cm$^{-2}$. (e) The generated frequency comb lines from an AlGaAs resonator under 23 μW power. (f) Frequency conversion for communication experiment by an AlGaAsOI waveguide.

**Current and Future Challenges**

Due to its short history, (Al)GaAsOI requires significant research and development in order to reach a similar level of integration as current quantum photonic circuits on silicon. Compared to the rich library of process design kits based on SOI waveguides, so far very few devices using (Al)GaAsOI have been developed to realize the necessary components in PICs, such as wavelength demultiplexing, Mach-Zehnder interferometers, modulators, and detectors. One challenge that must be overcome to support the growth of (Al)GaAsOI technology is to transition from research development level fabrication to commercial vendors and/or foundries. Another potential challenge for the scalability of this platform is the cost of III-V materials, which is one of the key advantages that silicon photonics holds over traditional native III-V platforms.

**Advances in Science and Technology to Meet Challenges**

Due to the exemplary properties of III-V semiconductors for photonics, plentiful functions can be implemented on the (Al)GaAsOI platform for quantum applications. Besides following the route of SOI in standardizing the basic passive components, (Al)GaAsOI provides several prospective advantages for the monolithic integration of active and passive components. In terms of quantum light sources, the large $\chi^{(2)}$ nonlinearity can lead to entangled and single-photon generation based on spontaneous parametric down-conversion (SPDC), which is expected to be even more efficient compared to SFWM, further easing the pump power requirements. Similarly, producing squeezed optical states—another essential resource for quantum optical information processing in the continuous variable regime—can be realized via either the $\chi^{(2)}$ or $\chi^{(3)}$ nonlinearities. In addition to the nonlinear processes, one key feature of the material system of (Al)GaAs and its alloys is the compatibility with InAs/GaAs quantum dots (QDs), which can be used for generating on-demand single photons and for achieving single-photon nonlinearities.

By leveraging III-V quantum well (QW) or QD epitaxial layers, electrically pumped lasers can be directly integrated along with the passive circuits on (Al)GaAsOI, which can serve as a tunable pump for quantum light generation and an on-chip local oscillator for interferometry. The existence of the electro-optic and piezo-electric effects potentially can enable high-speed modulators with low insertion loss. Importantly, those modulators are suitable to work under cryogenic temperature, which are required for operating QD-based single photon emitters as well as superconducting nanowire single-photon detectors (SNSPDs), which recently has also been successfully integrated on GaAs waveguides [10]. Therefore, a route exists to fully integrated quantum PICs by monolithically integrating all those functions onto (Al)GaAsOI, as shown in Fig.2.

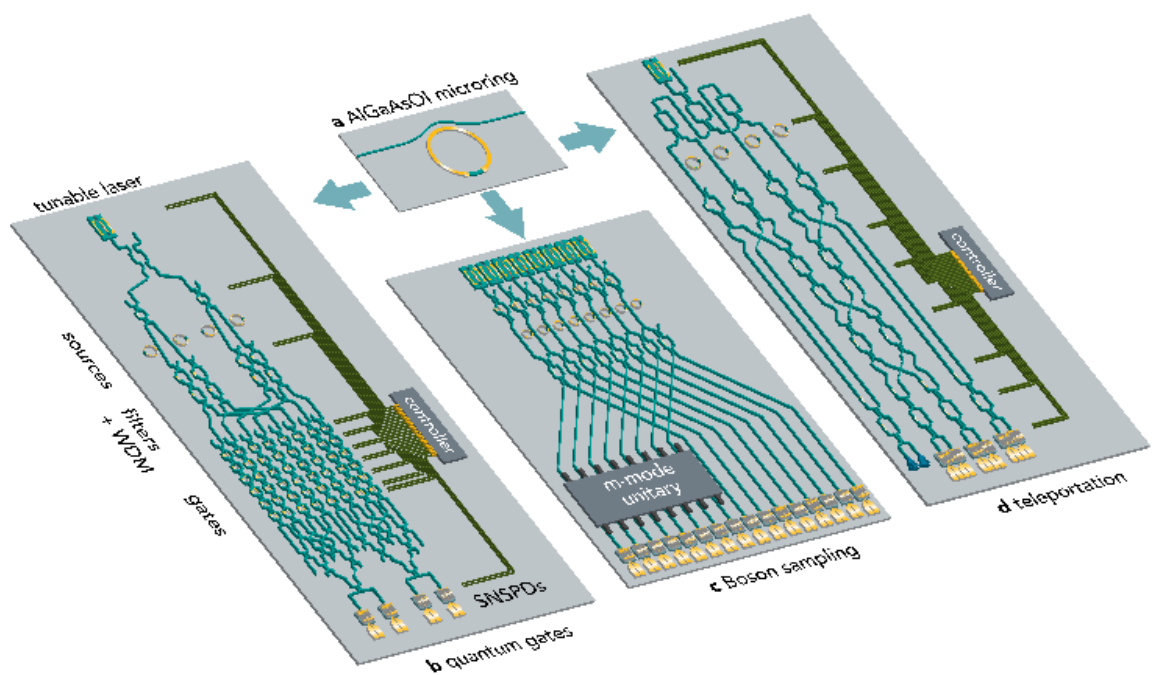

**Figure 2.** Concept of fully integrated quantum photonic chips based on AlGaAsOI.

To further improve the scalability, (Al)GaAsOI can be heterogeneously integrated onto SOI PICs to harness large-scale CMOS compatible production. Another strategy to lower the cost of III-V materials is direct epitaxial growth on Si wafers instead of native III-V substrates. Remarkable advances have been made along this direction for the (Al)GaAs material system, whereby the film qualities are comparable with layers on GaAs native substrates.

**Concluding Remarks**

With the remarkable performance of quantum light sources and the convenience of system-level integration, rapid growth of research in (Al)GaAsOI for integrated quantum photonics is expected. The efficiencies of current nonlinear devices will be further improved, and novel capabilities will be enabled by exploring different properties of the material system. The simultaneous realization of lasers, quantum light sources, modulators, and photodetectors will be the next key steps to accomplish fully integrated quantum photonics with this platform. These developments will also benefit other applications in integrated photonics and it will enable a wide range of new opportunities in classical and quantum computing, communications, and sensing.

**Acknowledgements**

*We gratefully acknowledge support via the UC Santa Barbara NSF Quantum Foundry funded via the Q-AMASE-i program under award DMR-1906325.  G.M. acknowledges support from AFOSR YIP Award No. FA9550-20-1-0150 and the National Science Foundation under award CAREER-2045246.*

## 6 – Integrated Quantum Photonics in Diamond

Niels Quack (EPFL), Christophe Galland (EPFL), Igor Aharonovich (UTS)

**Status**

The unique combination of extraordinary physical and optical properties of high-purity single crystal diamond has propelled this material far beyond its traditional use as gemstone in jewellery, fostering a growing body of scientific research in the field of diamond photonic integrated circuits (PICs) [1]. While PICs made of silicon or III-V materials have reached commercial applications, e.g. in telecommunications and interconnects, *quantum* PICs are still in their infancy [2]. They notably require on-demand generation of indistinguishable single photons, low-loss optical signal routing, on-chip single photon manipulation and detection, and engineering of specific quantum functionalities, such as quantum registers or quantum memories [3]. In this context, diamond's ability to host paramagnetic, optically active atomic defects [4], such as the nitrogen-vacancy colour centre [5], has led to its widespread recognition as a promising material for quantum information processing [6]. Diamond PICs have the potential to host large arrays of long-coherence electronic and nuclear spin qubits linked by flying photonic qubits and coherently controlled by microwave circuits [7]. In addition, diamond's unrivalled mechanical properties makes it promising for cavity quantum optomechanics, which provides additional opportunities for implementing quantum coherent frequency conversion and quantum memories in a same platform [8]. As an almost ideal quantum photonic material, diamond holds the promise to combine in a single platform all the required components for large-scale quantum PICs with the potential to operate at ambient temperature. To date, an impressive collection of experiments has been realized at the individual component level [9]. This progress has been enabled by the advances in chemical vapour deposition (CVD) growth of single crystal diamond with the level of purity and low defect density required for quantum photonics, and by advances in micro- and nanofabrication, allowing for precision shaping of diamond at the nanoscale [10]. This advancement has recently led to several small scale quantum photonic circuit demonstrations, using hybrid integration of diamond with other photonic material platforms [11]. However, despite these recent achievements, large-scale diamond quantum PICs, with thousands or millions of individual components, remain elusive. This roadmap aims at outlining the requirements for establishing such a platform.

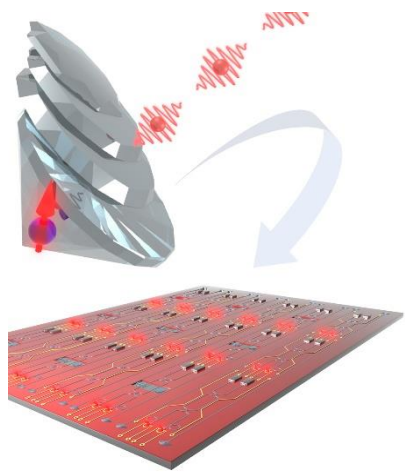

**Figure 1.** The outstanding material properties of single crystal diamond motivate the development of large-scale diamond quantum photonic integrated circuits hosting a wealth of quantum photonic components.

**Current and Future Challenges**

In order to establish mature and scalable diamond quantum PICs, several challenges need to be addressed, which can be divided into four major areas, namely (1) *components*, (2) *platform*, (3) *integration*, and (4) *quantum technology*. Over the past decade, an abundance of individual components for integrated diamond photonics has been investigated in academic research. For example, grating couplers, waveguides, phase shifters, power couplers, cavities, optomechanical resonators, single photon sources, light emitting diodes, Raman laser, supercontinuum generation and detectors have all been demonstrated [1]. These *components* have to be matured and their operation optimized for a selected wavelength range; for scalable quantum operation, their insertion loss has to be drastically reduced; finally, they must be combined into a standardized technology. The component-level challenges go naturally hand in hand with the establishment of a *platform* technology. Building on high quality thin film single crystal diamond as the photonic layer, a manufacturing process flow needs to be developed that can host at the same time passive photonic devices (e.g. couplers, waveguides) as well as active components (e.g. sources, detectors, modulators). The related micro- and nanostructuring techniques need to be transferred from academic settings to foundry-compatible fabrication processes for producing circuits with large quantities of components with high yield. For the construction of large-scale quantum circuits with coherent spin control in thousands to millions of qubits, the most suitable technology approach appears the tight co-integration of electronic integrated control circuitry. While electronic integrated circuits are by themselves a mature technology, this *integration* challenge requires not only the realization of the physical interface between the electronics and the diamond photonic quantum devices, but also adequate circuit control strategies, which is tightly linked to the development of overlaid quantum protocols and algorithms, and the abstraction from the physical layer into logic and software operation. This full stack integration from the quantum effects in the diamond photonic layer through the physical signal routing, the interface with electronic control towards the quantum protocol can be summarized as the *quantum technology* challenge.

**Advances in Science and Technology to Meet Challenges**
The outlined challenges to be met for the establishment of diamond quantum PICs require advancements ranging from scientific developments in the physical layer, significant innovations in manufacturing, to implementation of novel concepts in the abstract logical layer, as schematically represented in Figure 2. At the diamond quantum photonic *component* level, further improvement in the coherent control and single-shot readout of quantum degrees of freedom such as individual spin and photon states are required. Electrically driven single photon sources, low loss passives and on-chip single photon detectors have to be developed.

The closely related *platform* challenge requires significant technology developments. First, advances in material science and engineering are required to deliver high quality single crystal diamond thin films at wafer level. Quantum PICs will require diamond layers with a thickness of a few 100 nm with excellent wafer-level uniformity, low surface roughness, bow and intrinsic stress, as well as excellent crystal quality with low density of defects such as impurities, voids or dislocations. The required innovations in wafer scale growth or layer transfer, polishing, flattening, etching, bonding are far from trivial, and the technology needs to mature in order to achieve required yields. Second, diamond micro- and nanostructuring technologies have to advance, including diamond etch procedures, surface roughness and residual stress control, to achieve excellent dimensional control and photonic performance. Third, embedding optically active spin qubits requires novel approaches in manufacturing, in order to create individual colour centres with few nanometer positioning precision and high yield in a massively parallel way. It would be particularly efficient if future diamond quantum PICs can directly interface with existing telecom infrastructure, such as optical fibers or classical photonic components and integrated circuits, demanding research and engineering on new colour centres or optical frequency conversion.

*Integration* of the diamond photonic quantum components with their control circuits demands innovative approaches for the photonic-electronic interaction and the engineering of microwave circuits. Finally, the full *quantum technology* stack has to be addressed in a holistic manner, including abstraction to higher-level programming strategies and the development of optimized diamond photonic quantum algorithms addressing standardization and providing process design kits for accessibility, as has been proven successful in the electronic integrated circuit industry.

**Concluding Remarks**
Displaying unequalled gleam and internal fire, diamond also exhibits record material properties, that have been thoroughly explored for emerging quantum applications. The unique spin properties of the optically active colour centres in the diamond lattice provide unparalleled prospects for a platform integrating spin, photonic, and possibly phononic qubits. With recent advances in material science, increased availability of high-quality diamond wafers, and progress in micro- and nanofabrication techniques, diamond has now been established as a prime candidate for realisation of integrated quantum photonic circuits with emerging applications in quantum information processing, quantum sensing and single molecule detection. Decades of research generated significant knowledge of the fundamental properties of diamond colour centres and a wealth of optimized nanofabrication protocols. The time is ripe to expedite engagement with industry and foundries to scale these laboratory experiments up to market level.

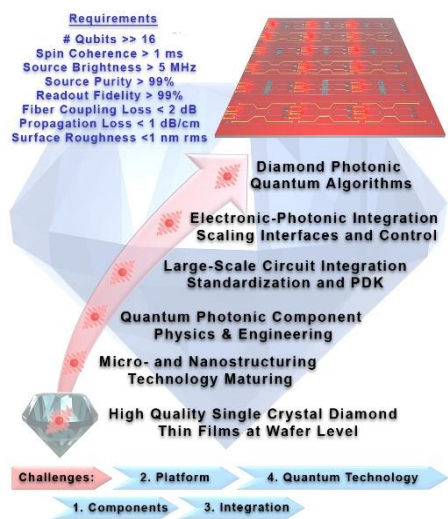

**Figure 2.** Roadmap for the transformation of single crystal diamond into large-scale diamond quantum photonic integrated circuits, involving efforts ranging from the physical diamond thin film up to the abstract quantum algorithm layers, highlighting future requirements and associated research and engineering challenges.

**Acknowledgements**

N.Q. acknowledges funding by the Swiss National Science Foundation under grants 157566 and 183717. C.G. acknowledges funding by the Swiss National Science Foundation under grant 170684. I.A. acknowledges the Australian Research Council (via DP180100077 and CE200100010).

## 7 – Integrated Quantum and Nonlinear Photonics with Tantalum Pentoxide
Martin A. Wolff & Carsten Schuck, University of Münster (Germany)

**Status**

Optical waveguiding in tantalum pentoxide ($Ta_2O_5$, also known as tantala) thin films was investigated as early as the 1970s but it was not until recently that $Ta_2O_5$-on-insulator is emerging as a promising platform for integrated nonlinear and quantum photonics. Ion beam sputter-deposited $Ta_2O_5$ thin films have been produced with excellent optical properties for several decades, enabling most demanding and timely applications from gravitational wave detection to mHz-linewidth lasers. In integrated photonics, low optical absorption and high refractive index contrast of $Ta_2O_5$ waveguides on oxidized silicon wafers have yielded propagation loss as low as 3 dB/m in CMOS compatible fabrication processes [1]. Notably, compact $Ta_2O_5$ micro-ring resonators with several million optical quality factors at telecom wavelengths [2] have already been demonstrated, underlining the great potential of amorphous oxides for matching or even exceeding current benchmarks for all waveguide-integrated devices set with significantly more mature dielectric material platforms.

Combining these achievements with the material's wideband transparency from 300 nm – 8 μm wavelength and a third order nonlinear Kerr coefficient of about three times that of $Si_3N_4$, makes an obvious case for exploiting $Ta_2O_5$ in integrated nonlinear photonics applications. The large bandgap of $Ta_2O_5$ (3.8–5.3 eV) here outweighs the lower nonlinear refractive index as compared to silicon waveguides, which however suffer from two-photon and free-carrier absorption. Initial work has focused on dissipative Kerr-soliton-based optical frequency combs in the IR spectrum [2] (see Fig. 1 a) and supercontinuum generation extending from the IR into the visible range of the spectrum [3] (see Fig. 1 b), which benefit from dispersion engineering in thick $Ta_2O_5$ layers that form without cracks due to the low intrinsic material stress [2]. While these results provide an excellent starting point for integrated nonlinear photonics, $Ta_2O_5$ has several other intriguing material properties that uniquely benefit integrated quantum photonics. Especially interesting in this regard are the record low thermo-optic coefficient [4], enabling stable operation of high quality-factor devices, in particular at cryogenic temperatures, and the extremely low intrinsic photoluminescence, which is an indispensable requirement when working at the single-photon level.

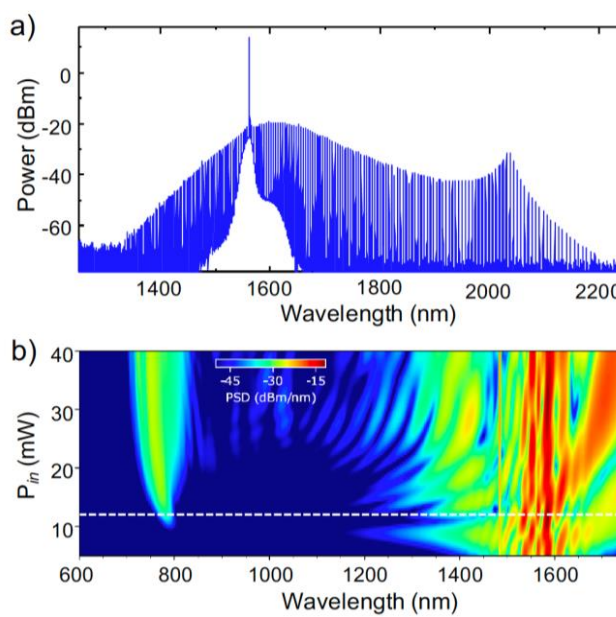

**Figure 1.** a) Single-soliton frequency comb output spectrum from an air-clad $Ta_2O_5$-micro-ring resonator (1.6 μm width, 570 nm height, 46 μm radius). b) Supercontinuum generation in oxide-clad $Ta_2O_5$-waveguides (1.45 μm width, 800 nm height, 5 mm length) as a function of input laser power. Reprinted with permission from [2] and [3] © The Optical Society.

**Current and Future Challenges**

The implementation of $Ta_2O_5$ photonic integrated circuits aims at realizing passive, active, and nonlinear functionalities. Moreover, $Ta_2O_5$ can host rare-earth ions, including erbium [5], which enables use as a gain medium, thus providing optical signal amplification and lasing capabilities on-chip. While ongoing efforts are mainly concerned with transferring established device concepts to the $Ta_2O_5$ material system and optimizing performance at the device level, future photonic integrated circuits will increasingly rely on combining several functionalities in a nanophotonic network. This is particularly relevant to integrated quantum photonics, where both scaling to large system size and detrimental effects from photon loss are outstanding challenges.

Essential building blocks of an integrated quantum technology platform will have to include quantum light sources, nanophotonic circuit components and efficient single-photon detectors. The exceptional material properties of tantalum pentoxide have allowed progress with all of these key components. While the generation of photonic quantum states has remained a challenge as yet, the attractive nonlinear properties of $Ta_2O_5$ offer exciting prospects for producing quantum optical frequency combs [6]. On the other hand, the low intrinsic photoluminescence and optical transparency at visible wavelengths also allows for integrating a wide range of solid-state quantum emitters, such as color centers in diamond [7], with $Ta_2O_5$ nanophotonic circuits (see Fig. 2 a), which had remained elusive for $Si_3N_4$. Moreover, superconducting nanowire single-photon detectors (SNSPD) seamlessly integrate with $Ta_2O_5$ waveguides (see Fig. 2 b) and provide an efficient, low-noise photon counting solution with excellent timing properties [8]. Importantly for realizing reconfigurable quantum photonic information processing systems, active and passive nanophotonic circuit components have been demonstrated with single-mode $Ta_2O_5$ waveguides [9]. Directional couplers and multi-mode interference devices with tunable splitting ratios as well as electrostatically actuated phase shifters, as those shown in Fig. 2 c), show suitability for controlling interference throughout complex, programmable photonic integrated circuits. Lastly, quantum communication scenarios between remote devices will require efficient optical interconnects, as those in Fig. 2 c) & d). Going from the device level to large system size, however, will require minimizing optical insertion and propagation losses across the entire network and accommodating heterogeneous processing techniques.

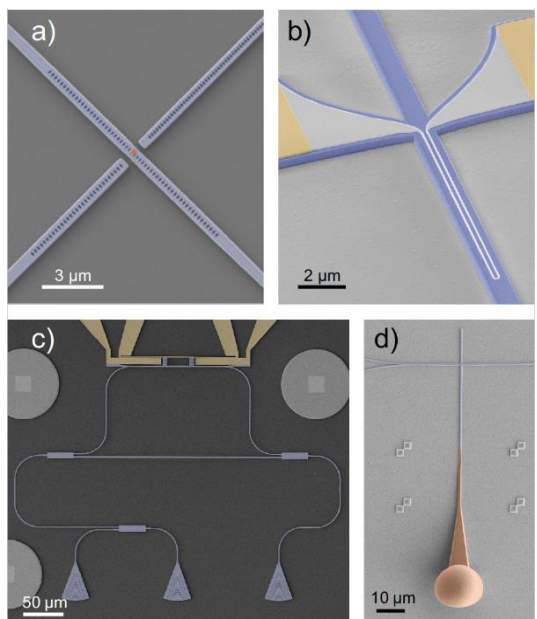

Figure 2. Nanophotonic circuit components for $Ta_2O_5$-on-insulator: a) Nanodiamonds (orange) containing quantum emitters coupled to 1D-photonic crystal cavities (blue) in $Ta_2O_5$-waveguides for optical excitation and single-photon collection. b) $Ta_2O_5$-waveguide-integrated superconducting nanowire single photon detector (white). c) Opto-electromechanical phase shifter (top) embedded in a Mach-Zehnder interferometer formed by multimode interference devices (center) accessible from optical fibers via grating couplers (bottom). d) Total internal reflection coupler (orange) produced in 3D direct laser writing, providing an efficient optical interface to $Ta_2O_5$-waveguides (blue).

**Advances in Science and Technology to Meet Challenges**

Leveraging the full potential of $Ta_2O_5$ photonic integrated circuits will require more detailed investigation of optical loss contributions from material absorption, impurities and inhomogeneities as well as the influence of design and thin-film processing techniques on waveguide scattering at both IR and visible wavelengths. Importantly for nonlinear optical processes, low loss should be achieved for strongly confined optical modes and $Ta_2O_5$ thin films of single crystal quality may here offer new perspectives. However, loss studies must also consider cladding materials for which new approaches to minimizing optical absorption may be required, e.g., deposition of deuterated $SiO_2$-claddings with reduced H-content, which may allow circumventing restrictions arising from the need for annealing $Ta_2O_5$ at approx. 600°C for enabling low loss waveguiding.

The longer-term potential of $Ta_2O_5$-photonic integrated circuits for quantum photonics will to a certain extent rely on the possibility of creating entangled photon pairs as a key resource across all disciplines of quantum technology directly on-chip. Spontaneous four-wave mixing has already been shown in $Ta_2O_5$-waveguides but nonclassical correlations have yet to be demonstrated. Similar to (quantum) frequency comb and supercontinuum generation, dispersion engineering will play a crucial role for exploring phase-matching conditions that strike an optimal balance between nonlinear parametric gain and loss.

In the short term, the range of functionalities currently available with $Ta_2O_5$-nanophotonic devices needs to be expanded considerably. On the one hand, the $Ta_2O_5$-toolbox of active and passive circuit components needs additions from established device concepts such as efficient spectral filters, e.g. for separating pump light from a desired single-photon signal. On the other hand, heterogeneous integration of quantum emitters, active devices and detectors, including operation under cryogenic conditions and from UV to IR wavelengths, will be necessary to enable the full range of quantum technology applications.

In the longer term, these functionalities need to be combined in larger numbers within chip-scale nanophotonic networks. Tantalum pentoxide has proven its suitability in semiconductor industry processes but in order to fully exploit the attractive material properties for realizing complex integrated quantum photonic circuits, a concerted community effort will be required for making foundry-type processing capabilities widely available.

**Concluding Remarks**

Tantalum pentoxide offers a combination of material properties that uniquely benefit key functionalities in integrated nonlinear and quantum technology. The fact that extremely low loss performance and a wealth of nonlinear optical effects could be demonstrated at an early stage of developing this material system for nanophotonic applications speaks to the tremendous potential for going beyond state-of-the-art photonic integrated circuit applications. It is particularly noteworthy, that key building blocks of integrated quantum technology are already available on $Ta_2O_5$-on-insulator chips with much room for future improvement following both intrinsic and hybrid material approaches. The prospect of realizing compact devices with Q-factors reaching billions, low thermo-refractive noise, operation from UV to IR wavelengths and excellent power handling capabilities may suggest that $Ta_2O_5$ can replace $Si_3N_4$ over a wide range of applications [10], however much work remains to be done to turn such visions into reality.

**Acknowledgements**

We thank Philip Schrinner, Thomas Grottke and Helge Gehring for their support and discussions on nanophotonic device designs. C.S. acknowledges support from the Ministry for Culture and Science of North Rhine-Westphalia (421-8.03.03.02–130428).

## 8 – Quantum Photonics with Thin-Film Lithium Niobate

Neil Sinclair[1,2] and Marko Lončar[1]

[1]John A. Paulson School of Engineering and Applied Sciences, Harvard University, Cambridge, Massachusetts 02138, USA

[2]Division of Physics, Mathematics and Astronomy, and Alliance for Quantum Technologies (AQT), California Institute of Technology, Pasadena, California 91125, USA

**Status**

Optical photons have many attractive properties for realization of quantum technologies [1]: they exist under ambient conditions, are generally impervious to environmental noise, and, to an extent, can be generated, manipulated and detected easily. Since they can also travel long distances without significant loss, individual photons are well-suited for quantum key distribution, which aims to secure messages between distant parties using quantum uncertainty. Yet, these properties of photons also introduce challenges to realize quantum technologies that require deterministic interactions between individual photons, e.g. for photonic quantum information processing.

Integrated photonics will play a crucial role in realizing long- (e.g. worldwide), medium- (e.g. metropolitan- or room-sized), and short-range (e.g. inter- or intra-chip) quantum networks. However, the performance of a photonics platform for quantum technology applications needs to be much better than, and in some ways different from, what is required for classical applications. For example, a quantum photonics platform needs to: (i) be ultra-low loss in order to preserve fragile quantum states; (ii) enable precise control of the temporal and spectral profiles of photons; (iii) allow fast and low-loss optical switches to route quantum information; (iv) be able to operate in visible and telecom wavelengths, where many single-photon sources and quantum memories operate, and low-loss optical fibers exist, respectively; (v) feature strong nonlinearities for efficient frequency up- and down-conversion, quantum transduction, and entangled photon pair generation; (vi) allow integration of photodetectors and operating electronics. Silicon and silicon-nitride, the leading integrated photonic platforms, do not meet these requirements due to the lack of a second-order nonlinearity, which restricts their functionality [1]. Although this could be addressed with crystal modification or heterogeneous integration, it remains to be seen what trade-offs, e.g. in terms of efficiency and scalability, this would encompass.

Thin film Lithium niobate (TFLN) has emerged as a promising quantum photonic platform. LN is transparent to optical photons (band gap of ~4 eV), possesses a strong electro-optic (EO) effect, allowing the phase of light to be rapidly varied using microwaves, and has a high second-order optical nonlinearity that can be engineered through ferroelectric domain modulation (i.e. periodic poling) [2]. Importantly, 4" and 6" TFLN wafers have recently become commercially available, which have stimulated interest in this exciting material platform.

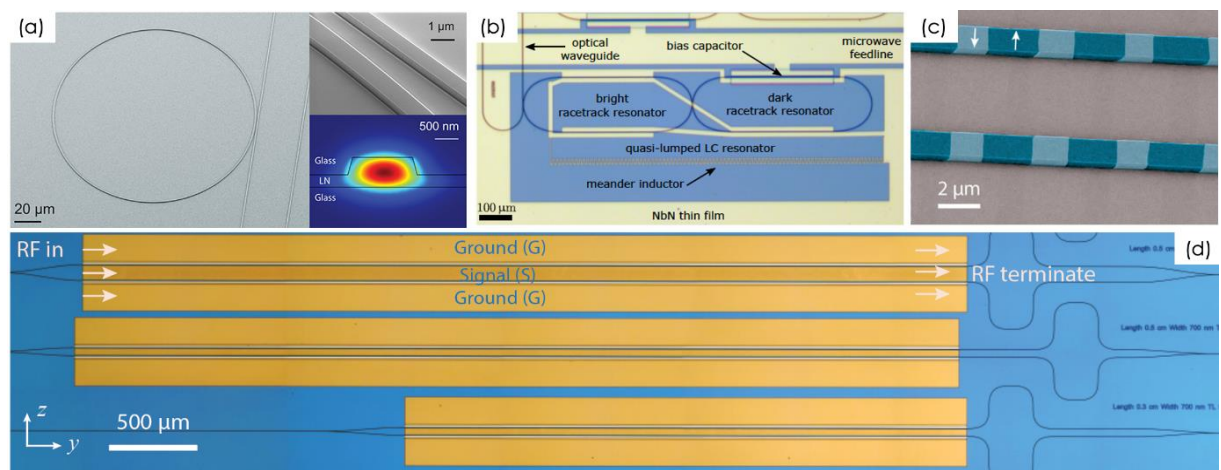

**Fig 1. Integrated TFLN Photonics.** (a) ultra-low loss (3 dB/m) optical waveguides and high-Q resonators (Q ~ 10,000,000). (b) Microwave-to-optical transducer. (c) Periodically poled TFLN frequency converter. (d) Efficient and wideband EO modulator. Reprinted with permission from [3], [9], [5], and [4], respectively.

**Current and Future Challenges**

Breakthroughs in TFLN nanofabrication [3] have enabled the creation of opto-electronic components with performance that surpasses those based on bulk LN, laying the groundwork for developing state-of-the-art photonic quantum processors. These components (Fig. 1) range from high-bandwidth EO modulators [4], which can enable rapid encoding of qubit states onto light or low-latency routers, to poled waveguides and resonators that convert the wavelength of a photon [5, 6] as well as generate non-classical states of light. These devices are particularly appealing for frequency-domain photonic processing applications that leverage wide bandwidth of optical photons but requires high speed modulators as well as frequency shifters/ beam splitters. The latter could be realized using an EO photonic molecule [7, 8], formed by coupling two electrically-driven resonators

Among several challenges, one common to all on-chip photonic platforms is propagation loss. TFLN has demonstrated ultra-low optical loss of less than 3 dB/m and on/off a chip coupling loss of 1.7 dB/facet using TFLN waveguide tapering [10]. The insertion loss of active EO devices must be considered, often taking into account trade-offs between efficiency and loss when designing the metal electrodes for example.

The bandwidths of TFLN EO modulators already exceed 100 GHz and is thus sufficient for many quantum applications. However, further reduction in drive voltages is important, especially when considering realization of multiplexed and dynamic switch networks and gates. These must also be interfaced with large-scale microwave circuitry for precision control and synchronization.

TFLN allows EO control of optical photons without any additional uncorrelated optical (noise) photons being induced by the (low-energy) microwave field. However, optical frequency conversion is accompanied by additional noise photons, e.g. due to Raman processes, which must be suppressed for single photon-level operation. Such conversion is important for interfacing telecommunication-wavelength photons with visible-wavelength devices, e.g. Si photodiodes and quantum memories. For that matter, TFLN circuits which operate at wavelengths out of the infrared must be further developed beyond that shown in proof-of-principle demonstrations. This would also benefit quantum state generation by way of nonlinear processes such as spontaneous parametric down conversion, which can convert a visible photon into two telecommunication wavelength photons, and squeezing.

Heterogeneous integration of devices with TFLN, an exciting prospect to realize high-bandwidth photonic interconnects between hybrid systems, is still in its infancy. Hybrid systems include acousto-optics, single photon detectors, or atomic interfaces on TFLN. The latter could be quantum dots, color centers, or rare-earth ions, of which could act as single photon sources or mediate photonic gates. Single photon detectors on TFLN must be improved beyond that shown in initial demonstrations, e.g. with respect to efficiency (46%) and timing resolution (32 ps) [11], while also sourcing feed-forward signals for control of later operations.

Finally, full integration of all functionalities must be realized (Fig. 2). This is especially challenging for quantum circuits that require high-contrast interference, meaning no distinguishing information about a photon may be introduced by the devices. Chips must be robust to environmental disturbances, e.g. heat generated at electrodes, and operate over long periods of time in order to allow statistically relevant measurements. The latter could be problematic due to photorefraction in LN [2], which is a change of refractive index with high laser intensities, e.g. during optical frequency conversion.

**Advances in Science and Technology to Meet Challenges**

Along with device demonstrations and improvements, photonic loss must be reduced to the levels of the material limit, which is 0.1 dB/m for LN. This could allow optical nonlinearities at the single photon level inside ultra-high-Q TFLN resonators. Large waveguides will help in this regard by ensuring the optical mode does not interact with the rough sidewalls that are introduced by etching, but this will reduce the interaction strength with LN, and hence reduce device performance. Sidewall loss can be mitigated with further optimization at the etch step, e.g. using purer chamber gases or improved removal of re-deposited material during dry etching. The role of surrounding materials must be carefully examined, including absorption or surface chemistry from deposited oxides. Further, imperfections from the TFLN production process, such as surface roughness or impurities can be addressed by polishing and annealing. Material improvement itself must be investigated, including stoichiometric or doped TFLN, including magnesium doping which will reduce the impact of photorefraction. These improvements must be consistent across the wafer and be repeatable.

Other fundamentals, such as improving the electro-optic bandwidth of modulators must be undertaken. For resonator-based modulators, this bandwidth is restricted to tens of GHz by the width of the optical resonance and resistance-capacitance limit of the electrode. Traveling-wave modulators allow bandwidths into the hundreds of GHz using optimized velocity and impedance matching between microwave and light. The role of piezoelectric loss should also be accounted for. Among other improvements, poling precision, uniformity, repeatability, and scaling must also be addressed. Shortfalls of TFLN must be overcome with suitable heterogeneous integration, e.g. integration of diamond color centers or by leveraging microwave-frequency quantum acoustics.

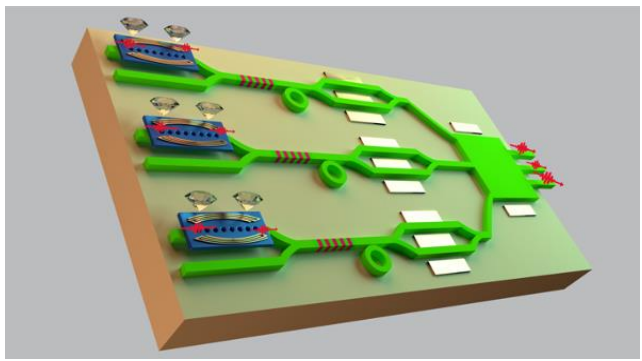

**Fig. 2**. Multiplexed quantum repeater node in LN that combines diamond quantum memories with low-loss nonlinear photonics.

**Concluding Remarks**

We are currently only beginning to see the potential that TFLN can provide to develop quantum photonics. Further steps must be taken to realize advanced and fully integrated quantum functionalities at scale. This will critically advance emerging quantum technologies, in particular quantum networks, and will result in improvements to devices used for classical photonics and telecommunications. Going forward, we expect light to play a significant role in quantum technology due to its ability to carry large amounts of frequency-multiplexed qubit information in a single spatial mode, which will allow the redundancy needed to overcome system loss and errors with a lower scaling overhead. Thus, focus should be placed on creating and improving light-matter qubit interfaces, a task well-suited to TFLN.

**Acknowledgements**
N.S. and M.L. thank Bart Machielse and Ben Pingault for manuscript comments, Boris Desiatov for the illustration shown in Figure 2, and the rest of the Lončar group for discussions. N.S. acknowledges support by the NSERC, INQNET research program, and by the DOE/HEP QuantISED and QCCFP programs. M.L. acknowledges support by NSF, DARPA, ARO, AFRL, ONR, AFOSR and DOE.

## QUANTUM AND CLASSICAL LIGHT SOURCES AND QUBITS

## 9 - Heterogeneous Integration of High-Performance Lasers for Quantum Applications

Tin Komljenovic, Nexus Photonics, Goleta, CA, 93117, USA; David Weld, John Bowers, University of California, Santa Barbara, CA, 93106, USA

**Status**

Quantum science is a field at a technological inflection point. Recent advances in control and readout of quantum bits have underscored the spectacular promise of quantum information processing [1], [2]. However, this promise will only be realized if the current laboratory successes mature into a fully integrated approach. In the development of classical information processing, the transistor represented a crucial advance, but equally important (and meriting a separate Nobel prize) were techniques for assembly and integration of heterostructures, leading to the integrated circuits which pervade our world today. Synthesis and assembly of integrated quantum platforms is a grand challenge which will require contributions from a diverse range of fields. Photonics will play a key role in this effort, thanks not only to the maturity and power of photon manipulation technologies but also to the unique quantum mechanical properties of the photon.

Visible light photonics is an area of particular importance, due to the typical energy scale of transitions in the atoms and ions that comprise a key element of quantum technology (see Fig. 1). An atom is the paradigmatic quantum mechanical system. The precise tools of atomic spectroscopy that gave birth to quantum mechanics have evolved into tools of control, allowing experimental access to all internal and external degrees of freedom and enabling, for example, the realization of states of matter in which millions of atoms occupy a single quantum state at nanokelvin temperatures. The precision and control available in such atomic physics experiments makes them a promising platform for quantum information processing and also an ideal tool for quantum-enhanced sensing, navigation, and timekeeping.

A key bottleneck in the development of atom-based quantum technology is the complexity of laser sources and associated optical hardware. Modern atomic physics began with the invention of the laser and has always depended upon it. As experiments grow more complex, involving for example molecules or multiple independently addressed atomic elements, the need for individually controlled laser sources at numerous distinct frequencies has limited scalability and restricted commercial applications. Integrated photonics will overcome this challenge and realize the potential of quantum technology [3], [4].

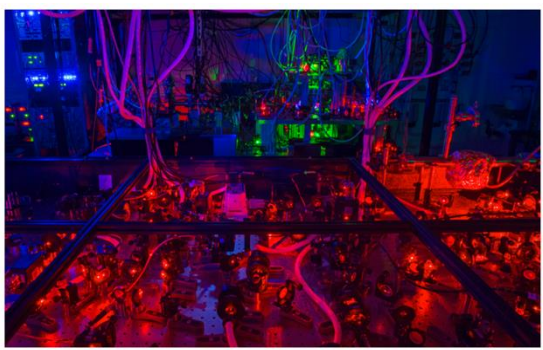

| Atom | Transition | Wavelength | Saturation Intensity | Linewidth |
|---|---|---|---|---|
| Sr | $(5s^2)^1S_0$-$(5s5p)^1P_1$ | 461 nm | ~40 mW/cm$^2$ | 32 MHz |
| Sr | $(5s^2)^1S_0$-$(5s5p)^3P_1$ | 689 nm | ~0.003 mW/cm$^2$ | 7 KHz |
| Li | $2^2S_{1/2}$-$2^2P_{3/2}$ | 671 nm | ~2 mW/cm$^2$ | 6 MHz |
| K | $4^2S_{1/2}$-$4^2P_{3/2}$ | 767 nm | ~2 mW/cm$^2$ | 6 MHz |
| Cs | $6^2S_{1/2}$-$6^2P_{3/2}$ | 852 nm | ~2 mW/cm$^2$ | 5 MHz |
| Rb | $5^2S_{1/2}$-$5^2P_{3/2}$ | 780 nm | ~2 mW/cm$^2$ | 6 MHz |

**Figure 1.** (left) Atom-based experiments in quantum science like this lithium quantum gas apparatus typically utilize benchtop equipment and free-space optics which dominate the cost and complexity and restrict scalability. (right) Properties of some atomic transitions of general interest.

**Current and Future Challenges**

A central challenge to wider deployment of quantum technology-based sensing systems is the need for tunable high-performance narrow-band lasers at a variety of frequencies which are used to cool, trap and manipulate atoms. Semiconductor lasers, due to their inherent size, weight, power and cost (SWaP-C) advantage are the preferred choice, but it has been challenging to make native lasers based on GaN or GaAs meet the requirements at atomically relevant wavelengths. The challenge is mostly due to a need for low-loss waveguides to form the laser cavity and enable the narrow-linewidth. The communications industry (historically mostly using InP based semiconductor lasers operating between 1.3 and 1.6 µm) had similar challenges regarding laser linewidth with the transition to advanced modulation formats, a challenge that to a large degree was addressed by combining multiple materials (e.g. InP for gain and Si for waveguide/resonator), either by hybrid integration or with heterogeneous integration. In both cases low-loss silicon waveguides reduce phase noise by 40 dB and enable linewidths in the range of 100 Hz while providing >100 nm of tuning range [5]. The heterogeneous integration approach brings many benefits including wafer scale processing and testing for improved uniformity and cost reduction, and is the path that many companies are pursing in Datacom markets e.g. Intel, Juniper and HPE[6]. Ideally such approach would be scaled down to wavelengths of interest for atom manipulation. Unfortunately, silicon waveguides cannot support wavelengths below 1200 nm so new materials have to be introduced. The leading candidates are silicon-nitride (SiN), tantalum pentoxide ($Ta_2O_5$), alumina oxide ($Al_2O_3$), aluminium nitride (AlN), lithium niobate ($LiNbO_3$) or similar high-bandgap materials providing good optical performance [7]. A second challenge is related to providing either direct generation active material supporting the wavelength range of interest or using non-linear materials to convert sources in more established wavelength ranges (1.2-1.6 µm) using various non-linear processes. Direct generation at the short end of the wavelengths (UV to green) can be supported with GaN based active devices, while wavelengths longer than ~630 nm can be supported with GaAs based active devices. There is a partial gap for direct generation between wavelengths supported by GaN and GaAs based active devices, that is currently an active area of research with recent demonstrations showing up to 33% wall-plug efficiency LEDs at 565 nm [8]. In the case of non-linear generation, if both frequency doubling and frequency tripling can be used, then telecommunication sources can be used to cover most of the wavelengths of interest. The challenge is then pushed to the design of non-linear elements to provide good phase matching and address other design constraints for better conversion efficiency. Very high on-chip efficiencies were demonstrated in high-contrast heterogeneous semiconductor platforms [9], although there are challenges in pushing such approach in UV to green range where the use of higher bandgap materials might be needed.

**Advances in Science and Technology to Meet Challenges**

Depending on the approach, multiple advances are needed to enable chip-scale optical sources meeting the requirements of next generation quantum technology-based sensing systems. Improvements in active materials (e.g. direct gain in green-yellow range) and passive materials (low loss, especially in the UV-blue range) can enable path for wafer-scale integration in full wavelength range. In the case of non-linear generation, improvements in heterogeneous integration to enable higher contrast waveguide geometries can increase efficiencies by reducing the mode volume. At the same time, processing has to be improved as phase matching generally requires very precise control of waveguide dimensions. The use of multiple materials can place additional restrictions on the processing thermal budget due to different coefficients of linear thermal expansion, which can be especially important when defining metal contacts. Once material related limitations are addressed, optical coupling between various materials has to be optimized. The main challenge is to facilitate efficient optical coupling between dielectric materials with refractive indices generally in the range of 1.45 to 2.2, and semiconductor materials that can provide gain and/or efficient non-linear conversion but typically have refractive indices in the range of 2.3 to 3.6. Such large difference in refractive indices, in case direct adiabatic power transfer is designed, requires narrow taper tips.

Many of said challenges are being actively addressed with recent demonstration of heterogeneous integration of GaAs lasers on SiN waveguides [10] where efficient coupling between materials with refractive index larger than 1.5 was demonstrated and is shown in Fig. 2.

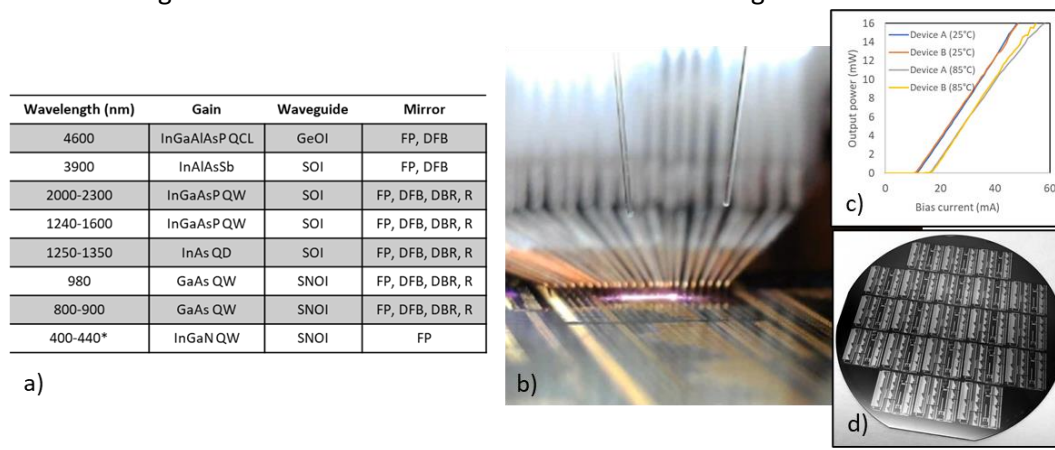

**Figure 2.** (a) Summary of laser technologies on silicon (GeOI – germanium on insulator, SOI – silicon on insulator, SNOI – silicon-nitride on insulator, R stands for one or more rings), (b) Automated wafer scale testing of lasers captured with operation Si camera showing light emission below Si bandgap, (c) Exemplary LIV curves showing very good laser performance (the wavelength of is 980 nm) and (d) Heterogeneous GaAs-on-SiN wafer with multiple active components including lasers, amplifiers, photodetectors and tuners.

**Concluding Remarks**

Photonic Integrated Circuits (PICs) are emerging as low-cost replacements for bulk and fiber-optic systems, where integration allows for a very large improvement in SWaP-C. First commercial PIC-based systems targeted telecom and datacom markets, but more recently there has been significant push to realize complex PIC-based sensors to leverage said SWaP-C benefits, especially in LIDAR space and now quantum applications. There are numerous tailwinds to accelerate said transition; primarily the increased activity in PICs in general leading to improved photonic fabrication capabilities in modern foundries that have transitioned to larger scale wafers and higher quality tools. Additional materials and processes will have to be developed to provide chip-size PICs supporting relevant wavelengths for atom cooling, trapping and manipulation.

**Acknowledgements**

*The authors thank Galan Moody, Steve DenBaars, Lin Chang, Weiqiang Xue, Chao Xiang, Hyundai Park, Chong Zhang and Minh Tran for useful discussions.*

## 10 - Chip-Scale Entangled-Photon Sources

Shayan Mookherjea, University of California, San Diego

**Status**

Entanglement is a useful resource in quantum networks. Entangled-photon sources play a crucial role in realizing such networks, similar to lasers in classical data networks. Similar devices can also be used to generate heralded single-photon states and squeezed light. Integrated photonics can generate and manipulate quantum states of light in a scalable and efficient way[1]–[4]. Entangled photons can be generated from quantum dots without fine-structure splitting, as well as from optically-pumped nonlinear devices which use spontaneous parametric down conversion (SPDC) or spontaneous four-wave mixing (SFWM). Figure 1 shows some of the popular photonic materials choices and devices formed using them towards the major goals of current research in chip-scale entangled-photon sources. In-depth progress reviews are presented elsewhere[5]–[8]. Here, we will mainly focus on SPDC and SFWM devices which can be easily realized at the micro-chip scale using simple dielectric or semiconductor materials, and the devices do not require cryogenic cooling.

When pumped with continuous-wave light, the signal and idler photons are entangled in a time-energy basis, if the joint spectral intensity (JSI) of the signal and idler photons is not separable. To realize a heralded source of pure, indistinguishable single photons, a separable JSI can be engineered through device design and adjusting the properties of the pump pulse. When the device is pumped with a sequence of closely-spaced short pulses, entanglement in a time-bin basis is created. By using various waveguide components such as beam splitters, polarization rotators and frequency shifters, entanglement can be created in the spatial, polarization, mode and frequency domains as well. Higher-order entanglement (dimension d>2) allows for more information (qudits) to be transported by photons, which are inevitably lost due to absorption, scattering or diffraction during propagation [9]. Since scalability of on-demand entangled-photon sources has not yet been realized, current practice relies on multiplexing probabilistic pair sources through linear optical gates and post-selection[10], [11], but other schemes have been proposed[12].

Recent advances in device fabrication techniques have led to the ability to lithographically create highly-confined waveguides and compact micro-resonators, which enables precise dispersion engineering[13]. For quasi-phase matching through periodically-poled thin-film materials, a combination of electron-beam lithography and optical lithography has been used to precisely define electrode structures for short-period poling[14], which has benefits in classical nonlinear optics as well. Table 1 compares two approaches for room-temperature entangled photon sources, those based on the $\chi^{(2)}$ optical nonlinearity and those using $\chi^{(3)}$, from an integrated photonics perspective.

Although the $\chi^{(2)}$ nonlinearity used in spontaneous parametric down-conversion (SPDC) is absent in un-strained silicon and silicon nitride, the weaker $\chi^{(3)}$-based spontaneous four-wave mixing (SFWM) nonlinearity can be enhanced by using micro-resonator and coupled-microresonator devices [15]–[17]. To achieve entanglement generation rates that are comparable to $\chi^{(2)}$ devices and comparable quantum dot technology, it is critical to achieve a high quality factor $Q_{loaded}$ >$10^5$ since the pair generation rate scales with the third power (approximately) of the quality factor, while also supporting a high pump power. The pump and entangled photon pairs can be placed within the telecommunications wavelength range, where low-noise pump laser diodes and filters are readily available. A dual-pump SFWM process can be used to generate frequency-degenerate signal and idler photons, with some additional considerations for noise management [18]. In contrast to fiber or silicon nitride sources, the Raman noise contribution in Si SFWM is narrowband, and spectrally well separated from the pump. Silicon microelectronic components can be incorporated in, or used with, silicon photonic structures, such as p-i-n diodes fabricated across the waveguide cross-section for resonance monitoring[19], and step-recovery diodes which generate short pulses using a continuous-wave RF input, and can drive an electro-optic modulator (EOM) for shaping optical pump pulses[20]. Out-of-band laser light transmission can be used for stabilization[11]. To help realize pure heralded single-photon states, different waveguide-resonator coupling coefficients are needed for the pumps, and for the signal or idler photons. Since the wavelengths are similar in SFWM, a more sophisticated coupling structure may be useful, such as the two-point interferometric couplers[21].

| Chip-scale Entangled-Photon Sources | | |
|---|---|---|
| Category | Nonlinear optics: SPDC and SFWM | Artificial atoms: Quantum dots and vacancy centers |
| Materials | Si, $Si_3N_4$, AlGaAs, LN, BBO, KTP, AlN, SiC, $As_2S_3$ | GaAs, InGaAs, InAsP, InAs Diamond, SiC |
| Devices | Waveguides, resonators, external coupling structures, tuning knobs, filters, switches, detectors. Linear gates and nonlinear operations. | |
| Architectures | (Quasi) On-demand heralded single photons High-dimension entanglement Graph, cluster and repeater states | |

**Figure 1.** Overview of chip-scale entangled-photon sources.

While it is possible to induce a weak $\chi^{(2)}$ effect in centrosymmetric materials, it is also possible to bring the stronger $\chi^{(2)}$ effect of SPDC in lithium niobate (LN) into silicon photonics ecosystem, for example, by using thin-film LN (TFLN) and bonding. SPDC requires a short-wavelength pump; however, low noise 775 nm pump diodes are common, although microchip-scale short-pulse lasers at these wavelengths are not yet widely available. Because of the small waveguide cross-sectional area, significant performance benefits can be achieved in comparison to traditional SPDC waveguides in LN[14], [22]. Waveguides in TFLN have a high group-velocity dispersion, and therefore require a shorter quasi-phase matching (QPM) poling period. Traditional SPDC requires ultrashort optical pump pulses generated by a mode-locked laser, which is difficult to integrate monolithically[23]. Although dispersion engineering over wide wavelength spans is challenging, TFLN and other $\chi^{(2)}$ materials can also be used in the micro-resonator configuration, achieving very high nonlinear efficiencies[24].

**Current and Future Challenges**
(A) **Improving the heralding efficiency for pure single-photon states:** The heralding efficiency of spontaneously-pumped integrated sources is low (typically <10%) whereas an on-demand single-photon probability >50% has been achieved using a quantum dot[25]. (A heralding efficiency of 50% has been quoted for a microring SFWM device using a different definition of heralding efficiency[11].) To bridge this gap, several advances will be needed. Accurate fabrication of integrated coupling structures is needed in resonant devices, such that the pump is coupled to the resonator with a different coupling efficiency than the signal/idler photons. Integrated filters and heralding detectors may avoid the insertion loss and experimental complexity of stand-alone modules. At present, only superconducting detectors show an adequately high detection efficiency and low jitter, but their integration with other photonic components can be difficult. High extinction ratio, low insertion loss integrated filters are required to isolate and possibly filter the signal and idler photons and divert the residual pump beam. Because fiber-chip coupling is imperfect, scattered light must be carefully managed.

(B) **Multiplexing:** Since the heralding efficiency of a spontaneous source is fundamentally limited to 25%, multiplexing is essential to improving the heralded single-photon generation probability (per pump pulse)[26]. A variety of multiplexing experiments have been demonstrated on a breadboard or optical tabletop, including frequency multiplexing schemes which require nonlinear wavelength-shifting operations[27]. To realize a microchip-scale, low size, weight and power (SWAP) and low-cost quasi-deterministic source of pure single photons will require engineering improvements in source brightness, indistinguishability, heralding efficiency, and (feed-forward) switching of single photons. Several hundred multiplexed sources might be required, depending on the level of loss and detection efficiency[28]. While this is not impossible to envision this level of integration using integrated photonics, a hybrid approach which uses some integrated photonics circuitry, and some discrete optical components may be more practical.

(C) **Compatibility of sources with quantum memories:** The bandwidths of long-lived quantum memories are typically much narrower than the bandwidths of photons generated by SPDC or SFWM, or for that matter, using quantum dots or other typical single-photon and photon-pair emitters. To bridge this gap, researchers are studying narrow-bandwidth SPDC using backward-wave quasi-phase matching, bandwidth compression through transduction, and construction of ultra-narrow linewidth cavities. Development of short poling periods is also benefitting research in quantum wavelength conversion between widely separated wavelengths. With the incorporation of feedforward logic and/or single-photon nonlinearity, efficient construction of graph states and entanglement purification circuits can also be realized using integrated photonics. These advances will benefit the development of quantum repeaters for interconnects as well as quantum sensing and quantum computing.

**Table 1.** Comparison of photon-pair and heralded single photon generation using silicon and lithium niobate (LN).

| Core material (core thickness) | Si (220 nm) | (thin-film) LN (300 nm) |
|---|---|---|
| nonlinearity | spontaneous four-wave mixing (SFWM) | spontaneous parametric down-conversion (SPDC) |
| type | Micro-resonator | Waveguide with QPM[a] |
| bandwidth | 1.4 GHz ($Q_{loaded} \sim 9.2 \times 10^4$) | 140 GHz |
| footprint | 20 μm x 20 μm | 5 mm x 2 μm |
| CAR x PGR[b] | $0.6 \times 10^9$ pairs.s$^{-1}$ CAR 532 at $1.1 \times 10^6$ pairs.s$^{-1}$ | $7.6 \times 10^9$ pairs.s$^{-1}$ CAR 668 at $1.1 \times 10^7$ pairs.s$^{-1}$ |
| $g^{(2)}_H(0)$[c] | $0.0053 \pm 0.0021$ at $N_H$ = 18 kilo-pairs/s | $0.0022 \pm 0.0004$ at $N_H$ = 15 kilo-pairs/s |
| Heralding (Klyshko) efficiency[d] | 4% | 2% |
| Energy-time entanglement visibility[e] | $98.2 \pm 0.9\%$, $97.1 \pm 0.5\%$ (at PGR = 68 kHz) | $99.3 \pm 1.9\%$, $99.5 \pm 1.8\%$ (at PGR = 1 MHz) |

(a) QPM: quasi-phase matching.
(b) CAR: Coincidences-to-accidentals ratio; PGR: Pair-generation rate. Generally, at high PGR, CAR $\propto$ (PGR)$^{-1}$.
(c) heralded self-autocorrelation at zero time delay; $N_H$ = (off-chip) detected herald rate.
(d) Including all on-chip and off-chip loss contributions, in Refs. [17] (Si) and [14] (LN).
(e) Measured using a Franson-type two-photon interference experiment, in two bases.

**Concluding Remarks**

Entangled photon-pair source devices are now available from research and commercial sources, similar to crystals for other nonlinear optics applications such as second-harmonic generation and wavelength conversion. However, they will not meet the needs for building a future scalable quantum network. Aside from the technical performance limitations, many traditional designs do not generally benefit from modern silicon-based microfabrication, which can realize hundreds or thousands of devices on a wafer. While silicon photonics is obviously (mostly) compatible with silicon microfabrication in industry, other integrated photonics approaches which offer better performance may also be compatible with the silicon industry, including $\chi^{(2)}$ materials such as lithium niobate. A quantum photonics designer can now benefit from a variety of engineering ideas, such as heterogeneous bonding, pick-and-place assembly and transfer printing, post-fabrication trimming and reprogrammable photonic circuits. Together, these advances will comprise a much richer set of tools with which to address the grand challenges of building low SWAP (size, weight, and power), low-cost, manufacturable *and* high-performance entangled-photon sources for large-scale entanglement distribution and error correction in quantum networks and quantum information processing.

**Acknowledgements**

This work was supported by National Science Foundation (NSF) (EFMA-1640968), NASA, Sandia National Laboratories, United Technologies Research Center, Keysight Technologies and IBM.

## 11 - Sources and Qubits - Emissive Centers in Silicon

Dr. Sonia M. Buckley, NIST

**Status**

Similar to the nitrogen vacancy (NV) center in diamond, there exist many different emissive centers in the silicon lattice. These silicon emissive centers (SECs) were characterized extensively in the 1980s [1], and underwent a resurgence in popularity in the 2000s by groups looking to develop silicon lasers for telecommunications applications [2]. While there was one report of the lasing G center, a carbon-based SEC, work on SECs for the development of lasers has largely been abandoned. Many SECs are only luminescent at cryogenic temperatures, which is unappealing to the telecommunications industry, but an accepted part of quantum technologies. SECs have been relatively underexplored for quantum information processing as they must emit below the silicon bandgap, where the most efficient APDs are insensitive. The increased availability of superconducting single-photon detectors has finally allowed the study of quantum SECs.

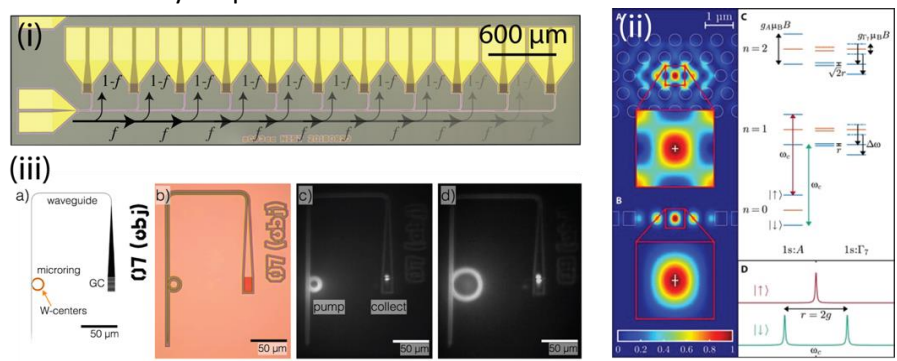

**Figure 1.** (i) Cryogenic optical link between electrically-injected waveguide-coupled W center LED and superconducting nanowire single-photon detectors [1]. (ii) Proposed spin-photon interface in silicon [6]. (iii) Demonstrated coupling of W centers to microresonators in silicon-on-insulator [10].

Many SECs form states within the bandgap of silicon, allowing electrical injection. Several electrically-injected all-silicon LEDs have been demonstrated, including an electrically-injected waveguide-coupled cryogenic optical link with a W center (silicon interstitial based SEC) LED [3] (Fig. 1(i)). The W and G centers exhibit some of the brightest emission of any SECs, and are promising as single-photon sources. In the past year, first measurements of single-photon emission from G centers have been reported [4], [5], and single SECs across the telecom wavelength range have also been observed in carbon-implanted silicon [9]. A silicon photonic spin-donor qubit quantum computing platform has recently been proposed [6], with the idea of using photonic coupling between chalcogenide spin qubits in silicon (Fig. 1(ii)). Spin qubits in silicon have exceptionally long coherence times due to the availability of isotopically pure silicon. Isotopically pure silicon has also been shown to favorably impact the linewidths of some SECs [7]. The spin-photon interface in silicon, could it be realized, would therefore have significant advantages both due to this property and the advanced fabrication capabilities available in silicon. Rare earth SECs, with very stable emission across crystalline hosts, have also been investigated for quantum applications. Rather than using optical emission, which has a very long lifetime, optical addressing of a single Er ion in silicon has been demonstrated, in conjunction with the injection of a single electron spin for manipulation [8].

**Current and Future Challenges**

Generation of SECs is usually performed by ion implantation followed by annealing. For many SECs, the process leaves significant damage in the crystal lattice, causing unintentional introduction of non-radiative recombination centers. The structure of many SECs and these non-radiative centers are also often poorly understood. The emission from single G centers that has been observed so far showed strong inhomogenous broadening. Ion implantation processes may generate many different types of SECs, offering opportunities but also challenges to develop processes that only create the SECs of interest [9]. For many applications, placement of individual SECs will also be necessary. For single-photon sources, there are several remaining challenges. The long radiative lifetimes of SECs lower the quantum efficiency due to competitive non-radiative processes, as well as reducing the maximum repetition rates. Single-photon emission has only recently been observed from G centers and SECs in carbon-implanted silicon, while single-photon emission still remains elusive for many other SECs interest. The most promising SECs may emit in the mid-IR, a technologically challenging wavelength range. From the point of view of an electrically-injected single-photon source, the ability to inject a single electron on-demand and generate indistinguishable single-photons is a holy grail, but work is still needed to develop the types of structures needed for this. The spin-photon properties of several types of SECs have been investigated, including Se, Mg, as well as the G, W and T centers (all based on interstitial silicon complexes either with or without carbon). It appears that the Se+ and T center are both promising as a spin photon interface, while the right energy level structure has not yet been observed in other SECs. A host of SECs remain unexplored, and it is likely that the perfect SEC for the application has yet to be identified. For a spin-photon interface, the emitters must be strongly coupled to a photonic cavity. This requires the development of a system of photonic cavities all on-resonance with the emitter zero phonon line, and with each other.

**Advances in Science and Technology to Meet Challenges**
Despite research in this area ongoing since the 1970s, the optimal conditions for generation of specific SECs, the physics of formation of many of the most common SECs, and the sources of non-radiative recombination are not well understood. Techniques such as atomistic simulations, variable energy positron spectroscopy, and Rutherford backscattering could help elucidate some of these properties. More study on the quantum properties of different SECs is ultimately needed. The main improvements in technology needed are improved detectors in the infrared, and ultra-low-loss integrated photonic elements in the mid-infrared. Coupling of W centers to microresonators has been demonstrated recently [10] (Fig. 1(iii)), but Purcell enhancement and strong coupling of SECs to photonic cavities are still awaiting experimental demonstration, despite indications that this is well within reach of silicon photonic microresonator technology. Purcell enhancement could overcome the long radiative lifetime that is characteristic of many SECs. It is unclear what effect the presence of the etched surfaces introduced by these cavities will have on the performance of SECs, and advances in our understanding of this effect are needed. For a photonic spin-qubit interface, a mechanism for overcoming homogeneous and inhomogeneous broadening of the emitters, such as electrical tuning, is needed, as well as tunable high-Q low mode-volume silicon photonic cavities for matching resonant wavelengths of cavities to each other and to the emitters. Placement of SECs with atomic-level precision will also be needed; this has been demonstrated with spin qubits in silicon based on phosphorus donors, but may be much more challenging for complexed SECs involving multiple atoms. For useful electrically-injected single-photon sources, a first demonstration of electrically-injected single-photon emission is needed, as well as a mechanism for resonant electrical injection for to ensure the generation of indistinguishable photons. The technologies used for single-electron injection and optical addressing of spin qubits in Er ions may prove useful in this case. With major technological developments, there is also the potential for SECs to contribute to quantum technologies in other ways, such as an electrically-injected cryogenic silicon laser to pump on-chip pair sources, or as an optical interface from cryogenic to room temperature.

**Concluding Remarks**

Emitters in silicon show promise for quantum technologies on several fronts. First, they may lead to silicon-compatible electrically-injected single-photon emitters, potentially enabling a wide variety of quantum information experiments. They may provide a spin-photon interface for a system with very long coherence time spin qubits, allowing a mechanism for long distance coupling between qubits. However, for these technologies to become a reality, the right defects must be identified and studied, and the photonic and electrical interfaces needed must be developed.

**Acknowledgements**

## 12 - Integrated Photonics with Silicon Carbide Color Centers

Marina Radulaski, University of California, Davis

**Status**

Silicon carbide (SiC) has attracted attention of the photonic and spintronic communities since 2010 when it was proposed as an alternative wide bandgap defect host to the prominently researched diamond, with an outlook of improved optical stability and industrial maturity. The advantages of SiC as a substrate included wafer-scale availability, presence of the second order optical nonlinearity, and easier opto-electronic integration, however, its color center properties were yet to be explored. Throughout last decade, research on divacancy and silicon vacancy centers (predominantly in 4H-SiC polytype) demonstrated microwave and phonon driven coherent control of the electron spin, millisecond spin-coherence times (even without isotopic purification), indistinguishable single-photon emission, and, moreover, photonic and optoelectronic device integration. These results set SiC on a promising path to defining the quantum technology backbone for applications in quantum communication and computation.

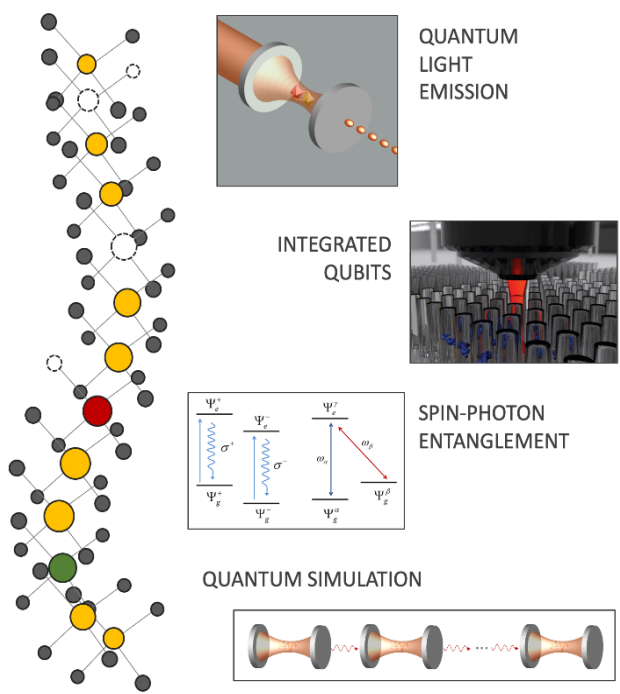

**Figure 1.** Silicon carbide lattice illustrated in a shape of a (technology) backbone hosting different types of defects with examples of applications in integrated nanodevices. Integrated qubits subfigure was adapted with permission from reference [1]. Copyright 2020 American Chemical Society. Spin-photon entanglement subfigure was adapted with permission from reference [10]. Copyright 2020 by the American Physical Society.

While greatly benefiting from being a solid-state platform in terms of scalability, the optically active defects take a hit in performance due to the interactions with lattice vibrations. Overcoming this limit, the integrated color center photonics in SiC leads to both an improved, as well as a paradigm shifting, performance of quantum emitters. Placing silicon vacancy into a nanopillar [1] has shown increased light collection needed for more precise optical readout of a spin-qubit, while its integration with nanocavities has resulted in Purcell enhancement of light emitted into the zero-phonon line [2, 3] needed for high fidelity indistinguishable-photon operation in quantum applications. Strong cavity quantum electrodynamical coupling of one or multiple color centers to a nanocavity has been proposed for high-speed quantum light sources, as well as for the generation of hybridized light-matter states that play a key role in photonic quantum simulation [4]. Due to their small inhomogeneous broadening and nanodevice scalability, color centers are poised to overcome the stalled progress in a variety of quantum photonic geometries whose realization has proved impractical to atomic and quantum dot systems.

**Current and Future Challenges**

The deployment of fiber-based quantum networks would benefit from a presence of telecommunications range quantum emitters. While silicon vacancy and divacancy in 4H-SiC provide longer operating wavelengths than diamond defects and quantum dots, emerging SiC color centers, such as the nitrogen-vacancy [5,6] and the vanadium center [7], are currently explored in the range 1,300-1,550 nm.

A major color center integration challenge is the incompatibility of well-defined defects with thin heteroepitaxial films. Most SiC polytypes do not grow on sacrificial substrates, except for cubic layers of 3C-SiC grown on silicon whose lattice mismatch prevents formation of narrow-linewidth color centers. This leaves no easy avenue for providing refractive index contrast between a structure and its surrounding, needed to confine light in nanodevices. Nevertheless, several successful directions have been pursued: photo-electrochemical etching [2], SiC thinning and bonding onto an insulator [3], and Faraday-cage angled etching [8]. These promising approaches are yet to be developed into wafer-scale arbitrary substrate processes needed for commercialization of quantum technologies.

Another exciting and complex challenge is the realization of a high-fidelity spin-photon interface using a color center with long spin-coherence time, essential to quantum repeater development and cluster-entangled state generation [10]. This task requires 1) a color center with a suitable entanglement scheme, 2) optimized coherence achieved via intrinsic symmetry of the defect, isotopic purification of the substrate, magnetic field nuclear bath dilution, or dynamic decoupling control, 3) integration into a low-loss nanocavity with a favorable dipole orientation which would strongly promote the no-phonon radiative emission, 4) efficient light coupling into an objective lens or an optical fiber.

To meet the challenges in photonic quantum simulation, collective coupling of nearly-identical color centers to a common nanocavity is to be demonstrated. Nanofabrication techniques that can both produce low-loss nanocavities and maintain low strain in the substrate are needed to achieve light-matter coupling in the cavity-protected regime which would enable exploration of polariton physics [4].

Finally, fabrication of silicon carbide color center photonic devices with superconducting detectors on the same chip would enable fully integrated photonic quantum circuits with applications in quantum computing. The advantage of such approach compared to the competing methods are the long coherence, connectivity and scalability of qubits.

**Advances in Science and Technology to Meet Challenges**

Promising experimental results on nitrogen vacancy center generation and coherent manipulation [9], as well as theoretical proposals for their spin-photon interfaces have been demonstrated in silicon carbide [10]. Integration of these telecommunications range emitting centers into nanophotonic devices, as well as an increase of their spin-coherence, is a work in progress. Applications in the proposed high-speed long-distance quantum communications, and measurement-based quantum computing, are likely to follow.

To meet the wafer-scale and arbitrary substrate nanofabrication needs, ion beam angle-etching approaches, similar to those developed in the diamond platform, are to be explored. These methods can be used to accelerate the progress toward achieving high-Purcell enhancement, cavity quantum electrodynamical interaction, as well as bipartite and cluster entanglement.

A proof-of-principle coherent control of a Purcell enhanced divacancy in 4H-SiC was recently demonstrated in a suitably doped substrate [2]. The generalization of this method to other substrates and color centers potentially offering longer spin-coherence times and alternative wavelengths needed for versatile quantum hardware development will make for an impactful step toward quantum hardware integration.

Demonstration of indistinguishable photon emission from a single silicon vacancy [11] represents an excellent steppingstone toward both multi-emitter entanglement and cluster-entangled state generation. Extending this experiment to color centers integrated into nanocavities would increase the fidelity of entangling processes. Localized wavelength tuning implemented via strain, electric field or Raman scattering would enable entanglement of photons emitted from different color centers.

Recently, the first superconducting nanowire single-photon detectors on 3C-SiC substrate were implemented [12]. Further exploration with substrates that host color centers, as well as the integration with prefabricated photonic circuits, would help meet the quantum computing frontier. Due to the silicon carbide's commercial presence in power electronics and MEMS, expansion of industrial involvement into quantum applications, including commercial availability of isotopically purified samples, SiC on insulator platforms and *in-situ* doping of desired defects, is expected to propel further research and development.

**Concluding Remarks**

A decade of investment into the exploration of optically active defects in silicon carbide has borne a frontrunning platform for commercial integration of quantum hardware. With further advancements in substrate processing, photonic design, and optical and microwave control, silicon carbide photonics and spintronics are poised to form the quantum technology backbone that supports key processes across all flavors of quantum information processing.

**Acknowledgements**

This work is supported by the National Science Foundation (CAREER-2047564).

## 13 - O-band and C-band Single Quantum Emitters
Stephan Reitzenstein, Technische Universität Berlin

**Status**

Quantum light sources, which emit photons in the telecom windows at 1.3 µm (O-band) and 1.55 µm (C-band), are basic building blocks for fiber-optic-based quantum communication (QC) over long distances. Generally, the single-photon character can be approximated comparatively easily by attenuated lasers. However, the emission statistics of attenuated lasers is intrinsically of classical nature, which prevents their use in advanced QC approaches. In contrast to this, solid-state quantum emitters can generate individual photons in principle on demand [1]. In this context, high quality semiconductors quantum dots (QDs) are of great interest and can be produced in a self-organized manner using modern nanotechnology processes. In fact, GaAs and InGaAs QDs emitting at around 780 nm and 930 nm are among the best single-photon emitters today. InGaAs QDs can also be designed for emission at telecom wavelengths. For this purpose, the difference in lattice constants between InGaAs and GaAs must be reduced by suitable measures. For emissions at 1.3 µm, the method of strain reducing layers has been established and optimized [2], while metamorphic buffer layers are used for emission at 1.55 µm [3]. Alternatively, InGaAs QDs at 1.55 µm can also be realized using InP substrates [4]. In all cases, single-photon sources (SPSs) with narrow linewidths and high multi-photon suppression could be demonstrated. It was also possible to develop electrically driven entangled photon pair sources based on 1.55 µm QDs and to use them for fiber-based QC [5].

**Current and Future Challenges**

Despite the enormous technological advances in the development of telecom quantum emitters, there are still many challenges and open questions that must be solved in order to open up the envisaged fields of application in photonic quantum technology. With respect to device fabrication, it is important to go beyond proof-of-principle results and realize single quantum emitter SPSs in a controlled manner and with a high process yield. This is a challenging task because the used self-organized growth process leads to randomness in the position and emission wavelength of individual QDs, which leads typically to process yields in the few percent range. With respect to the brightness of the sources, it is important to increase the photon extraction efficiency significantly from <40% achieved to date [6] to values beyond 90%. A promising approach for this is the (deterministic) integration of QDs in circular Bragg gratings (CBGs) in combination with a gold mirror on the back which promises photon extraction efficiencies >95% and Purcell factors close to 30 [7]. In addition, for the future application of SPSs as modular building blocks in photonic quantum technology, efficient on-chip coupling to single-mode fibers is essential. It is technologically very demanding to implement such a coupling: It requires not only a robust mechanical alignment with sub-µm accuracy, especially for the often-required cryogenic temperature operation, but also an optical mode-matching between SPS and fiber. In addition, increasing photon indistinguishability to values beyond 90% is of great importance for QC applications. In fact, so far the photon indistinguishability of telecom QDs has been limited to less than 20% without temporal postselection [2, 6] – a very problematic issue which is attributed to the non-ideal crystal quality of the QDs and their surrounding semiconductor matrix and material interfaces in case of heterogenous device concepts. Finally, a very precise control and stabilization of the quantum emitter's wavelength on the order of the homogenous linewidth is essential for applications that are based on entanglement swapping through Bell-state measurements.

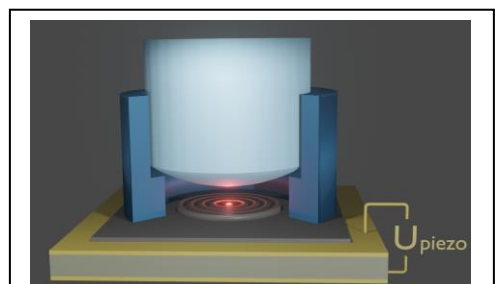

**Figure 1.** Schematic view of an "ideal" telecom wavelength SPS. The on-chip fiber coupled quantum device includes a single quantum emitter integrated deterministically into a circular Bragg grating structure with backside gold mirror and a piezo element for spectral control. By courtesy of Sven Rodt, Technische Universität Berlin.

**Advances in Science and Technology to Meet Challenges**

Extensive technological developments are necessary in order to solve the diverse challenges towards ideal telecom SPSs. First, it is of paramount importance to maximize the optical quality of the emitters themselves in terms of their emission linewidth, coherence and indistinguishability. In the field of semiconductor quantum dots, this primarily affects epitaxial growth. Here it is certainly necessary to sensitively optimize the layer design and the growth parameters in order to increase the optical quality of the QDs via structures with reduced defects – or to develop revolutionary novel growth concepts for telecom QDs. Here, Hong-Ou-Mandel measurements on photon indistinguishability, which is a very sensitive probe regarding the detrimental impact of defect states etc., can provide important feedback for iterative optimization of the QDs. Beyond that resonant excitation schemes like resonance fluorescence and two-photon resonant excitation need to be established and optimized as experimental tools for the maximization of the photon indistinguishability. Concerning the randomness associated with the typical self-assembled realization of many types of quantum emitters and the resulting poor process yield, it is important to develop and apply deterministic nanofabrication technologies. Among them in-situ electron beam lithography is very attractive and has to be adopted and optimized for the deterministic device integration of telecom quantum emitters. First results in this direction have recently been published in Ref. [2], but further refinements are required to deterministically realize for instance telecom QD-SPSs that are based on CBG structures. In connection with this, it is important to develop efficient and precise techniques for the spectral control of the quantum emitters. In this context, the spectral strain tuning via attached piezo elements that are electrically controlled is extremely attractive. This technology has already been used very successfully for QDs with emission <1µm and could recently also be demonstrated for deterministically fabricated QD-SPS in telecom O-band. In recent years, very interesting technological approaches for on-chip fiber coupling of QD-SPSs have been developed. Published results rely either on optical alignment in combination with gluing the fiber [8] or on micromechanical fiber holders and micro-optics [9]. Based on these initial successes, further technological developments must follow in order to significantly increase the photon coupling efficiency from a few percent achieved to date to values beyond 90%, for example using CBGs [7].

**Concluding Remarks**

Quantum light sources with emission in the telecom O- and C-band are among the most important building blocks for the implementation of fiber-based quantum networks, which can pave the way for the global quantum internet. Based on existing results, QD-SPSs are very promising candidates for meeting the diverse requirements of these application scenarios. In the future, it will be crucial to develop modular stand-alone SPSs with the highest photon coupling efficiency and the best possible quantum properties, which can be conveniently integrated into quantum networks.

**Acknowledgements**

Financial support from the European Regional Development Fund of the European Union via the FI-SEQUR project and from the German Research Foundation through CRC 787, Re2974/23-1 and Re2974/24-1 is gratefully acknowledged.

## 14 - Diamond Integrated Quantum Photonics

Benjamin Pingault1,2*, Bartholomeus Machielse1,3*, and Marko Lončar1
1 School of Engineering and Applied Sciences, Harvard University, Cambridge MA, USA
2 QuTech, Delft University of Technology, Delft, The Netherlands, EU
3 Department of Physics, Harvard University, Cambridge MA, USA
*These authors contributed equally.

**Status**

Over the past decade, color centers in diamond have been explored with the goal of utilizing their optically accessible spin for the implementation of quantum photonics and quantum networks [1,2]. These networks could be constructed from a series of nodes comprising color centers embedded in nanophotonic cavities, connected to each other via telecommunications fibers. The nodes act as quantum repeaters and processors, and enable novel, optically mediated entanglement topologies over distances for which photon loss would make direct transmission of entanglement impossible. Realization of such a network would enable technologies such as long distance QKD, distributed quantum computing, and enhanced long-distance quantum sensing [2].

Several breakthroughs towards this goal have already been achieved utilizing two primary color centers, the negatively charged nitrogen-vacancy (NV) center (see Fig. 1a) and the negatively charged silicon-vacancy (SiV) center [1] (see Fig. 1b). The NV center has been used to demonstrate loophole-free, photon-mediated entanglement between distant spins [3], as well as deterministic delivery of entanglement at a rate which exceeds the decoherence rate of the entangled state [4]. In parallel, the coherence properties of the NV spin have been improved to over a second [5]. Neighboring nuclear spins have also been harnessed through the NV spin as long-lived ancilla qubits and memories culminating in the realization of a 10-qubit quantum register and entanglement of 7 qubits [6].

More recently, the SiV center has attracted interest due to its superior optical properties [1]. In particular, the SiV maintains its charge and spectral stability when integrated into nanophotonic cavities. This property has enabled the realization of an integrated, SiV-controlled single photon switch [7], coherent interaction of two SiV spins through the optical mode of a photonic cavity [8], and memory-enhanced photonic Bell-state measurements [9] (see Figs. 2a and 2b). The cooperativity of the spin-photon interface enabled by the nanophotonic cavities now reaches above 100, thus indicating reliable coupling between itinerant photons and the SiV spin memory. Tapered waveguides allow the extraction into an optical fiber of more than 90% of SiV photons emitted into waveguides (see Figs. 2a and 2b). Integration of nanophotonic devices comprising SiV color centers with mechanical devices [10] and nonlinear optics platforms has also been demonstrated [11].

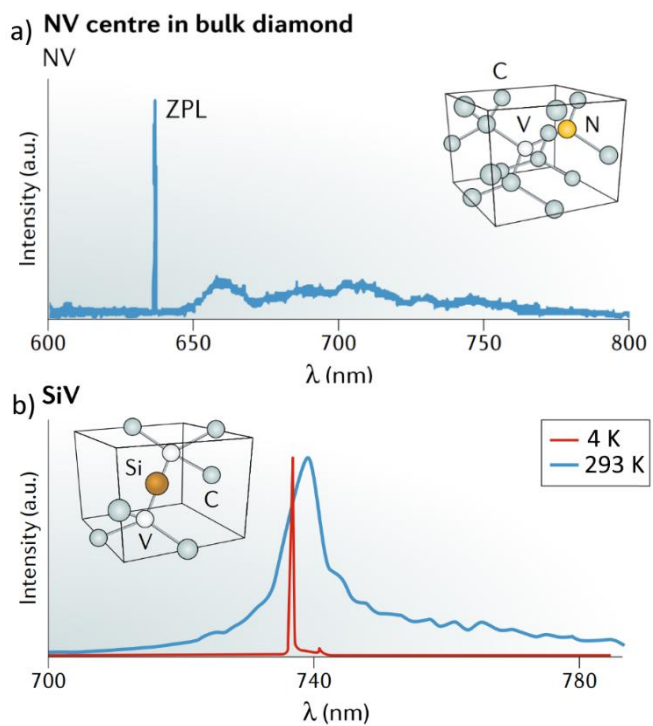

**Figure 1.** Fluorescence spectra of the two most prominent diamond color centers. **a)** The negatively charged nitrogen-vacancy (NV) center displays a sharp zero-phonon line (ZPL) amounting to 4% of the total emission at 4K, with the rest corresponding to the broad phonon sideband. **b)** The negatively charged silicon-vacancy (SiV) center has 70 to 80% of its emission into the ZPL. Among the dozens of known diamond color centers these two are the most studied, as they offer long-lived, optically-accessible spin qubits. The structures of the color centers (insets) explain the differences between their properties. The inversion symmetry of the SiV reduces its sensitivity to stray electric fields but also introduces increased sensitivity to thermal phonons. Other group IV color centers such as the germanium- (GeV) and tin- (SnV) vacancy centers share the structure of the SiV, making them promising candidates for further study [10]. a) and b) are reproduced with permission from [1].

**Current and Future Challenges**

Both the NV and SiV platforms have demonstrated the individual components required for the creation of quantum networks with quantum repeaters (see Fig. 2c), and are now pursuing integration of the required technologies and enhancement of optically mediated entanglement rates.

The highest rate of photon-mediated entanglement between two NV center spins is currently 39 Hz [2,4]. This rate is limited by the 4% probability that the NV emits photons into its zero-phonon line (ZPL) (see Fig. 1a) and by a limited collection efficiency. This can be addressed by embedding NVs into optical cavities, which in the case of fiber cavities has enabled a 10-fold Purcell enhancement [12]. Further integration requires the use of nanophotonic cavities fabricated directly in diamond, which is hampered by spectral instability of the NV near surfaces [13].

Such nanophotonic cavities, as successfully used for SiV experiments (see Fig. 2b), remain difficult to fabricate due to the lack of techniques for heteroepitaxial growth of pure, single-crystal diamond. Instead, several techniques have been used to undercut devices in bulk diamond [13]. Among these, crystallographic and angled etching are the most promising, but can produce only specific device topologies.

Furthermore, to enable large spin-photon coupling, photonic cavities rely on small mode-volumes. Reproducibility remains a challenge for such cavities due to the limited precision with which color centers can be localized through implantation and due to fabrication-induced damage. Crystal damage also causes spectral instability and inhomogeneity between centers [13]. This is a limiting factor in scaling up to larger networks, as photons most easily couple emitters with the same transition energies.

The design of larger networks will also require color centers to be integrated into or coupled to other photonic platforms for photon routing, manipulation, and conversion to telecom wavelengths [11,13]. Further integration with other quantum computing platforms, such as superconducting qubits or trapped ions, will be beneficial to hybrid approaches to quantum information processing.

Finally, much progress could be made through the search for novel color centers combining superior spin and optical properties [14]. These properties can include high photon yield into the ZPL with near unity quantum efficiency, long spin coherence times with optical addressability at elevated temperatures, or low sensitivity to strain, charge and magnetic noise.

**Advances in Science and Technology to Meet Challenges**

Ongoing theoretical and experimental research efforts have led to several proposed solutions for the challenges facing diamond color center-based quantum networks.

Of primary interest is improving the understanding and control of the elements that impact the inherent properties of color centers. This is especially important for the NV center, but would also contribute to the development of new surface treatments, implantation techniques, annealing parameters, growth conditions, and novel strategies such as shallow implantation followed by overgrowth that could improve the properties of all forms of color centers [13].

Narrowing the range of novel emitters that need to be explored experimentally requires a systematic understanding of the desired properties of emitters and further improvement of computational methods for evaluating candidates [15]. The related efforts into reducing the cost and improving cooling power of cryogenic technologies are also important for increasing academic and industrial access to new color center technologies.

The improvement of cavities for spin-photon interfaces and their integration with new technologies requires progress in fabrication techniques. The most developed fabrication strategies rely on undercut photonic devices that can still be optimized further, while novel techniques involving the use of suspended diamond films that could enable more straightforward device fabrication require more investigation. Future diamond platforms should also continue to integrate microwave, acoustic, and electromechanical functionalities as control mechanisms for color centers [10,13]. Improving access to ultrapure diamond substrates and facilities for overgrowth of existing substrates remains an ongoing and important effort.

The parameter space of existing techniques for nanophotonic device design also remains underexplored. Larger mode volume cavities could be coupled to NV centers, overcoupled or ultrasmall mode volume cavities could enable higher cooperativity or more efficient spin-photon gates, and inverse design techniques could be used to enable more sophisticated device properties [13,16].

Finally, permanent packaging of diamond photonic devices that allow simultaneous addressing of multiple color centers will be essential to large scale integration. This could be accomplished by permanent splicing of fibers onto existing devices, pick-and-place integration with other platforms [11], or flip-chip technology [13].

**Concluding Remarks**

Driven by progress in nanofabrication, theoretical understanding, and experimental sophistication, diamond color centers coupled to nanophotonic devices have emerged as a leading photonics and quantum communication platform. Technology surrounding the NV center has progressed furthest, resulting in demonstrations of long distance entanglement. The entanglement rates could be boosted to technologically relevant levels with the development of techniques for integration of the color center with cavities. The SiV center has been successfully integrated with nanofabricated cavities, has been used to demonstrate enhancement of quantum communication rates, and is progressing rapidly towards similar demonstrations of long distance entanglement. Meanwhile, ongoing research into other color centers could lead to the discovery of emitters which can combine the best properties of existing platforms. The challenges for these platforms lie primarily on the road to large scale integration, as the individual components required for near term application for quantum networking technologies have been demonstrated.

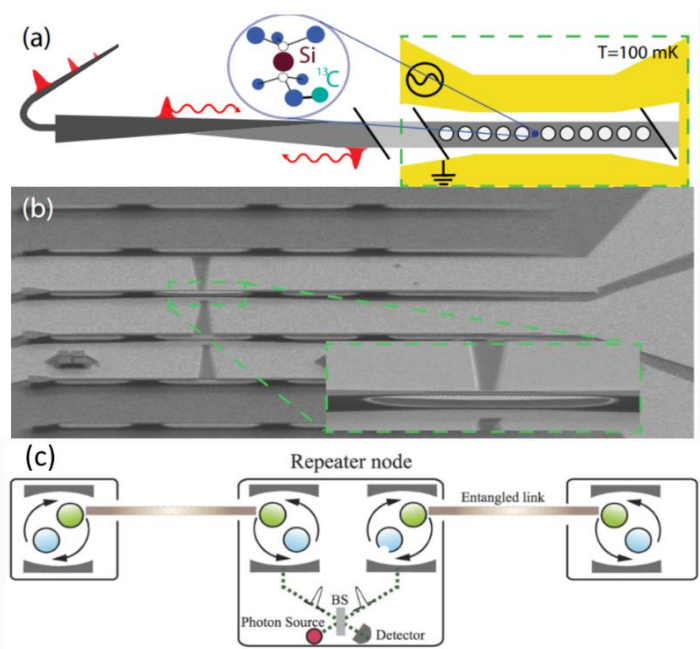

**Figure 2:** Example structure of a quantum network node. **a)** Schematic of the experimental setup used in Ref. [5]. SiV centers are embedded in a nanophotonic cavity, surrounded by gold coplanar waveguides for spin control. Photons produced by the color centers are collected through the tapered fiber interface at the end of the device. **b)** Scanning electron microscopy (SEM) image of the device shown in panel a) Inset shows the nanophotonic crystal cavity. **c)** Schematic representation of a general quantum repeater node. A pair of quantum emitters inside cavities are placed at each node. The nature of the emitters and cavities varies depending on the platform used. After successfully heralding entanglement with adjacent nodes, Bell-state measurements are performed between the memories in the two cavities to teleport entanglement to the distant nodes. a) and b) are reprinted with permission from C.T. Nguyen, D.D. Sukachev, M.K. Bhaskar, B. Machielse, D.S. Levonian, E.N. Knall, P. Stroganov, R. Riedinger, H. Park, M. Lončar, and M.D. Lukin, "Quantum network nodes based on diamond qubits with an efficient nanophotonic interface", *Phys. Rev. Lett.*, vol. 123, pages 183602, 2019. https://doi.org/10.1103/PhysRevLett.123.183602. Copyright 2019 by the American Physical Society. c) is reprinted from K. Nemoto, M. Trupke, S. J. Devitt, B. Scharfenberger, K. Buczak, J. Schmiedmayer, and W. J. Munro, Photonic Quantum Networks formed from NV− centers, *Scientific Reports*, vol. 6, pag. 26284 2016, under a Creative Commons Attribution 4.0 International License http://creativecommons.org/licenses/by/4.0/

**Acknowledgements**

*We thank N Sinclair and R Riedinger for helpful conversation. B.M. thanks NSF GRFP for funding provided via research fellowship. B.P. acknowledges financial support through a Horizon 2020 Marie Skłodowska-Curie Actions global fellowship (COHESiV, Project Number: 840968) from the European Commission. ML acknowledges support from NSF, ONR, DoE, and AFOSR.*

## 15 - Quantum Optics in Integrated Waveguides and Photonic Crystal Fibres

Debsuvra Mukhopadhyay (1), Alexey Akimov (1,2,4), Aleksei Zheltikov (1,3,4), Girish S. Agarwal (1),

Texas A&M University, College Station TX 77840; (1); PN Lebedev Physical Institute, Moscow 119991, Russia (2); M.V. Lomonosov Moscow State University, Moscow 119992, Russia (3); Russian Quantum Center, Moscow Region, 143025 Russia (4).

**Status**

Waveguide QED offers a complementary paradigm to cavity QED for studying and regulating light-matter interaction which forms the cornerstone of state-of-the-art chip-scale photonics. Nanofiber-controlled fluorescence from an evanescently coupled atom was first demonstrated by Le Kien et al [1]. The radiative behavior of multi-emitter configurations is governed by the relative location of the emitters, allowing more efficient control on photon transport and possible adoption as a prototypical element for quantum information. As quantum emitters, cold atoms have largely led the way in the engineering of strongly integrated atom-waveguide architectures, due to their perfect coherence properties. Enhanced Bragg reflection off a chain of identical atoms trapped in an optical lattice in a waveguide was demonstrated in [2]. An ensemble of emitters interfacing with a waveguide, as shown in Figure 1, exhibits a myriad of exotic phenomena such as single-photon super- and sub-radiance, nonreciprocal photon transport, and asymmetrical Fano lineshapes. When the lattice periodicity equals an integer multiple of the wavelength, the cooperative emission from the atomic array retains its Lorentzian profile with a linearly scaling decay rate, as has been demonstrated in [3, 4]. By stimulating interaction between photons, a multi-spin cluster can serve as an optical switch or nearly any possible quantum gate. Possibility of transparency due to single-photon transport across differentially detuned two-level emitters has been theoretically demonstrated, without applying any control field [5]. Directional dependence of the emission can entail asymmetric relaxation rates to the left- and right-propagating modes, which significantly modifies the transport. The interaction mediated by an optical fiber wields flexible control on both dispersive and dissipative couplings, level repulsion and attraction, and on sensing capabilities [6]. All these ideas can be contextualized in a variety of experimental models of topical interest such as resonators on a transmission line, cold atoms near a fiber and quantum dots coupled to plasmonic excitations in a nanowire or photonic crystal waveguides [7, 8]. Among solid-state candidates, some promising ones are epitaxy grown quantum dots, color centers in diamond and solids like silicon carbide, and rare-earth defects in glasses and crystals. As a versatile platform for photonic-state engineering, spontaneous four-wave mixing (FWM) stands out as a unique resource for executing a broad class of quantum phenomena in a fiber-optic format [9, 10].

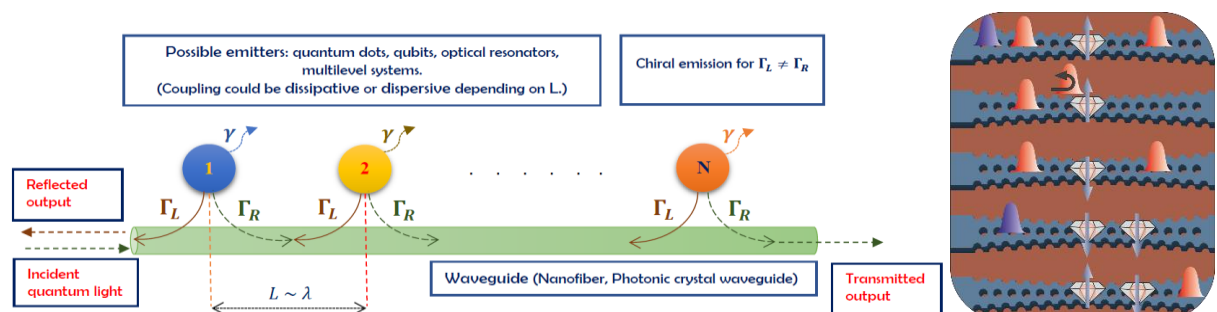

**Figure 1.** (LEFT) Model of a string of N quantum emitters side-coupled to a one-dimensional continuum. (RIGHT) An on-chip integrated-waveguide quantum gate with color-center nanodiamonds; the blue and red colours represent single photons of different frequencies, with output dependent on the state of the color center.

**Current and Future Challenges**
Currently, the prime challenge is to develop a universal platform for integrated waveguide devices, which would allow easy scalability for industrial applications. The propagation of nanofiber-guided light through a periodic array of identical emitters has been extensively studied. While these works lay out a propitious roadmap for the novel possibilities with waveguide QED, a majority of them, starting from the demonstration of transparency to sensing of anharmonicity, premise their analyses on loss-free or low-decoherence models. To realize these effects with better precision requires better control on extraneous decoherence and overlap in the spatial modes of incident and scattered waves. Indeed, if one considers current information technologies, the only photons used now are at telecom frequencies, while the majority of trailblazing results are being unveiled in the visible and near-infrared domains, where there are plenty of convenient laser sources and familiar high-quality emitters. The conversion of optical to telecom frequency at the level of single photons, which is interesting for quantum information processing, still poses a challenge. Remarkable progress has been achieved with diamond-based waveguides. No less challenging is optimizing the choice of quantum emitter itself. An obvious trend is to deploy solid-state emitters, but unavoidable decoherence and strong scattering off phonons brings down the quality of these emitters, even at cryogenic temperatures. Nevertheless, the impact of decoherence channels in generic waveguide-QED models has been theoretically studied from rigorous, first-principle considerations [11]. Nitrogen-vacancy color centers in diamond can boost the coherence at cryogenic temperatures but defects intrinsic to nanostructures can be an issue. The problem of inadequate coupling could be overcome to some extent by integrating photonic cavities to waveguides, but cavities are unfortunately bulkier. Previously mostly ignored for their very weak optical transition, rare-earth ions in crystals are showing functionalities via their telecom transitions and possibility of strong enchantment in photonic crystal cavities. Furthermore, quite generally, chiral waveguides provide an entirely different setup for manipulating photon scattering. Therefore, the controllability of chiral characteristics and the feasibility of unidirectional emission merits further investigation. Finally, with regards to quantum-state generation, nonlinear methods like FWM in optical fibers are becoming successful as recent efforts have shown [9, 10].

**Advances in Science and Technology to Meet Challenges**
Although experimental investigations on photon transport in waveguide-integrated setups have picked up speed over the last two decades, it is imperative to make continual upgrades to bridge the chasm between theory and experiment. The subwavelength-sized mode volumes in cavity QED made it possible to reach even ultrastrong atom-photon coupling, however many waveguide-based platforms are still lagging behind in implementing efficient photon-emitter interaction required for fabricating chip-scale optical devices. Interestingly, the cold-atom community is now investigating not only multi-atomic systems coupled to waveguides, but also into possibilities of constructing waveguides from atomic arrays themselves [12]. With diamond-based platforms, several amazing results involving silicon vacancy centers, on strong coupling, all-optical switching of photons and realization of quantum gates in all-optical cavities in diamond-made cavity were demonstrated [13]. While color centers in silicon carbide show clear promise in moving to industry-friendly applications, they have a lot of catching up to do in comparison to diamond-based structures. There has been a number of advances in positioning and even fabrication of the nanocrystals themselves, which has created potential for these systems to compete with pure diamond cavities or silicon-carbide cavities [14]. Plasmonic structures recently focused on development of low-loss surface plasmon polaritons as well as growing low-loss plasmonic materials. When implemented in optical fibers, spontaneous FWM provides a vast arsenal of methods and an ample parameter space for engineering photon entanglement in spectral, temporal, spatial and polarization modes. Advanced methods of fiber-dispersion management, on the other hand, may prove instrumental for entanglement-time engineering [10]. As one of the recent trends, FWM with cross-polarized pump and sidebands finds growing use as a powerful resource of quantum entanglement, enabling creation of efficient fiber-optic sources of strongly antibunching heralded single photons [9] and high-brightness entangled photon pairs [10].

Fiber-mediated coherence between emitters is instrumental to the sensing potential of these devices and can also generate long-range inter-emitter entanglement. In order to deftly explore topological phenomena in qubits coupled to fiber waveguides, more technical sophistication based on modern nanofabrication methods is needed to increase the qubit-waveguide coupling.

**Concluding Remarks**

With the sheer amount of progress being made in waveguide QED, waveguide-integrated circuits and networks are continually expanding the frontiers of research in purely optics-based quantum communication. The Fano-like interference between multiple emission pathways in the photonic interaction with an array of spatially separated emitters makes the photon transport strongly nonreciprocal and sensitive to the spatial separation. In addition, regulation of chiral couplings grants tremendous information processing potential to fiber-integrated multi-emitter devices. The dissipative interaction mediated between the emitters enables one to not merely entangle these emitters but also to explore exceptional point induced phase transition, without resorting to extrinsic gain as is usually the case with directly coupled systems. Finally, optical fields tailored in waveguide and fiber-optic systems lend a vast parameter space, defining the spatial, temporal, spectral, polarization, and spin modes of photons, thus opening new avenues for quantum information science, quantum metrology, and ultrahigh-resolution quantum imaging.

**Acknowledgements**

The authors acknowledge the support of The Air Force Office of Scientific Research [award no FA9550-20-1-0366], The Robert A. Welch Foundation [grant nos. A-1243; A-1801;], Herman F. Heep and Minnie Belle Heep Fellowship, Texas A and M University (X-grant 497); National Science Foundation (PHY-1820930), Texas A&M College of Science STTR grant, Russian Foundation for Basic Research (projects 18-29-20031, 19-02-00473), Russian Science Foundation (project 20-12-00088), Ministry of Science and Higher Education of the Russian Federation (project 14.Z50.31.00400).

# QUANTUM FREQUENCY CONVERSION

## 16 - Quantum Frequency Conversion using Integrated Nanophotonics

Kartik Srinivasan, Joint Quantum Institute, NIST/University of Maryland and Physical Measurement Laboratory, National Institute of Standards and Technology

**Status**

Quantum frequency conversion (QFC) generally refers to the spectral translation of a quantum state of light, such as a single-photon state, between frequency bands. Its role in quantum information science and technology is well-defined in the context of quantum networks (Fig. 1), where it can enable telecom-wavelength fiber optic links between visible wavelength local quantum systems (e.g., quantum memories). More generally, QFC is needed to make interconnections between quantum systems operating at different wavelengths, for example, to connect a trapped ion quantum processor to a neutral atom ensemble quantum memory. QFC may also have relevance in being able to convert photons to wavelengths for which single-photon detectors have the highest performance, though ever-improving superconducting single-photon detector technology has now made efficient and low-noise detection from the ultraviolet to the near-infrared possible.

The two main requirements for QFC are high conversion efficiency, so that a large fraction of the input photon flux is frequency converted to the target output frequency, and low added noise, so that the output frequency channel is dominantly composed of light originating from the input source (rather than spurious photons). In the 30+ years since QFC was first discussed explicitly in the literature [1], efficient and low-noise QFC demonstrations have been powered by advances in nonlinear optical technology, in particular based on sum- and difference- frequency generation in quasi-phase-matched nonlinear crystals [2] and four-wave mixing in optical fibers [3], though multi-level atomic systems can also be used to mediate QFC, as has been demonstrated with cold atomic gases [4], for example. QFC with true single-photon sources has by now been demonstrated many times [5], and in recent years has been used to connect disparate systems such as a cold atomic gas (at 780 nm) and a solid-state crystal (at 606 nm) [6] and to enable entanglement of Rb quantum memories (at 780 nm) across a >20 km fiber optic link [7].

For many such experiments requiring discrete QFC elements, a blueprint for operation has been established. The dispersion of a material creates a phase-mismatch between the fields involved in a nonlinear process and without compensation, limits the maximum achievable conversion efficiency. Materials such as lithium niobate combine the ability to realize quasi-phase-matching by periodic poling [2] – a particularly effective approach for overcoming dispersion – with a strong nonlinear coefficient and broadband optical transparency, resulting in near-unity internal conversion efficiency being shown in some cases. Waveguide geometries provide a few centimeter interaction length while retaining bandwidths that are sufficiently wide for a broad range of quantum sources and memories. Low-noise operation typically involves having wide spectral separation between the pump source that mediates frequency conversion (whose power can easily be >$10^{13}$ that of the input quantum light source) and the input signal frequency and target output frequency. Together with appropriate spectral filtering, this limits the extent to which broadband processes (e.g., Raman scattering, parametric fluorescence, and defect-mediated fluorescence) can produce noise that overlaps with the input and output frequency channels, particularly when a long wavelength pump can be selected [8].

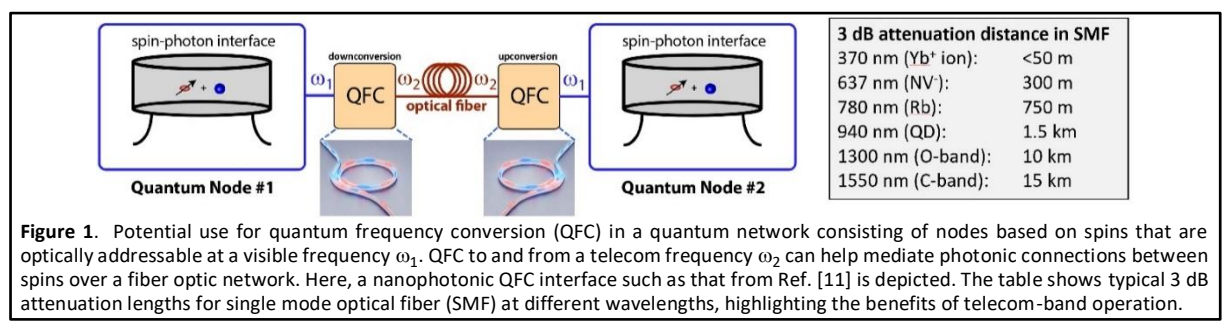

**Figure 1**. Potential use for quantum frequency conversion (QFC) in a quantum network consisting of nodes based on spins that are optically addressable at a visible frequency $\omega_1$. QFC to and from a telecom frequency $\omega_2$ can help mediate photonic connections between spins over a fiber optic network. Here, a nanophotonic QFC interface such as that from Ref. [11] is depicted. The table shows typical 3 dB attenuation lengths for single mode optical fiber (SMF) at different wavelengths, highlighting the benefits of telecom-band operation.

**Current and Future Challenges**

Although existing QFC technology is thus already being used to help construct elementary quantum networks, there are many areas in which further development is needed. QFC across large spectral gaps – for example, from the visible (or ultraviolet) to the telecom remains a challenge, as the frequency separation of >300 THz ensures that in a single three-wave mixing process, the pump will need to be placed between the input and output frequency channels, making simultaneous high conversion efficiency and low added noise difficult to achieve. Wider frequency separations also tend to make poling (and other phase-matching techniques) more difficult. A challenge specific to heterogeneous quantum networks is the need to not only match photon frequencies, but also bandwidths (more generally, temporal profiles) to ensure optimal storage of a frequency-converted photon in a quantum memory, for example. Nonlinear optics provides some mechanisms by which such temporal shaping can be realized [9], though this issue can also be addressed by other quantum network components (e.g., in the quantum light generation or quantum memory stages).

One significant challenge is in making QFC systems extensively deployable. Along with the nonlinear medium and pump laser(s), QFC typically requires optical fiber coupling, wavelength (de)multiplexers, and narrowband optical filters to create a full setup (these components also reduce the overall conversion efficiency from near-unity to ≈ 50 %). This combination of lasers with linear and nonlinear optics is accessible in table-top setups, but the size, weight, and power requirements of such systems are not ideal for deployment in all environments. Integrated photonics provides a compelling route to combine such functions and create full QFC systems in a deployable and scalable platform.

**Advances in Science and Technology to Meet Challenges**

Integrated photonics is being developed in a wide range of materials of relevance to quantum photonics [10]. This section identifies several important aspects of this technology as it pertains to QFC.

*Materials for QFC* — Although the maturity of silicon nanophotonics provides strong motivation for its adoption when possible, applications often require functionalities such as gain, broadband optical transparency, and fast, low-loss switching, that are not easy to achieve in silicon. This has motivated research on platforms that have more suitable properties but are still amenable to similar nanofabrication processes. For nonlinear optics, lithium niobate remains a preferred choice for second-order processes, and its availability in thin film form offers new opportunities for dispersion control, power reduction, and integration. Aluminium nitride is another wide-bandgap, second-order nonlinear material available in thin-film form. For third-order nonlinear processes, silicon nitride and tantalum pentoxide offer broadband optical transparency and a nonlinear coefficient that is more than an order of magnitude larger than silica optical fibers, which have been the dominant four-wave mixing platform for decades and have been used in QFC experiments based on the four-wave mixing Bragg scattering process [11]. With an appropriate choice of material composition, III-V materials from the InGaP and AlGaAs families can provide both sufficiently broad optical transparency and a very large optical nonlinearity (both second-order and third-order), while diamond may be a natural choice for third-order QFC applications that involve its color center quantum emitters and/or spins. To that end, we also note that QFC approaches involving multi-level single quantum emitters coupled to nanophotonic elements have also been proposed [12].

*Control of nonlinear processes* — Nanophotonic geometries provide stronger optical confinement than conventional geometries such as optical fibers or silica planar lightwave circuits. One consequence of this confinement is its impact on dispersion, as the fraction of the field that resides within the waveguide core strongly depends on wavelength and core geometry. As a result, intrinsic material dispersion can often be compensated by precise tailoring of a nanophotonic geometry [13], thereby providing phase-matching that has enabled QFC configurations not easily accessible otherwise. For example, four-wave mixing QFC between 980 nm and 1550 nm (on-chip efficiency ≈ 60 %) was recently shown in silicon nitride microrings [14] and constituted a wider spectral gap than previous demonstrations in optical fibers; efficient and low-noise visible-to-telecom QFC has also been proposed using similar devices [15]. Geometric dispersion engineering can also be combined with quasi-phase-matching [16]. Moreover, nanophotonics can provide access to other frequency shifting mechanisms, for example based on acousto-optics, that can be used in some scenarios [17].

*Low power QFC* — A typical pump power for efficient QFC in conventional quasi-phase-matched waveguides is about 100 mW, while in optical fibers, Watt-level powers are often needed. Nanophotonic geometries can reduce such powers, particularly when microresonator geometries are employed, due to the combined effect of confinement and long photon lifetimes. The aforementioned silicon nitride microring QFC [14] achieved high efficiency at the ≈ 10 mW power level, and reduction to the 1 mW level should be possible through improved cavity performance (higher quality factors) or a more nonlinear material (e.g., a wide-bandgap III-V system). Such powers can easily be accessed by chip-integrated lasers; alternately, it would also allow a single off-chip pump laser to service multiple QFC elements. While a resonant cavity limits the photon conversion bandwidth, the ≈1 GHz bandwidth typical of high quality factor integrated microcavities is adequate for most high-performance quantum emitters, including InAs/GaAs quantum dots (as used in nanophotonic QFC in [18]).

*Understanding and mitigating noise* — One important task for nanophotonic QFC platforms is an understanding of added noise sources. While it can be expected that the main noise categories will include Raman scattering, spontaneous parametric processes, and fluorescence, each must be understood and characterized in the context of the material platform and geometry under investigation. Moreover, although a nanophotonic geometry can provide enhancement of the target QFC process, it may also enhance unwanted noise processes. Moving beyond creation of spurious photons, thermorefractive noise may limit the frequency stability of converted photons, which can be of importance when working with narrow linewidth sources.

*Integration:* One of the most compelling reasons for pursuing nanophotonic QFC is the potential for seamless integration with other photonics technologies, which can simplify the deployment of full QFC systems in applications. Along with combining the QFC element(s) with linear components such as filters and couplers, integration with pump lasers and other quantum photonic elements, such as single-photon sources and solid-state spins, is feasible, particularly in the context of hybrid photonic platforms that combine the advantageous characteristics of dissimilar materials [19]. Figure 2 gives two examples of nanophotonic QFC systems that can be envisioned based on current technology.

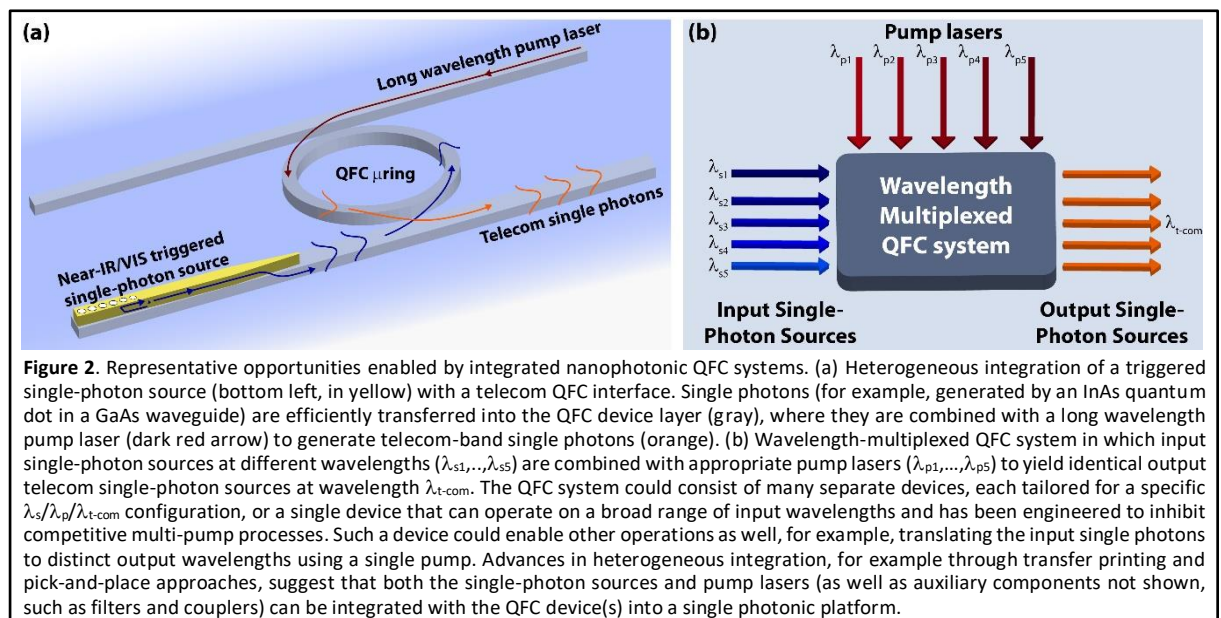

**Figure 2**. Representative opportunities enabled by integrated nanophotonic QFC systems. (a) Heterogeneous integration of a triggered single-photon source (bottom left, in yellow) with a telecom QFC interface. Single photons (for example, generated by an InAs quantum dot in a GaAs waveguide) are efficiently transferred into the QFC device layer (gray), where they are combined with a long wavelength pump laser (dark red arrow) to generate telecom-band single photons (orange). (b) Wavelength-multiplexed QFC system in which input single-photon sources at different wavelengths ($\lambda_{s1},..,\lambda_{s5}$) are combined with appropriate pump lasers ($\lambda_{p1},...,\lambda_{p5}$) to yield identical output telecom single-photon sources at wavelength $\lambda_{t-com}$. The QFC system could consist of many separate devices, each tailored for a specific $\lambda_s/\lambda_p/\lambda_{t-com}$ configuration, or a single device that can operate on a broad range of input wavelengths and has been engineered to inhibit competitive multi-pump processes. Such a device could enable other operations as well, for example, translating the input single photons to distinct output wavelengths using a single pump. Advances in heterogeneous integration, for example through transfer printing and pick-and-place approaches, suggest that both the single-photon sources and pump lasers (as well as auxiliary components not shown, such as filters and couplers) can be integrated with the QFC device(s) into a single photonic platform.

**Concluding Remarks**

Quantum frequency conversion (QFC) is an important physical resource for enabling photonic interconnects in quantum networks. While existing QFC devices based on conventional nonlinear waveguide geometries are already being used to build early-stage quantum networks, integrated nanophotonic platforms can play an important role in future construction of more complex and functional networks. They offer clear advantages with respect to control of nonlinear processes, size, weight, and power consumption, integration with pump lasers and other photonic components, and scalability of the underlying manufacturing processes. While much work remains to be done in understanding noise generation in these platforms, they offer the potential to make QFC a truly plug-and-play component for quantum information applications.

**Acknowledgements**

The author acknowledges funding from the NIST-on-a-chip and DARPA LUMOS programs.

## 17 - Single-Photon Nonlinearity in Integrated Quantum Photonics

Juanjuan Lu and Hong X. Tang

Department of Electrical Engineering, Yale University, New Haven, Connecticut 06511, USA

**Status**

Quantum photonic integrated circuits have received growing interests since such platforms offer the stability and integrability towards solid state quantum applications. By encoding quantum information into the optical photons, the quantum information processing, quantum communication and quantum metrology would benefit from the merits of the bosonic carriers, including the high propagation velocity, long propagation distance and infinite-dimensional Hilbert spaces. Single-photon nonlinearity enabled by quantum two-level systems is essential for future quantum information technologies, as they are the building block of quantum photonics logic gates, deterministic single-photon sources and for coupling distant nodes to form a quantum network.

Microwave photons could be processed by superconducting quantum circuits, where high fidelity quantum operations approaching error-correction thresholds are achieved by the lossless nonlinearity inherent to Josephson effect at cryogenic temperatures [1]. However, at optical frequencies, the absence of the single-photon nonlinearity has been a major roadblock in developing quantum photonic circuits. While the photon-photon interactions required to realize quantum gate can be mediated through light-matter interaction with atomic or atom-like solid-state emitters, their cryogenic temperature requirements greatly reduce the prospects for portable devices and routes to scaling such systems remain challenging [2].

Recent progress in nonlinear optical materials and micro-/nano-resonators has brought single-photon non-linearity employing bulk optical nonlinearities into the realm of possibility [3]. One significant advantage of this approach is its great potential for emitter-free, room temperature quantum photonics applications. In Fig. 1(a), we illustrate the single-photon nonlinearity in a $\chi(2)$ cavity, which is intrinsically attributed to the strong single-photon coupling rate $g$ surpassing its dissipation rate $\kappa$ and thereby is quantified by the figure of merit FOM = $g/\kappa$. Figure 1(b) summarizes the coupling rates, dissipation rates as well as the corresponding single-photon nonlinearity FOMs in various integrated $\chi(2)$ photonics platforms, which are elaborated in Ref. [4]. We note that the state-of-the-art $g/\kappa$ is approximately 0.01 and remains two orders of magnitudes smaller than unity for practical applications.

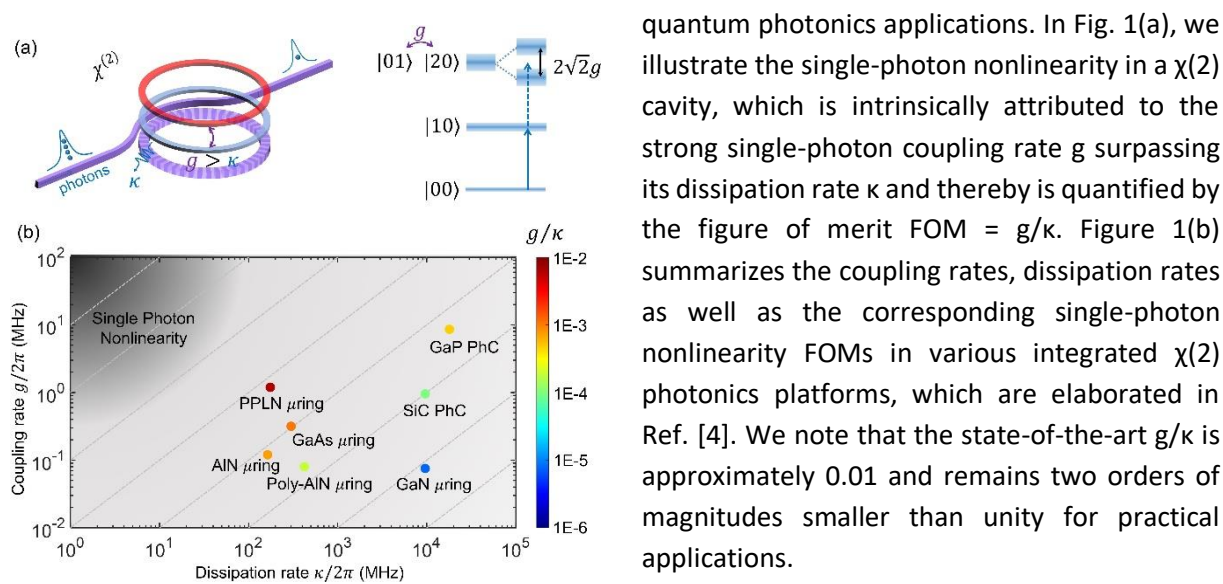

**Figure 3.** (a) A $\chi(2)$ cavity with single-photon nonlinearity based on a periodically pole lithium niobate microring and its schematic energy level diagram (right panel). The anharmonicity of the nonlinear system is determined by the energy level splitting relative to their widths. (b) The coupling rate g, dissipation rate κ as well as the corresponding single-photon nonlinearity FOM demonstrated in various integrated $\chi(2)$ cavities. μring, microring; PhC, photonic crystal.

**Current and Future Challenges**

Significant engineering efforts are required to maximize the nonlinear coupling strength $g$ while minimizing the cavity loss κ simultaneously and thereby advance the directions and applications laid out in the previous section via single-photon nonlinearity. Theoretically, in a $\chi^{(2)}$ cavity, $g \propto \chi^{(2)}_{eff}\gamma/V$ and $\kappa \propto 1/Q$ give the single-photon nonlinearity FOM $\propto \chi^{(2)}_{eff}\gamma Q/V$ and, where $\chi^{(2)}_{eff}$, $\gamma$, $V$, and $Q$ denote the effective nonlinearity, spatial mode overlap, mode volume and optical quality factor, respectively. While individual parameter seems readily optimized in different platforms, several technological challenges should be tackled to ensure an ultimate optimization of FOM in one single platform as outlined below:

**Trade-off between Q and V:** A high Q/V resonator is always demanding to realize single-photon nonlinearity. However, cavity Q factor typically decreases as its dimension scales down. For example, $Q$ of $10^6$ has been demonstrated for micro-resonator (V ∼ 100(λ/n)³) while $Q \sim 10^4$ is typical for photonic crystal nano-resonator (V ∼ (λ/n)³), which is mainly limited by the scattering loss induced in the fabrication and radiation loss as the device dimension is extremely small rather than intrinsic material properties (*n* is the refractive index). Therefore, a significant reduction of loss is expected in the near future via the careful device design and optimized fabrication techniques.

**Phase matching for integrated cavities**: The coupling rate g is non-zero when the phase-matching condition (momentum conservation) is satisfied, which is demanding in the case of microcavity and even nanocavity. Substantial progress including modal-, quasi- and cyclic-phase matching have demonstrated in the microrings. However, in the case of photonic crystal nanocavity, its ultra-small mode volume suffer from a large index dispersion (momentum mismatch), and require more stringent phase matching condition, whose effective nonlinearity χ⁽²⁾ and spatial mode overlap γ are far from optimized so far.

**Advances in Science and Technology to Meet Challenges**

To circumvent the aforementioned challenges, many advances would greatly benefit the future of single-photon nonlinearity in integrated quantum photonics employing bulk optical χ⁽²⁾ nonlinearity for potential deployment. Some advances include the following:

**Material platforms with ultralow-loss** Several integrated platforms based on different materials have been exploited for nonlinear and quantum applications. Each material has its advantages and disadvantages. However, due to their significant χ(2) nonlinearity and demonstration of ultra-low loss, two platforms are particularly promising towards single-photon nonlinearity:

*LN on insulator platform.* Due to the high electro-optic coefficient ($r_{33}$ = 32 pm/V), significant quadratic nonlinearity ($d_{33}$ = 30 pm/V), and wide transparency window (350 nm to 5 μm), thin-film LN has recently risen to the forefront of chip-scale nonlinear and quantum optics research since its demonstration as an ultralow-loss integrated photonics platform [5]. Most recently, with the high-fidelity domain poling, an state-of-the-art second-harmonic generation efficiency of 5,000,000%/W is demonstrated in a periodically poled lithium niobate microring resonator, whose single-photon nonlinearity approaches 1%. Moreover, a photonic molecule using two coupled periodically-poled microrings is theoretically proposed to facilitate single-photon generation [4].

*(Al)GaAs on insulator platform.* Compared to commonly used dielectric materials for integrated nonlinear and quantum applications, (Al)GaAs has the highest nonlinear coefficients in both χ⁽²⁾ ($d_{14}$ = 119 pm/V) and χ⁽³⁾ ($n_2$ = 1.6 10–13 cm2/W) and its high refractive index provides high optical confinement and small mode volume [6]. Along with its recent advance in high-Q microring fabrication, ultra-low threshold optical parametric oscillator and efficient second-harmonic generation have been obtained in (Al)GaAs on insulator platform, showing its great potential for strong nonlinearity at the single-photon level [7].

**Novel photonic structures** Due to the trade-off between the *Q* and *V* in the traditional microring resonators, some novel photonic structures such as micro-post and grating-slab cavities, and heterostructure photonic crystal slab cavities utilizing bound states in the continuum [8, 9], are theoretically proposed to ensure the high *Q/V* and large mode overlap factor *γ* simultaneously, which is critical for approaching the unitysingle-photon nonlinearity FOM.

**Concluding Remarks**

Although recent advances on the ultralow-loss demonstration of thin-film LN and AlGaAs integrated circuits along with their significant $\chi^{(2)}$ nonlinearities show very encouraging performance, research on the single-photon nonlinearity in integrated quantum photonics employing bulk optical nonlinearities remains at its early stages and experimental implementation has yet to be demonstrated. Further breakthroughs on high Q/V integrated resonator fabrication and strong $\chi^{(2)}$ coupled device design are required to reach the single-photon nonlinear regime, and pave the way for the emitter free, room-temperature quantum photonic applications.

**Acknowledgements**

We acknowledge financial support from Department of Energy, Office of Basic Energy Sciences, Division of Materials Sciences and Engineering under Grant DE-SC0019406.

## 18 - Quantum Transducers

Jiang, Wentao, McKenna, Timothy P., and Safavi-Naeini, Amir. H.

Department of Applied Physics, Ginzton Lab, Stanford University, Stanford, California 94305, USA

**Status**

The ability to control quantum coherence and entanglement in several physical platforms has now reached the level where significant industrial resources are being applied to realize the promises of large-scale quantum information processing. One such platform consists of superconducting circuits operating at microwave frequencies. Microwave frequency photons are used to store, process, and control the internal quantum dynamics of these systems. The most striking recent manifestation is the result achieved by Google in 2019 on a processor consisting of 53 microwave-frequency qubits [1]. Though this experiment demonstrated that a quantum processor can perform tasks beyond the reach of classical computing technology, it is nonetheless possible that much larger-scale systems consisting of many more qubits will need to be realized before "useful" quantum computing algorithms can be routinely used to solve problems of real scientific, industrial, and commercial interest. One approach to facilitate scaling is to implement quantum transducers that can act as transparent interfaces between distant microwave-frequency nodes in a quantum data-center. The advantage of such an approach is that it would enable increased modularity in the system design, enable implementation of specialized (and possibly heterogeneous) subunits that can perform certain tasks more efficiently, and enable larger scales by distributing computational tasks across many cryogenic systems which can be operated independently. A natural approach to this problem is to realize a microwave-to-optical converter that efficiently interconverts between microwave and optical photons. The most important characteristic of a microwave-to-optical transducer is the rate of entanglement generation between optical and microwave photons that is possible given the constraints on dissipated power and other deleterious effects caused by absorption of light. Other important factors are the bandwidth, efficiency, whether the converter can support heralded operation, and the ease at which it may be scaled up to more channels. Fundamentally, a nonlinear process must occur to change the photon frequency. At this stage, the hardware remains the primary challenge and is the focus of this roadmap. We consider here the approaches based on the direct electro-optic conversion, and approaches that use an optomechanical resonator as an intermediary. There are also promising approaches based on rare-earth dopants or other solid-state emitters, which we will not discuss here due to lack of space, but note that they share many of the same implementation challenges [2].

**Current and Future Challenges**

**Figures of Merit:** In most approaches, the effective interaction rate between light and microwaves can be captured in a single parameter called the coupling rate $g_0$. In an electro-optic transducer, a microwave resonator has an oscillating voltage V which drops across the electrodes of an integrated electrooptical resonator with a tuning coefficient of $g_V$ Hz/V. The coupling rate $g_0$ is the optical frequency fluctuation of the electro-optic cavity caused by vacuum fluctuations of the microwave voltage, $V_{zp} = \omega \sqrt{hZ/2}$. This number is typically on the order of a few microvolts for standard microwave resonators, while high impedance resonators utilizing super-inductors have been able to achieve $V_{zp}$ on the order of a hundred millivolts [3]. Electro-optic tuning rates $g_V/2\pi$ are on order of 1 GHz/V in many systems, leading to coupling rates of $g_0/2\pi$ in the range of a few 1-30 kHz in integrated cavity electro-optic devices. In the integrated optomechanical transducers, a closely related

parameter signifying the optomechanical coupling can reach values in excess of 1 MHz. The interaction rate is only one part of the story. Other important parameters include the linewidths of the optical and microwave modes, $\kappa_a$, $\kappa_b$. Together, with an incident optical field, an internal efficiency of $4C/(1 + C)^2$ is achieved where $C = 4g_0^2 n_a/(\kappa_a \kappa_b)$ is the cooperativity. The total efficiency is the product of the internal efficiency with the efficiencies of the various coupling junctions the signal experiences, including the chip-to-fiber coupling. The conversion bandwidth $C\kappa_b$ scales with pump power. Since bandwidth scales with power, the energy per bit scales simply as $E \approx \hbar\omega_a(\kappa_a/g_0)^2$ [4].

**Electro-optic Conversion:** The direct electro-optic approach to microwave-to-optical transduction uses an electro-optic material for photon frequency conversion via a three-wave mixing interaction. An incident microwave photon creates an electric field that modulates the index of refraction of the electro-optic material. When pumped with a strong optical beam, this coupling scatters a microwave photon to a sideband optical photon with the pump photon acting to conserve energy. This three-wave mixing process is bidirectional and enables quantum state transfer between microwave and optical photons. The process can also be configured to create entangled pairs of microwave and optical photons. The efficiency of these systems is often enhanced by incorporating low loss superconducting microwave and optical resonators in a single device. The three waving mixing process is enhanced by confining the interacting photons to a small volume – and by bringing the electrodes as close as possible to the optical field; however, the metal used for the microwave resonator causes optical loss. Superconducting circuits can also suffer increased loss from quasi-particle generation due to incident optical pump photons, so proper codesign is clearly required. Recent demonstrations in aluminum nitride [5], lithium niobate (LN) [6, 7], and electro-optic polymer have demonstrated total efficiencies roughly ranging from $10^{-9}$ to $10^{-2}$. A key challenge is integrating high-Q optical and microwave resonators on the same chip while achieving large overlap with EO material. For example, in the LN platforms, the microwave modes can have significant unintended losses due to the piezo-electric properties of the EO substrate and care must be taken in the integration [6, 7]. Extrapolating what is achievable with the current approach, and assuming that improvements in system performance will enable demonstrations to reach the basic energy-per-qubit limit of roughly 10 pJ based on the material EO properties and best demonstrated quality factors ($Q > 10^7$, $g_0/2\pi \approx 10$ kHz), we expect achievable communication rates on the order of 100 kHz if we are limited to 10 microwatts of dissipated power, as is typical in a dilution cryostat. Much higher rates may be achieved by either using materials with effective EO coefficients that are much higher (*e.g.* the piezo-optomechanical approach described below), or working with higher frequency electromagnetic fields (100-300 GHz) that can operate at a higher temperature with no added thermal noise, and where much higher dissipated optical powers are manageable. The latter scheme calls for a two-stage converter to move information between microwaves (10 GHz) and mm-waves (100-300 GHz) [4], and mm-waves and optical photons (193 THz).

**Optomechanical Conversion:** Optomechanical converters utilize a two-step process for the microwave-to-optical quantum transduction. A microwave photon is first converted to a mechanical phonon with the same frequency, the phonon then scatters with a pump optical photon in an optomechanical cavity where they combine into the converted photon. The first step is enabled either by piezoelectric coupling between the electric field from the microwave resonance and mechanical deformation or using a parametric coupling between the two [3], while the second step relies on optomechanical interaction where mechanical deformation introduces a frequency shift of an optical resonance via photo-elastic effect as well as radiation pressure, realizing a three-wave mixing process. Several challenges need to be resolved to achieve an efficient piezo-optomechanical frequency conversion. High coupling rates between both the microwave and mechanical resonances, as well as between the mechanical and optical resonance are desired to overcome the intrinsic loss and enable larger conversion bandwidth. Efforts towards higher coupling rate include miniaturizing the mode volume and utilizing new materials with higher interaction strength. Optomechanical crystals (OMC) offer wavelength-scale confinement of both the optical and mechanical modes with a simultaneous photonic-phononic crystal structure, with $g_0/2\pi \sim 1$ MHz in high-index materials such as Si and GaAs. This improvement in $g_0$ would increase the quantum communication rates by $10^4$ over an EO converter with $g_0/2\pi \sim 10$ kHz – if the rest of the converter system can be suitably implemented. One challenge is frequency matching microwave mechanics and electromagnetic fields, though approaches with tuning the microwave resonances have recently succeeded in this by using magnetic fields and specially patterned superconducting resonators [8]. Another challenge is that materials with strong piezoelectricity are favored for the microwave-mechanics coupling – these materials, such as LN and AlN, are typically lower index and have lower optomechanical coupling. Combining the two components remains challenge and requires careful design and engineering of the full system [9, 10]. Piezo-optomechanical frequency converters have been demonstrated with different piezoelectric materials, including GaAs [11], AlN [12, 8, 13], and LN [9]. Recent focus of development has been shifted towards heterogeneous integration of different materials such as lithium niobate on silicon and aluminum nitride on silicon [10, 13], as the two steps involved in the conversion process favors different material properties. Heterogeneous integration allows combination of large optomechanical coupling in silicon OMC and high piezoelectricity of LN or AlN, enabling considerably higher optomechanical coupling or higher microwave-to-mechanical conversion efficiency compared to single material converters.

**Controlling Coupling and Noise:** An important factor in operating a converter is the fiber-to-chip coupling efficiency. Efficiently coupling light into and out of optical structures is a common challenge to all device types. Since quantum microwave systems operate at extremely low temperatures (< 1 K), optical coupling schemes also need to be robust against thermal stresses induced by thermal cycling. Tapered fiber coupling that features a tapered fiber that is aligned to evanescently couple to an on-chip waveguide has achieved broadband operation with > 90% efficiency at cryogenic temperatures. Edge coupling has also been demonstrated at cryogenic temperatures but also requires in-situ alignment. In general, schemes that require in-situ alignment do not scale well enough to enable larger systems involving many optical devices. A common coupling technique at room temperature utilizes gratings patterned on the surface of a chip to scatter off chip light into a waveguide. A recent demonstration glued an angle polished fiber to a grating coupler and cooled the device down to < 10 mK. The relatively high tolerance of the technique to misalignment due to thermal stresses allowed the technique to achieve 25% coupling efficiency [7]. Far more efficient approaches will be needed to achieve deterministic quantum operations, and generalizable solutions are currently lacking. While major efforts have been focused on improving the efficiency of quantum frequency conversion between microwave and optical frequencies, state-of-the-art converters suffer from combinations of added noise or high energy consumption. To minimize the added noise in the conversion process, efficient thermalization of the involved microwave/mechanical mode is required to keep its thermal occupation as low as possible. For a piezo-optomechanical converter, the absorption of pump photons generates heating and adds noise to the mechanical mode, which is the predominant noise source especially for one dimensional optomechanical crystals. Pulsed pump has been widely adopted in cryogenic optomechanical and piezo-optomechanical measurements with sufficient intervals for the mechanical mode to relax to the ground state. Better thermalization has been achieved with 3He buffer gas and two-dimensional optomechanical crystals [14], the latter has shown an effective quantum cooperativity above unity. On the other hand, the microwave circuit is prone to quasiparticles generated by stray pump photons. Cares need to be taken to minimize such negative impact [10].

**Concluding Remarks**

We discussed the current state of the art electro-optic converters, which use either the direct electro-optic effect in a material with a high EO coefficient such as LN, or use an indirect piezo-optomechanical approach. Achieving highly efficient conversion remains a daunting task requiring optimization and ultra-high performance from every part of a whole heterogeneous system. With several paths forward on all remaining technical challenges, we expect transducers to become reality in the next decade and to enable a variety of new and exciting applications in quantum computing, communications, and sensing as well as forming part of the technological bedrock of an emerging quantum internet.

**INTEGRATED DETECTORS**

## 19 - Waveguide-Integrated Superconducting Single-Photon Detectors

Stephan Steinhauer, Ali W. Elshaari, Val Zwiller, KTH

**Status**

Photonic integrated circuits (PIC) have found wide-spread applications in classical optics and quantum information technology. To take full advantage of the powerful capabilities offered by integrated photonics (e.g. high circuit complexity, integration density, stability), effective schemes for on-chip coupling of high-performance photodetectors with PICs is required. In this regard, superconducting nanowire single-photon detectors (SNSPDs) devices are a most promising technology: they are capable of light detection at the single-photon level from the visible to the mid-infrared, can be fabricated in a straight-forward manner (superconducting layer patterned in a single lithographic step) and allow for efficient coupling of the detectors to the evanescent field of the guided optical modes. After first demonstrations of waveguide-integrated SNSPDs around a decade ago, research and development efforts resulted in devices with excellent performance metrics such as on-chip detection efficiencies > 90 %, < 20 ps timing jitter, low noise with dark count rates in the mHz range and ultrafast recovery times enabling maximum count rates exceeding 1 GHz [Ferrari2018]. Superconducting materials typically used for fiber-coupled as well as waveguide-integrated SNSPDs include nanocrystalline (e.g. NbN and NbTiN) and amorphous (WSi and MoSi) thin films that are most commonly deposited by magnetron sputtering. Thin film deposition and nanowire device fabrication are compatible with the major integrated photonics materials, which allowed the successful realization of waveguide-integrated SNSPDs on silicon-on-insulator, silicon nitride, gallium arsenide, aluminium nitride, lithium niobate and diamond platforms. Major advances have also been made in the deterministic integration of high-performance detectors, e.g. via pick-and-place approaches [Najafi2015] or by photonic circuit fabrication on top of pre-selected nanowire devices [Gourgues2019]. Despite the outstanding detector properties and the versatile options for PIC realization, several challenges still need to be addressed to realize complex PICs comprising a large number of SNSPDs (>> 1000) that can be reproducibly fabricated and reliably operated with minimized trade-off or compromise in performance.

**Current and Future Challenges**

Ongoing research and developments efforts aim at further improvements of the SNSPD performance metrics at the single-device level. SNSPD devices with enhanced properties are needed as several quantum information processing schemes place stringent requirements on certain detector parameters, in particular detection efficiency and maximum count rate. Photon number resolution (PNR) is another important capability required for many quantum protocols and linear optical quantum computing [Slussarenko2019], which is typically not offered by SNSPDs. Alternatively, transition-edge detectors would provide the possibility for PNR, but they often need to be operated at low temperatures in the 100 mK range and are linked with disadvantages of slow reset times in the microsecond range and limited timing resolution. Significant progress has been achieved in the development of SNSPDs with pseudo PNR, relying on spatial multiplexing in arrays of superconducting nanowires [Mattioli2015].

Figure 1 shows different device concepts for single-detector, multi-detector and cavity-enhanced SNSPDs integrated with photonic circuits, highlighting their advantages and shortcomings. To achieve GHz detection rates, it is required to minimize the reset time of the detectors to the sub-nanosecond range. Although decreasing the device length reduces the kinetic inductance and hence the detection pulse decay time constant, this approach places limitations on the on-chip detection efficiency due to reduced waveguide-detector interaction length. This challenge has been successfully addressed by integrating SNSPDs with one-dimensional and two-dimensional photonic bandgap cavities, boosting the on-chip detection efficiency for short nanowires. However, it remains a major challenge to simultaneously optimize several key detection parameters and combine them with (pseudo) PNR capabilities.

**Figure 1** Device concepts for superconducting nanowire single photon detectors (SNSPDs) coupled to photonic integrated circuits (PIC). U- or W-shaped SNSPD coupled to a single waveguide (yellow). Multiple SNSPDs on a single waveguide or an array of waveguides (blue). SNSPDs coupled to one-dimensional and two-dimensional cavities (green).

| | Single SNSPD | Multiple SNSPDs | Cavity-enhanced SNSPD |
|---|---|---|---|
| Complexity of design and fabrication | low | medium | high |
| Active detector area | large | large | small |
| Max count rate / efficiency | trade-off | trade-off | no trade-off |
| Detection wavelength range | broad | broad | narrow |
| Photon number resolution (PNR) | none | pseudo-PNR | none |

Moreover, scalability and yield, i.e. the ability to reproducibly fabricate waveguide-integrated detectors with tight control over their properties such as critical current, on-chip detection efficiency, and timing resolution, are at the forefront of current challenges. Additionally, realizing radically up-scaled systems with many independently addressed waveguide-integrated detectors can pose challenges on the readout scheme. Although a free-space coupled kilo-pixel SNSPD array suitable for imaging systems was recently realized using a row-column multiplexing architecture [Wollman2019], demonstrating even larger systems in conjunction with complex networks of integrated quantum photonic devices will require sophisticated interfacing and multiplexing solutions.

**Advances in Science and Technology to Meet Challenges**

Reports in recent literature have highlighted the importance of engineering microwave signal propagation in SNSPD devices. Multi-photon coincidence detectors based on two-terminal devices consisting of sixteen detector elements were accomplished by slow-wave transmission lines for time-domain multiplexing via signal delays [Zhu2018]. The detectors were designed for integration with waveguide arrays and are thus suited for on-chip quantum information processing applications. Another example is the demonstration of PNR capabilities of a single SNSPD enabled by tapered transmission lines for measuring Hong-Ou-Mandel interference [Zhu2020]. The reported taper design combines high impedance load for the SNSPD and 50 Ω matching with the readout electronics, while showing multiple advantages including single-layer fabrication process, simple readout scheme and suitability for high packing density. However, significant engineering efforts will be needed to enable reliable operation of large-scale PICs employing waveguide-integrated SNSPDs as a multitude of factors needs to be considered, e.g. on-chip footprint, noise and electrical/optical crosstalk, cryostat heat load via cabling and optical fibers, efficient optical coupling and interfaces to electronic circuitry. Hence, it will be crucial to identify application-specific requirements and devise optimized tailored solutions. It can be expected that these developments will profit from advances in multiple fields, including microwave engineering, cryogenics and photonic packaging.

To address current challenges related to large-scale fabrication and yield, it will be important to establish a better understanding of the relationship between nanoscale material structure / chemical composition, nanofabrication processes and the resulting superconducting device properties. To date, it appears that amorphous superconductors are favourable for the high-yield fabrication of SNSPDs with consistent properties. However, a recent report on NbN detectors realized by plasma-enhanced atomic layer deposition demonstrated excellent yield and reproducibility [Cheng2019], which highlights the promising prospects for SNSPD technology upscaling via improving thin film deposition. Further advances in materials science and fabrication technology combined with developments in characterization and metrology tools for defect analysis would also allow the local identification of SNSPD failure mechanisms, which would in turn provide crucial input for the rational development of industrial detector fabrication processes to achieve superior device reliability.

**Concluding Remarks**
Waveguide-integrated SNSPDs are central to quantum photonics as they enable scalable measurements of quantum states and their interference directly on-chip. There are ongoing efforts in the community to push the limits of single-detector metrics. The overarching goal is to realize complex quantum photonic circuits co-integrated with SNSPDs that provide unprecedented performance and functionality for emerging quantum information processing applications. As the complexity of photonic quantum circuits is continuously growing, innovative methods to interface and address individual photonic and superconducting elements will become increasingly crucial. Exciting future avenues include combining on-chip detection with dynamic reconfiguration of photonic circuits using electro-optic materials, for instance lithium niobate and aluminium nitride. Realizing such devices will enable on-chip qubit rotations through active feed-forward architectures and fast interconversion of quantum and classical information, which would for instance eliminate the need for post-selection in Bell-state measurements and teleportation [Elshaari2020].

**Acknowledgements**
S.S. acknowledges support from the Swedish Research Council, Starting Grant (Grant No. 2019-04821). A.W.E. acknowledges support from the Swedish Research Council, Starting Grant (Grant No. 2016-03905).

## 20 – Integrated Waveguide-Coupled Single Photon Detection Near Room Temperature

Paul S. Davids, Nicholas Martinez, and Michael Gehl, Sandia National Labs, Albuquerque, NM.

**Status**

Detection at extremely low-light levels is of fundamental importance for imaging and remote sensing applications and lies at the heart of quantum information systems and interconnects. Single photon detectors (SPD) are a key building block device of quantum photonics, which integrates quantum optical circuits onto a densely integrated chip-scale platform that leverages advanced Si-CMOS processing and heterogenous integration.

Commercially available single photon detectors are available in a number of technologies. The highest performance detectors are based on superconducting nanowires, providing high single photon detection efficiency and low dark counts, with the drawback of required cooling below 4K, making them impractical for many applications. An alternative technology is single photon avalanche photodiodes (SPAD) utilizing charge multiplication to convert a single photogenerated carrier into a detectable current pulse. Sillicon based SPADs provide high efficiencies in the visible and near infrared wavelengths, while InGaAs versions provide operation out to telecom wavelengths. The ability to operate near room temperature increases the appeal of this technology. Implementing this technology in a scalable CMOS fabricated quantum integrated photonic platform is desirable to realize deployable, low power quantum devices. One path towards this integration is to utilize epitaxial growth of Ge on Si in a separate absorption and charge multiplication avalanche photodiode (SACM-PD). This device, demonstrated near room temperature, takes advantage of the low noise avalanche gain provided by intrinsic silicon, combined with narrow band absorption of germanium[1,2]. Figure 1a shows a schematic and fabricated cross-section of the waveguide-coupled Ge absorption layer and lateral Si multiplication APD device. These single photon detectors work in the telecom bands and operate with moderate cooling near room-temperature and can be deployed in quantum key distribution chip-based transceivers that do not rely on high SPD efficiency [3]. Figure 1b shows the impact of temperature and Geiger mode gate duration on dark counts and illustrates the need for significant reduction in dark count rates. Thermal generation in semiconductor photodetection is intrinsic due to the bandgap of the photo-absorbing material and can be mitigated through extreme volume reduction of the absorber material to reduce the impact of bulk and surface defects and by cooling of the device. The single photon detection efficiency and the dark count rate are complex functions of the Geiger mode operation of the device and depend critically on the gate duration, repetition rate, over bias, temperature, and device size. For advanced quantum interconnect applications, dramatic improvement in the performance of uncooled integrated SPAD's is required due to the preciousness of each photon.

Moreover, non-cryogenically cooled single photon detectors with selected wavelength absorbers can be used in a host of applications; such as to extend the range of LIDAR systems, improve scintillation for radiation detection, reduce power consumption in focal-plane array imagers, and for quantum sensing for chip-based atomic sensors. These semiconductor-based devices are compatible with advanced CMOS manufacturing processes and with advances in heterogenous integration can be integrated into complex photonic circuits. Further research and development on integrated SPAD devices is needed to reach the full potential of these devices for quantum interconnects.

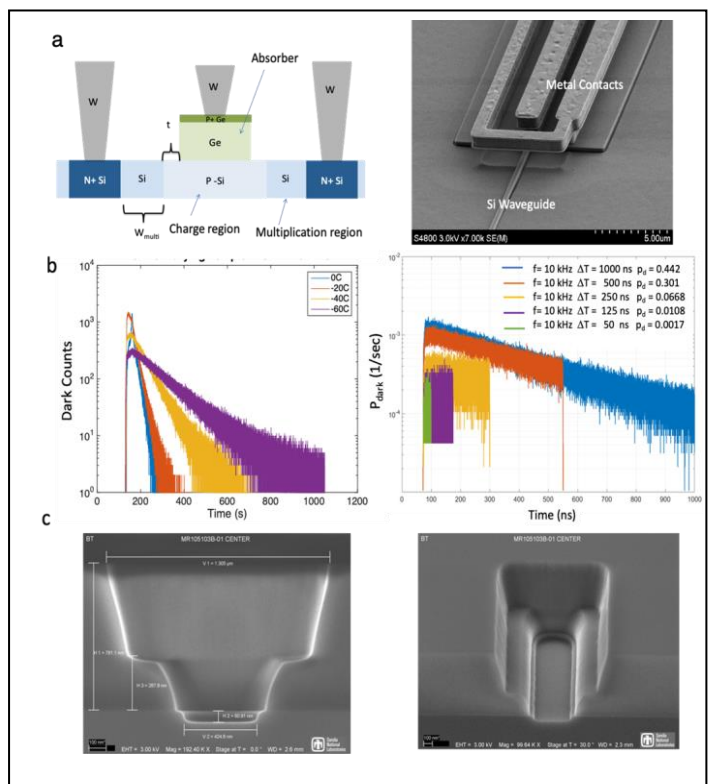

**Figure 1.** **a** (left) Schematic of lateral Ge on Si SACM APD and (right) top-down view of waveguide coupled SACM with oxide cladding removed. **b** (left) Measured dark counts of SACM APD for various operating temperatures and (right) dark count probability for fixed repetition rate with variable gate duration. **c** SEM template for aspect ratio trapped growth of Ge on Si for reduction of defects and improved dark count rate.

**Current and Future Challenges**

Single-photon detectors based on avalanche photodiodes (SPAD) work by amplifying a weak photogenerated current from a single absorption of a photon into a detectable signal while supressing dark detection events due to thermal or defect driven processes in the device. These two processes seem to be in opposition, indeed the high amplified device currents needed can create localized heat sources and emit out of band photons which can influence dark counts in nearby detectors. However, by reducing the active volume of the narrow bandgap absorbing material and reducing defect density at interfaces we can mitigate thermal generation effects that contribute to dark counts in these devices. Continued advances in semiconductor processing and integrated control circuits can be used to reduce these effects.

A major challenge in semiconductor single photon detectors is to reduce spurious dark counts while maintaining high quantum efficiency. Dark count rates can be manipulated with temperature but other methods, specifically fabrication, exist to reduce these noise sources. In selectively grown epitaxial structures on Si, defects exist at interfaces due to lattice mismatch resulting in strain which can act as traps and release carriers in strong fields. Graded strain relief layers can be used to reduce the interfacial defects. Furthermore, by reducing the selective growth interface area where defects from the lattice mismatch occur, the crystalline quality of the selectively grown material is greatly improved through aspect ratio trapping, and defect areal density is reduced. Figure 1c illustrates the modified selective epitaxial template for aspect ratio trapping growth of Ge. Another device technique to reduce dark counts is by operating the SPAD in gated Geiger mode. This high-speed Geiger mode gating acts as a system level trick to reduce the influence of dark counts by reducing the probability of triggering a dark event in a small gate time window which is similar to lock-in techniques. However, this method requires that the arrival of the single photon signal to be in a synchronous timing window for detection. Gated operation gives rise to capacitive feed-thru response and various circuit subtraction methods can be used to reduce the triggering threshold [4]. Furthermore, careful consideration and design of the location of high electric field regions within the device can be tailored to avoid high defect surfaces using doping, symmetrically placed electrodes, strain engineering and other semiconductor processes. Similarly, photonic design principles can be used to improve light trapping in the waveguide coupled structure and to remove parasitic absorption that does not lead to viable photogeneration.

## Advances in Science and Technology to Meet Challenges

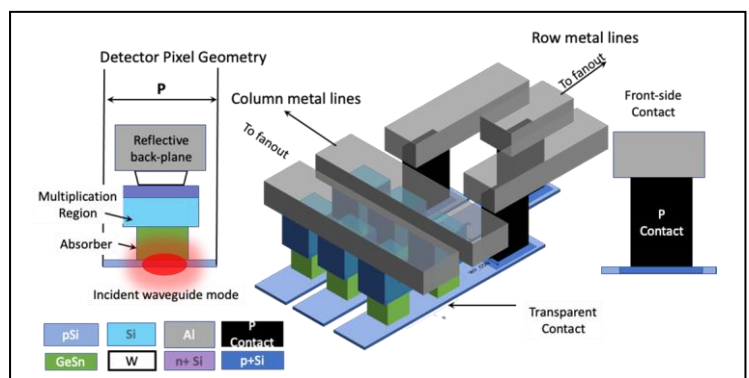

**Figure 2.** Resonant GeSn waveguide-coupled SACM APD design with Si epi multiplication region on top. Right figure shows GeSN APD array for through Si illumination or waveguide coupling.

Low-defect selective epitaxial growth of Ge or related materials on Si by UHV-CVD processes with in-situ doping in compact templates would be a significant scientific advance that would benefit semiconductor-based single photon detectors. Novel photonic light trapping structures integrated into SACM-APD devices that resonantly confine light to subwavelength volumes offer potentially efficient single photon detection with moderately cooled devices. Furthermore, integrated fast electronic gating and signal analysis or hybrid APD coherent detection schemes can be used to improve moderately cooled single photon detector performance. One key advantage for the separate absorber charge multiplication APD design is the flexibility to change wavelength absorption band by changing the absorber material but retaining the advantages of the Si multiplication. High confinement dielectric resonant absorber designs allow for compact device structures that result in lower bulk thermal generation (See fig. 2). Recently, advances in selective epitaxial growth of strained $Ge_{1-x}Sn_x$ alloys have led to a new high-refractive-index semiconductor alloy with a tuneable direct bandgap for varying Sn concentration (x=5 to 20%) [5]. This allows for strong absorption in the short-wave infrared (SWIR) spectral range with tuning between wavelengths in the range of 2-2.5 microns while maintaining CMOS compatible processing. In the SWIR band, Si remains transparent but with improved free-carrier dispersion effect for active device modulation and a reduction in two-photon absorption allowing for interesting possibilities for non-linear optical phenomena on-chip. $Ge_{1-x}Sn_x$ single photon SACM APD's would allow for the building of a new complete quantum photonics platform with potentially new quantum and classical applications [6-8].

## Concluding Remarks

Superconducting nanowire single photon detectors (SNSPD's) with background limited performance have been demonstrated with the major drawback that they need to operate at cryogenic temperatures near 4K which may not be practical for some applications [9]. Recent developments in semiconductor based single photon detectors based on compact Ge on Si separate charge absorption and multiplication region APD's present opportunities for large scale integration and improved performance through fabrication advances and designs in applications that are non-cryogenically cooled. Remember that the human-eye is a few-photon visible light sensor that operates above room-temperature!

## Acknowledgements

Sandia National Laboratories is a multimission laboratory managed and operated by National Technology & Engineering Solutions of Sandia, LLC, a wholly owned subsidiary of Honeywell International Inc., for the U.S. Department of Energy's National Nuclear Security Administration under contract DE-NA0003525.

APPLICATIONS

## 21 – Photonics for Trapped-Ion Quantum Systems

John Chiaverini[1,2], Karan K. Mehta[3]

(1) Lincoln Laboratory, Massachusetts Institute of Technology, Lexington, MA 02421, USA
(2) Massachusetts Institute of Technology, Cambridge, MA 02139, USA
(3) Department of Physics, Institute for Quantum Electronics, ETH Zurich, Zurich, Switzerland

**Status**

Atomic ions held in electromagnetic traps and manipulated with lasers and/or radio-frequency and microwave fields form one of the most promising physical implementations of quantum information processing (QIP) [1]. Current systems employed in industry, academia, and national laboratories can precisely manipulate up to a few tens of ions in 1D arrays with sufficient control to perform basic quantum algorithms. However, optical systems currently used for trapped-ion control, mainly based on bulk optics, are susceptible to substantial noise and drift that can introduce errors even in few-qubit operations. Furthermore, they are challenging to make extensible as required to surpass NISQ-era scales [2] and thereby enable practically relevant quantum computation.

An approach instead relying on photonics tightly integrated within ion trap chips offers a path to addressing these challenges. In this architecture, as illustrated in Fig. 1, the multiple visible/UV wavelengths required for ion control and readout are routed on-chip and emitted, via diffractive elements, toward ions held in surface electrode traps typically 50-100 μm above the electrode surfaces. Single-photon detectors integrated in such a platform would allow for qubit state readout in largescale arrays via site-specific fluorescence collection. Additionally, waveguide splitting and appropriate active devices would allow frequency, phase, and amplitude control of optical fields as required for parallel modulation of a number of channels in excess of the number of optical inputs to the device.

Experiments to date have demonstrated the core addressing element of this architecture [3], at all wavelengths relevant to particular species [4], as well as in low-error multi-qubit operations [5]. Realization of the full set of components and functions required for optical control and readout of precise, large-scale trapped ion systems for quantum computation, simulation, and precision measurement motivates basic research and development along multiple lines in the years ahead.

**Current and Future Challenges**

While low-loss (~1 dB/cm) single-mode waveguide propagation has been achieved at wavelengths as low as 370 nm [8], pushing operation wavelengths further into the UV is an outstanding challenge. While some species, such as Ca+, Sr+, and Ba+, require only longer wavelengths for full functionality, several others require yet shorter wavelengths; for example, logic operations in the Yb+ ion are often implemented using pulsed laser light near 355 nm, and Be+ and Mg+ have transitions near or below 300 nm. Realizing low-loss propagation, as well as high-confinement waveguides with sufficiently low spurious scatter, is a particular challenge at such wavelengths.

The requirement to precisely address individual ions within a larger ensemble motivates beam-forming optics capable of delivering micron-scale spots with < 10−4 relative intensities a few microns from the beam center. While early work has shown promise on this front [9], efficient methods for optimizing beam profiles in flexible geometries have received relatively little attention. Optimization routines for tailored free-space emission profiles, e.g. flat-top envelopes for uniformly addressing multiple ions, or with carefully positioned nulls to minimize crosstalk at certain locations, are likely to be of value for both QIP and atomic clock applications.

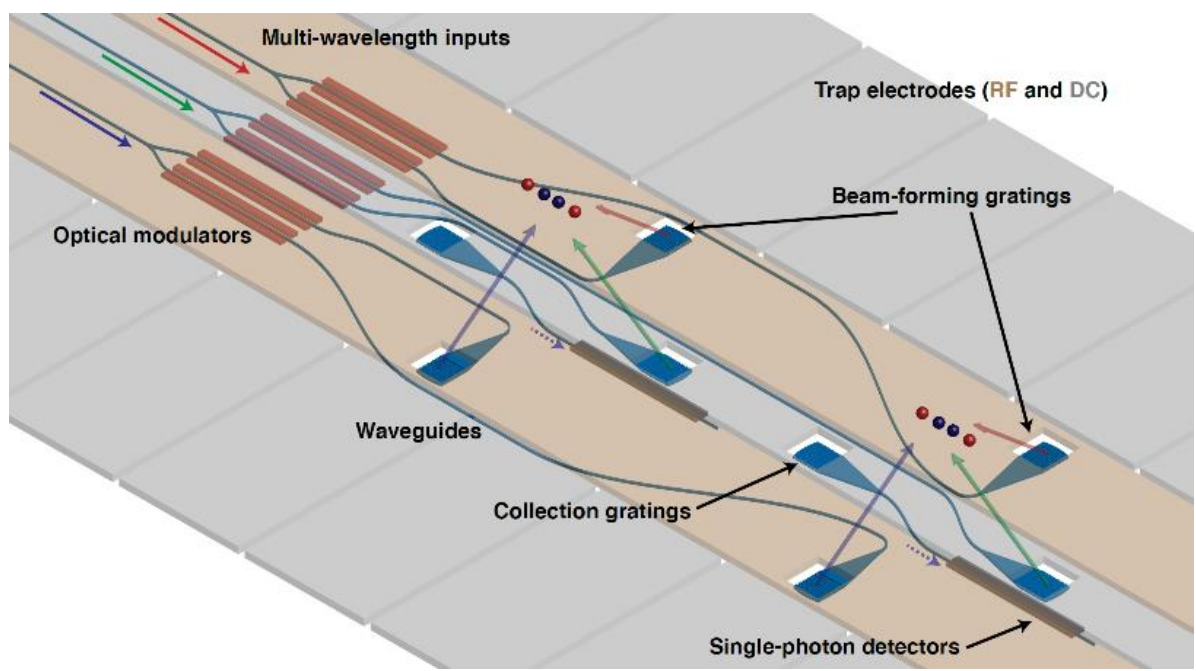

**Figure 1.** Integrated optical architecture for trapped-ion quantum control, as described and with some elements realized in refs. [3-7]. Tan (gray) electrodes, transparent here for clarity, are for production of RF (DC) components of the trapping potential. Photonic elements in layers below the electrodes include single-mode waveguides and splitters, modulators for switching and frequency control, grating couplers for control-light emission or photon collection (possibly into multi-mode waveguides), and single-photon detectors for ion-qubit state determination. Control electronics could be in a lower layer or in a separate, bonded substrate.

Ion experiments with integrated photonic addressing to date [3–5, 10] utilize multiple inputs passively routed on-chip to arrays of ions, with modulation implemented off-chip. As systems scale to beyond tens of ions addressed in parallel, input-fiber number will become a practical limit and it will be valuable to supply light for multiple channels from a single input, and then to split and modulate multiple channels independently. Integrated active components for visible and UV wavelengths are at a relatively early stage of development, and the high extinction (of order 60 dB) and precise frequency/phase control required for high-fidelity operations, together with the requirement for compatibility with the multilayer metal and insulator structures employed for trap-chip fabrication, are significant challenges. This will motivate research on new electro- and acousto-optic materials and device architectures for visible and UV wavelengths, with major impacts for trapped ion quantum systems and other photonic applications at these wavelengths.

On-chip photodetection with single-photon efficiency and site-specific collection over an array can enable larger-scale experiments not limited by light collection via external bulk optics. As most QIP applications do not require ion imaging, but only collecting fluorescence from each ion into a single detector, single-pixel detectors local to each ion site can be more quickly read out than an image of a typically sparse ion array obtained by a macroscopic lens. While impressive performance for single-ion readout has been achieved with integrated superconducting single-photon detectors [7], implementing high-quality detectors in a platform capable of appropriate shielding to minimize unwanted coupling to the ion qubits has not been addressed.

As these multiple elements are developed and brought together, careful characterization within trapped-ion quantum experiments will be critical to understand any additional noise introduced by such components. In this regard, experiments to date are encouraging in some respects, demonstrating that such approaches in fact reduce some typical sources of noise, e.g. vibrations and phase fluctuations [4, 5]. However, electric-field noise, both at megahertz frequencies where it heats ion motion and at much lower frequencies where it leads to trap potential drifts, can be more problematic in complex traps with integrated functionality [4, 5, 10]. Feedback on such issues from experiments will in general be critical as complexity grows.

**Advances in Science and Technology to Meet Challenges**

Relatively few materials candidates, even for passive waveguides at blue/UV wavelengths, have been thoroughly explored in the community, and this is likely to constitute a fruitful area. Careful characterization of nonlinearities and power limits will play a critical role in informing which platforms can support the routing of optical powers required for large-scale arrays. On-chip optical amplifiers could obviate the need for high power in a single bus waveguide, but the feasibility of integration of such devices in the visible/UV is relatively unexplored.

Both electro-optic and acousto-optic configurations are of interest for related active beam-control devices. This broadly motivates effort into assessing materials candidates for the wavelengths of interest, their integrability in larger systems, power handling, and susceptibility to photodamage, including at cryogenic temperatures. Materials of interest include but are not limited to $LiNbO_3$, BBO, AlN, and GaN. Surface-, guided-, or bulk-acoustic wave configurations could enable established materials for short-wavelength operation, e.g. $Al_2O_3$ to serve as the modulation interaction region when paired with appropriate transducers and coupled optical/acoustic mode design; this challenge points to a vast design space with much room for optimization.

While integrated, high-efficiency, single-photon detection has been demonstrated with superconducting detectors (SNSPDs), integration with standard CMOS is somewhat challenging due to the materials required, and many candidate superconductors have somewhat impractical transition temperatures, i.e. below 4 K [7]. Avalanche photodiodes (APDs) can achieve similar single-photon efficiencies in the blue and UV, particularly in silicon, with the opportunity for monolithic integration into CMOS-based platforms [6]. Challenges with this technology include minimizing defects that cause dark counts and addressing after-pulsing through active biasing and readout circuitry. Both detector technologies suffer from susceptibility to RF pickup from trap electrodes; scatter from light routed in nearby waveguides may also lead to excess background. Careful electrical and optical shielding is likely needed to mitigate these effects. Alternatively, the use of diffractive elements for collection of ion fluorescence into waveguides can potentially address some of these challenges in interoperability of photodetectors by allowing the latter to be somewhat removed from the ions' locations.

Integrated collection optics, when further combined with single- or few-mode waveguide beamsplitters, may also enable high-speed remote-entanglement generation via single-photon interference [11] in a parallelizable implementation. Multiplexed collection using arrays of ions would enable parallel heralded generation of entangled pairs at a distance within a single chip, or between separate chips; such pairs can subsequently be used for gate teleportation between distant ions, in conjunction with more traditional Coulomb-based multi-qubit gates.

No less important than advances in individual components as described above will be the development of a platform that can bring together the multiple required elements—passive/active photonics, detectors, integrated electronics, and appropriately shielded multilayer trap structures—in a robust and accessible fashion. As requirements and constraints become more clear, a CMOS-based photonics platform with some custom layers and steps is likely to be of interest in this regard [12–14].

**Concluding Remarks**

Advances in photonics, particularly at visible and UV wavelengths, are likely to play a critical role in extending trapped-ion quantum computing systems beyond NISQ scales. Resulting systems, leveraging solid-state photonics and electronics intimately coordinated with naturally pristine atomic-ion qubits, promise a strong foundation for quantum applications more broadly, including portable, precise systems for metrology. The components motivated by this pursuit furthermore are likely applicable to quantum systems based on neutral atoms and solid-state defect centers, as well as more broadly in visible/UV photonic applications.

**Acknowledgements**

This material is partially based upon work supported by the Department of Defense under Air Force Contract No. FA8702-15-D-0001. Any opinions, findings, conclusions, or recommendations expressed in this material are those of the authors and do not necessarily reflect the views of the Department of Defense. KKM acknowledges support from an ETH postdoctoral fellowship.

## 22 – Photonic Qudits on Chip

Jacquiline Romero[1,2]

[1]Australian Research Council Centre of Excellence for Engineered Quantum Systems (EQUS)

[2]School of Mathematics and Physics, University of Queensland, Queensland 4072, Australia

**Status**

Qudits [1]—the *d*-level version of the more ubiquitous 2-level qubit—should figure greatly in future quantum technologies, whether they be for communication, sensing, or computation. The most compelling reason is that quantum systems are *naturally* high-dimensional. The restriction to qubits is not necessary, particularly when engineering coupled systems together is difficult and qudit entanglement is physically possible. Qudits have the potential to outperform their qubit counterparts by providing access to a larger state space for storing and processing quantum information. Integrating qudit systems—like those that exploit multiple degrees of freedom—offers more favourable scaling than multi-particle qubit approaches because detection rate significantly decreases as the number of entangled particles becomes larger. Qudit advantages have been shown for quantum communication and quantum computation, e.g. qudits lead to enhanced robustness against eavesdropping and simplified gates. More importantly, qudit entanglement has been shown to overcome noise in entanglement distribution, which is important for any future quantum communication network [2].

It is well known that a photon possesses several properties that naturally accommodate a qudit description, e.g. path, frequency, time-bins, photon number, and transverse spatial mode such as those associated with orbital angular momentum (OAM). Long room-temperature coherence times, ease of transmission via free space or optical fibre, and availability of efficient detectors also make photonic platforms excellent for implementing quantum technologies based on qudits.

Photonic qudits on chip is a nascent field but there have been some notable results in recent times. The most mature photonic qudit on chip platform uses path encoding: in [3] Wang et al. demonstrated 15 x 15 dimensional entanglement using a silicon photonic device that integrated photon-pair sources, multimode interferometers, phase-shifters, beamsplitters, waveguide crossers, and grating couplers; in [4] Llewellyn et al. demonstrated quantum teleportation between two silicon chips. Using time and frequency encodings (Figure 1), both Imany et al. [5] and Kues et al. [6] demonstrated two-qudit gates acting on photons generated via spontaneous four-wave mixing (SFWM) in a microring resonator. Exploiting the transverse spatial modes, an on-chip source of polarisation and OAM modes have been demonstrated on a silicon platform, and quantum interference has been demonstrated using photons generated from spontaneous parametric down-conversion (SPDC) and silicon nitride waveguides that act as mode multiplexer and mode beamsplitter [7] (Figure 2).

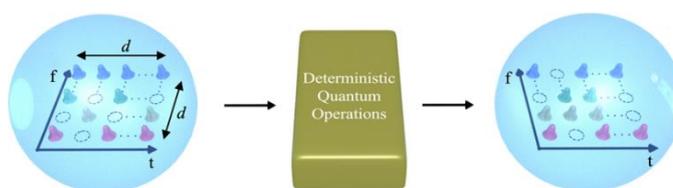

**Figure 1.** Two *d*-dimensional qudits can be encoded a single photon using time bins and frequency. Arbitrary superpositions of different time and frequency bins can be encoded (unused time–frequency slots are shown as dashed circles). Deterministic quantum operations can be performed on the two-qudit state such that the orientation of the time–frequency superpositions is changed, hence creating a new two-qudit state. This approach gets rid of the multiphoton interaction usually required for this kind of operation. (Figure reprinted from Imany et al, Ref [4], Springer Nature Limited).

**Current and Future Challenges**

A key hurdle that applies to the study of qudits in general—not just for qudits on chip—is that qudits attract less attention than qubits. A Hilbert space of dimension *d* is associated with $d^2-1$ distinct unitaries which could be seen as unnecessary complication—both for theory and experiments. Qudit versions of the various quantum computation models are thus seldom explored. Nonetheless, photonic qudit information processing on chip is an active area of research, with path, time-bin, and frequency encodings the most developed in experiments.

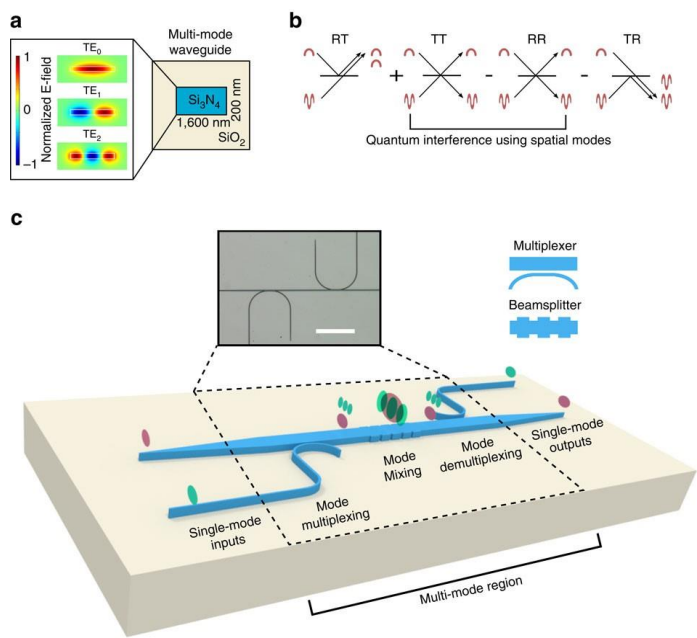

Like in any other platform for quantum computation, scaling up is a huge challenge. Current estimates of the number of logical qubits necessary for doing quantum simulations—arguably the first practical application of a quantum computer—is in the order of at least hundreds. This number significantly increases when quantum error correction is considered with estimates going as high as ~$10^6$ physical qubits. Scaling up is challenging whether we consider the circuit-based or measurement-based models of quantum computation. In both cases we need to integrate a significant number of photon sources and detectors, in addition to the requisite quantum gates and measurements necessary. Integration also minimises loss, which is especially evident when coupling in and out of chip (where one can have ~10s dB of insertion loss).

**Figure 2.** Transverse spatial modes in a multi-mode waveguide (a) can exhibit Hong-Ou-Mandel two-photon interference (b). A chip implementation (c) is achieved through spatial mode multiplexing via asymmetric directional couplers and spatial mode beamsplitters via nanoscale gratings. The mode order within the multi-mode region is indicated by the colours (red is TE0, green is TE2) indicate the mode order of the device. The inset shows a microscope image, scale bar is 160 µm. (Figure reprinted from Mohanty et al, Ref [6], Springer Nature Limited).

Universal quantum gate sets for qudits are more challenging to implement, regardless of the degree of freedom used. Specific to path encoding is the need for thermo-optic phase-shifting that often requires cooling. Thermal phase shifting is even more of a constraint when there are several transverse modes in a single path because more heat is needed to introduce a phase shift between the various modes. For frequency modes, conversion is often achieved using nonlinearities and achieving both high selectivity and high efficiency remains a challenge. However, recent results using ring resonators in lithium niobate have shown ~99% efficiency and also programmability [8]. This could be important for quantum communication since frequency is compatible with current optical fibre networks.

**Advances in Science and Technology to Meet Challenges**

There is significant work to be done on the theoretical side in order to motivate focused efforts that use qudits for quantum technologies. For the same dimensionality, we have to compare (1) ideal qudits to ideal qubits, and (2) noisy qudits to noisy qubits. The latter is especially important given the rapid advance of noisy intermediate-scale quantum (NISQ) technology. We have to show that the advantages gained from using qudits persist even in realistic noise scenarios. For example, [9] shows that using qutrits achieves logarithmic depth decomposition (cf. linear depth for qubit-only counterpart) of the generalised Toffoli gate with no ancilla and that the qutrit circuit remains reliable even in noisy regimes.

There are two approaches to scaling up qudits on chip. One is to increase the dimension of the qudit. This approach has to be weighed carefully with the capability to do operations in a larger Hilbert space for although it is straightforward to increase the dimension—due to the theoretically infinite state space of photonic qudits—implementing *d*-level operations is not. Another approach is to use lower-dimensional qudits (e.g. qutrits) while increasing the density of source, gates, and detectors.
For example, there could be multiple microring resonators in a single chip to produce either heralded single photons or entangled photon pairs. Single photon detectors should also be integrated to the chip to limit losses. The availability of photon-number-resolving detectors should also be exploited, either to use photon-number as a qudit or to account for multiple photon pair events which limit quantum interference in parametric processes.

Implementing operations for qudits on chip depends on the degree of freedom used. For path encoding, phase-shifters based on electromechanical actuation will be more efficient—because they do not need cooling—and precise—because there will be no thermal crosstalk with other paths. For other degrees of freedom like transverse spatial mode, designing from first principles often lead to more complex devices (e.g. increased number of interferometers). For such cases, inverse design based on desired output properties can prove to be advantageous as dimension increases. Hyper-entanglement, wherein multiple degrees of freedom are entangled, is readily achievable in photons. Hyperentanglement should be developed not just for dense coding but also for generating cluster states for quantum computation [10]. Implementing two-qudit gates is also likely to require nonlinearities, which can also be useful in interfacing with memories that are naturally multilevel. Lithium niobate will possibly be a valuable material for future photonic chips at is has a strong $\chi^{(2)}$ nonlinearity and can be used to provide precise timing control via electro-optic modulation [11].

**Table 1. Examples of qudits-on-chip demonstrations**

| Qudit | Path [3] | Frequency [5] | Transverse mode [6] | Time-frequency [9] |
|---|---|---|---|---|
| Source | SFWM on-chip | SFWM on-chip | SFWM on-chip | SPDC off-chip |
| Operation | unitaries (Reck-Zeilinger) on-chip | phase gates via programmable filters and electro-optic modulation off-chip | unitaries (Reck-Zeilinger) on-chip | controlled phase gate via fibre Bragg mirrors and electro-optic modulation off-chip |
| Detection | off-chip | off-chip | off-chip | off-chip |
| Quantum demonstration | Bell-CGLMP inequality violation; randomness expansion | Bell-CGLMP inequality violation; | Hong-Ou-Mandel interference | one-way qudit cluster state processing |

**Concluding Remarks**
Qudits hold great potential for improving current quantum technologies and opening up research avenues not possible with qubits. Photons are endowed with naturally high-dimensional properties that can be used as qudits—there have been many impressive qudit experiments using bulk optics. Qudit on chip provides a better way of scaling up, especially when compared to multiple photons where state purity and detection rate decrease dramatically with increasing number of photons.

Qudits on chip are expected to provide access to increased information capacity and information processing capability not achievable with bulk optics.

Qudit on chip platforms are most mature for path, time, and frequency degrees of freedom, Table 1 summarises current progress. However, there is still much room for improvement especially in precise control of these systems and in integrating source, gates, and detectors into one device. Qudit gates are the least developed of these, although recent advances in programmability for frequency [8] and path [12] holds promise for the quantum regime.

**Acknowledgements**
The author thanks: Westpac Bicentennial Foundation; L'Oreal-UNESCO For Women in Science Foundation; Australian Research Council Centre of Excellence for Engineered Quantum Systems (EQUS, CE170100009); Stephen Barnett and Barry Sanders for useful discussions; Andrew G. White and William J. Munro for reading an early draft of this roadmap.

## 23 – Frequency Bin Quantum Photonics

N. B. Lingaraju, A. M. Weiner (School of Electrical and Computer Engineering and Purdue Quantum Science and Engineering Institute,
Purdue University, West Lafayette, Indiana 47907, USA)

**Status**

Quantum frequency combs are the single- or few-photon analog of classical frequency combs. As one might presume, these states populate a superposition of modes that are equidistant from one another in frequency. What makes them particularly attractive is that they can support many temporal and frequency modes; high dimensionality and hyperentanglement can potentially offer a large quantum resource over just a few photons. Quantum frequency combs were first observed in optical parametric oscillators pumped below threshold. More recent interest in the field was spurred by the generation of two-photon quantum frequency combs in CMOS-compatible optical microresonators [1], [2]. Owing to their compact geometry, these sources support free spectral ranges in the 10s of GHz – a range well suited to commercial telecom equipment. Recent accomplishments [3] include characterization of high-dimensional frequency-bin entanglement, single-photon two qudit gates, generation of four-party 32-dimensional Greenberger-Horne-Zeilinger states, and nonlocal delay metrology featuring picosecond-scale resolution with potential relevance to the synchronization of quantum networks [4].

Progress in the field also extends to the emergence of a paradigm for quantum information processing with comb-like states. Like any platform, one needs to access individual modes to apply prescribed phase shifts, as well as mix a combination of modes in any desired ratio. The former is readily accomplished with programmable filter banks or Fourier transform pulse shapers. The latter, on the other hand, poses a real challenge as there is no off-the-shelf component to mix frequency modes in an arbitrary manner. To overcome this limitation, a team at Oak Ridge National Laboratory devised a protocol capable of arbitrary spectral transformations using electro-optic phase modulators (EOMs) and Fourier transform pulse shapers [5]. This concatenation, called a quantum frequency processor or QFP for short, was used in a series of demonstrations over the last couple of years, culminating in completion of the gate set for frequency [3].

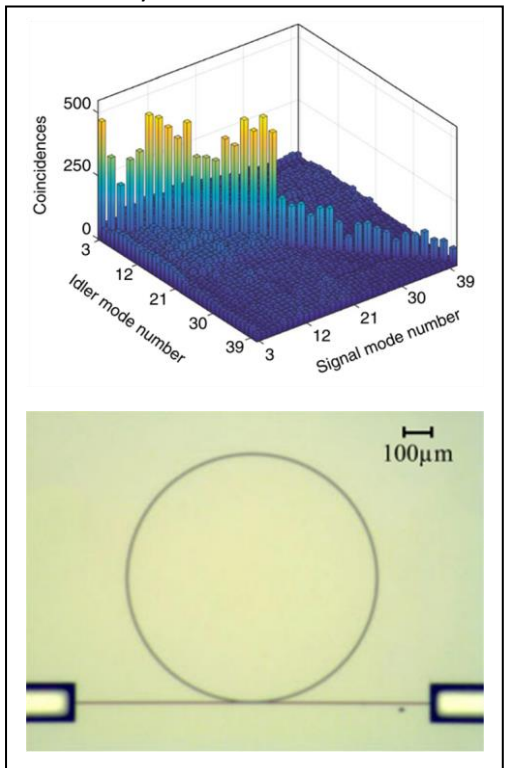

**Figure 1.** (top) Joint spectral intensity showing spectral correlations between signal and idler photons across 30+ modes. (bottom) Microscope image of the microring and U-grooves to support fiber coupling. Source: Ref [2].

This progress in realizing arbitrary spectral transformations is perhaps most relevant to the development of quantum interconnects, which would benefit from the ability to implement high-dimensional gates, work with spectrally multiplexed quantum channels, and carry out in-band quantum frequency conversion.

**Current and Future Challenges**

Outside of photon pair sources, experiments to date have relied on off-the-shelf telecommunications equipment to manipulate and retrieve quantum information. However, such experiments have some important limits. One drawback to working with discrete components is that they have high insertion losses (4 dB for EOMs and 5 dB for pulse shapers). Reducing these losses is critical for scalable quantum information processing and deployment in quantum networks.

Another limit is that adjacent modes must be separated by at least 20 GHz – set by the resolution of commercial pulse shapers – in order to avoid crosstalk. Consequently, if one wants to efficiently mix even just adjacent modes, EOMs need to be driven with at least a 20 GHz radiofrequency (RF) drive tone. However, to connect more distant modes, or implement high-dimensional quantum gates using a QFP, one needs access to not just a 20 GHz tone but also to higher RF harmonics. For example, to permit the full range of 2D quantum operations with both high fidelity (> 0.9999) and high success probability (> 0.95) using just two modulators and a pulse shaper, the RF drive needs to include at least two harmonics. Similar behavior holds for higher dimensions as well. This points to a unique advantage of the QFP protocol – gate complexity can be scaled without any increase to the number of physical elements! This interplay between the modulator bandwidth ($BW_{RF}$) and the fundamental mode separation (FSR) suggests a rough gauge of operation complexity. To first order, the ratio $BW_{RF}$/FSR corresponds to the highest dimensional gate set one can synthesize using a three-element QFP. Thus, even with a generous RF bandwidth of 60 GHz, one is limited to just 3-dimensional gates owing to the 20 GHz pulse shaper resolution.

For these reasons, realizing the full potential of quantum information processing in the frequency domain will require the migration of functionality to an integrated optical platform. In addition to a reduction in optical loss, integrated photonics will enable more complex operations, either by increasing the number of RF harmonics accessible to the system or by permitting realization of QFPs that comprise concatenations of additional pulse shaper and modulator elements.

**Advances in Science and Technology to Meet Challenges**
Quantum comb sources have benefited from advances in the development of microresonator-based (classical) comb sources. While further refinements, like reducing the comb FSR, are desirable, there is a greater need for tools to manipulate frequency-encoded quantum information on chip. This includes pulse shapers for channel-by-channel phase control and EOMs for frequency mixing.

Pulse shapers can be realized using banks of tunable ring resonators [6] and are not tied to any one material system. Low waveguide propagation loss (< 1 dB/cm) has been reported in a variety of platforms at both academic facilities [7] and at commercial foundries [8]. These developments, coupled with astute design choices, have made it possible to realize sub-GHz filters with minimal drop loss [8]; such elements will enable more closely spaced frequency modes.

On-chip phase modulators compatible with low loss and large-scale photonic integration have traditionally been problematic, but recent research provides new possibilities. Especially notable are developments with thin-film lithium niobate, which has a native electro-optic efficient. By relying on the Pockels effect rather than on free carrier dispersion or some other intrinsically lossy mechanism, devices capable of high-speed operation (> 100 GHz) and low propagation loss (0.2 dB/cm) can be realized [7]. While the headline numbers are impressive, it is important to recognize that average loss alone does not tell the full story. A better prism through which to view modulator performance is the voltage-loss product expressed in VdB. Quite simply this is the loss incurred when the modulator is scaled to support a half-wave voltage of 1V. Thin-film lithium niobate leads the pack ($\alpha V_\pi L$ = 0.5 VdB), but indium phosphide (0.9 VdB) [9], as well as more exotic approaches like silicon-organic hybrids (1.0 VdB) [10], are not far behind and have advantages in terms of platform maturity and economies of scale.

At the device level, progress in coupled-cavity modulators [10] or so-called "photonic molecules" offers a complement to the QFP protocol. These devices comprise two evanescently coupled optical microresonators and, owing to mode hybridization, can scatter an input light field between two, *and only two*, closely separated spectral modes – an operation potentially equivalent to a frequency qubit rotation.

**Concluding Remarks**

It is notable that frequency bin quantum photonics has synergies with fields that are the subject of significant research and development interest – Kerr microcombs, radiofrequency photonics, and high-performance computing. Quantum frequency combs, like their classical counterparts, can be generated in nonlinear microresonators with low anomalous group velocity dispersion. Scalable production of these sources would make this platform readily accessible to the wider research community. Similarly, progress in high-performance computing and wideband communications hinges on the development of high-speed, low-loss optical switches, which differ from low-loss phase modulators by only a scaling of the device length. While frequency encoding has some catch up to do, the natural advantages of this degree of freedom – stability over standard optical fiber, compatibility with telecom infrastructure, and support for multilevel quantum information – coupled with recent developments in integrated photonics, point to a bright future in the development of practical quantum networks.

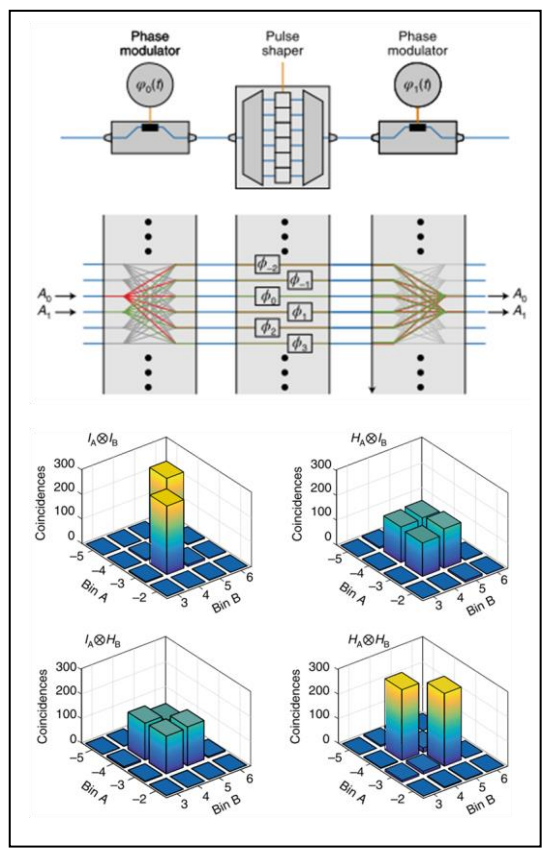

Figure 2. (top) Concept illustration for a three-element quantum frequency processor. The physical configuration comprises phase modulators and a central pulse shaper. (bottom) Measured correlations of a two-qubit frequency-bin entangled state, where each qubit is operated on by either the identity (I) or Hadamard (H) gates independently. Source: Ref [3]

**Acknowledgements**

*The authors acknowledge funding from the National Science Foundation (2034019-ECCS, 1839191-ECCS). N. B. Lingaraju acknowledges support from a QISE-NET Triplet award (NSF 1747426-DMR).*

## 24 – Continuous Variables Computing with Lithium Niobate Photonics

Daniel Peace, Robert Cernansky and Mirko Lobino
Centre for Quantum Computation and Communication Technology and Centre for Quantum Dynamics, Griffith University, Brisbane, QLD 4111, Australia

**Status**

Continuous variable (CV) quantum technologies encode information in specific states of the electromagnetic field quadrature operator $\hat{q} = (\hat{a}^\dagger e^{i\theta} + \hat{a} e^{-i\theta})/\sqrt{2}$ whose spectrum has continuous values. This approach offers several advantages compared to polarization and photon number encoding, both with a discrete set of eigenvalues, but is more susceptible to losses. Specifically, the generation of large entangled states and high efficiency detection are open problems and technological challenges for discrete variable (DV) systems. In contrast, quadrature entanglement and squeezing can be generated deterministically, and homodyne detectors (HD) can measure quadrature with efficiency greater than 99%.

Two-dimensional cluster states, which are necessary universal resource for CV computation, have been demonstrated in experiments over several thousands of nodes [1] [2], highlighting the scaling potential of the CV approach especially for measurement-based quantum computation (MBQC) (see Fig. 1). Furthermore, proposals based on the Gottesman–Kitaev–Preskill (GKP) state encoding [4] recently established a fault tolerance threshold for CV-MBQC. But while a basic computation of ~100 steps has been demonstrated with 1D cluster [3], the implementation of non-Gaussian resources, which is necessary for universal quantum computation, is still an open problem.

Lithium Niobate (LiNbO3, LN) is one of the most widely used materials in photonics. This is because of its wide transparency window (400nm – 5μm), high second order nonlinearity ($d_{33}$ = 25 pm/V at 1550 nm), and large electro-optic coefficient, which enables fast amplitude and phase modulation. These properties are essential for the realization of a MBQC platform and have been recently demonstrated in a single integrated device where CV squeezing and entanglement was generated, manipulated, and measured [5].

While the experiments in [5] used in-diffused LN waveguide, which are 6-10 μm wide and have a low index contrast, lithium-niobate-on-insulator (LNOI) waveguides offer scaling properties similar to silicon photonics, but with the cited benefits of LN [6]. Propagation losses as low as 2.7 dB/m [6], ultra-fast amplitude modulation, and single pass second harmonic generation (SHG) efficiency more than an order of magnitude higher than their diffused counterparts [6]. So far, uses of LNOI waveguide for quantum photonics have been limited to generation of single photons [6].

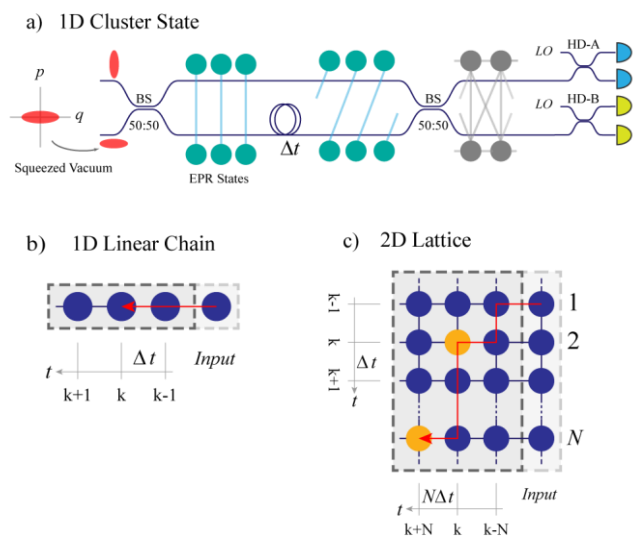

Figure 1. a) Schematic for the generation of a 1D cluster state [3], where $p$ and $q$ squeezed vacua are interfered creating a series of EPR states every time constant $\Delta t$. One spatial mode of the EPR state is delayed by one time constant forming a dual rail cluster state which is represented by a b) linear chain of temporal modes. Extending the 1D scheme with a second set of squeezers and delay of length $N\Delta t$ forms a c) 2D cluster state with a square lattice shape, which is the primary resource for CV-MBQC. Circuit for 2D cluster state shown Figure 2. Blue circles indicate temporal modes, Yellow circles are GKP states inserted into the cluster for error correction or non-Gaussian operations. Red arrow shows the flow of information encoded in the input states down the chain (cluster) for 1D (2D) cluster states. BS - beam splitter, HD - homodyne detector, LO - Local Oscillator.

**Current and Future Challenges**

The realization of a universal CV-MBQC machine requires technological innovation for the LNOI platform and theoretical advances for the efficient generation of non-Gaussian resources. The error correction threshold for CV 2D cluster based on GKP encoding requires ~15-17 dB of squeezing, independent on the level of anti-squeezing [7]. This value caps the maximum amount of total loss tolerable in a given circuit to a few percent and includes both propagation and coupling losses in addition to detection efficiency. To date the largest level of squeezing reported is 15 dB, obtained from a semi-monolithic optical parametric oscillator (OPO) operating in the continuous-wave regime [8].

On-chip generation of squeezed vacuum is currently limited to 6 dB from a ridge waveguides in ZnO-doped LN with a 5 μm x 6 μm cross section in a single-pass configuration [9]. Broadband single-pass sources are more suited for time-multiplexed 2D cluster states since they enable shorter temporal modes thereby increasing the size of the cluster state for a given time window resulting in a larger computational power. The strong nonlinearity and low losses of LNOI waveguides can provide a crucial advantage over other material platforms [6].

A time multiplexed MBQC is characterized by two main timescales: a short time scale which separates two neighbouring nodes of the cluster, and a long one which set the qubit width of the computation [1],[2]. While a fully integrated approach offers clear advantages (see Fig. 2), it is possible that CV-MBQC machines will use off-chip delays to increase their computational power and that high efficiency in/out coupling photonic structure will be necessary in order to perform within the fault tolerant threshold.

Large-scale cluster states and Gaussian operations are not enough for universal quantum computation. Non-Gaussian states, operations or measurements must be included. Utilising GKP encoding is optimal for universality and error correction, but experimentally they are yet to be realised with photonics. For a large-scale system, a high rate of GKP states is required, which remains one of the main fundamental challenges for the realization of a MBQC universal machine.

**Advances in Science and Technology to Meet Challenges**

Recently, spurred by advances in the commercial availability of up to 6" wafers, the thin film LNOI platform has gained significant interest, with demonstrated low-loss ridge waveguides and small bending radii (<100 μm), ultrafast light manipulation, and large nonlinearity, highlighting its potential for scalable quantum technologies in communication, metrology and computation.

With propagation losses as low as 2.7 dB/m at 1550 nm and measured nonlinear conversion efficiency of 4600% $W^{-1}cm^{-2}$, a periodically poled LNOI waveguide (ppLNOI) can theoretically reach the 15 dB squeezing threshold for fault-tolerance if pumped with ~ 130 mW of power in fifteen millimetre long devices. So far ppLNOI waveguides have underperformed in terms of propagation losses when compared to the 2.7 dB/m mark, and improvements in the fabrication process are necessary.

The low loss threshold of CV quantum computation requires the integration with high efficiency detectors. Integrated superconducting single photon detector on LNOI have been recently demonstrated [10] and while the 46% efficiency is not the state of the art, this is an important result. High efficiency photodiodes for HD already exist off chip, an integrated version is yet to be demonstrated. Designs for coupling structures such as grating couplers tapers and mode converters have been proposed showing theoretical efficiencies near 80%, however experimental realisations typically half that are still far from the few percent tolerable losses. As such work is still required to further reduce these losses, possibly with the use of inverse design tools and improved fabrication processes.

The efficient generation of a GKP state is the main theoretical challenge for the implementation of a full scale, error corrected universal CV quantum computer. This state is a coherent superposition of a comb quadrature eigenstates that could be generated from optical Schrödinger cat state strongly coupled to a qubit system. No experimental generation of GKP states using photonics has been reported so far. Theoretical proposals based on photon-number-resolving detectors (PNR), squeezed states and linear optical circuits [7] offering an approximate, non-deterministic, heralded path for experimentalists to follow.

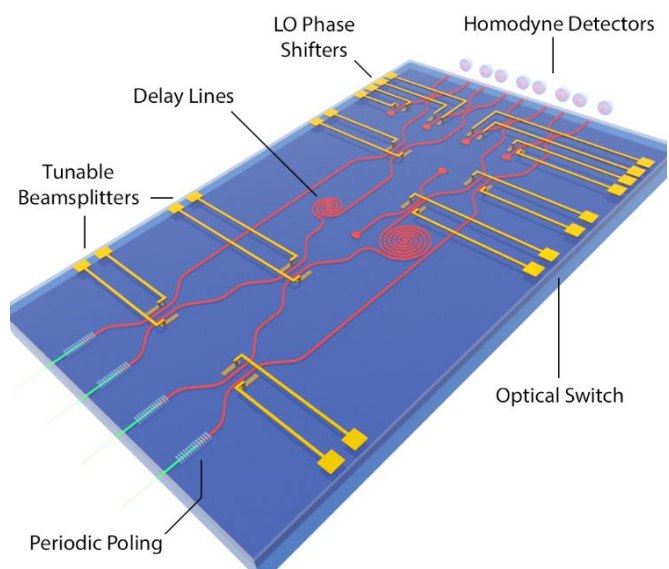

Figure 2. Artistic impression of a photonic integrated circuit for the generation of 2D cluster state based the design of [1]. Photonic circuit features periodic poling for generation of squeezed vacua, tunable beamsplittters, on-chip delay lines and homodyne detection. Using an optical switch prepared GKP states can be injected into the cluster to provide error-correction and non-Gaussian operations.

**Concluding Remarks**

The development of the LNOI photonic platform with low propagation losses, large nonlinearity, and ultrafast electro-optic reconfigurability embed several key properties necessary for MBQC. This technology, together with the recent generation of large CV cluster states and theoretical progress for fault tolerant computation hold great potential for the realization of a photonic universal quantum computer. While a LN device has demonstrated the generation, manipulation and detection of CV quantum states, future technological development will focus on the integration of efficient homodyne and single photon detectors, and low loss delay lines*.*

Theoretical progress is still needed for the efficient generation of non-Gaussian resources necessary for universal quantum computation. While here we focused on GKP state encoding because it is the one used for the derivation of the fault-tolerance threshold, other proposals to solve the encoding and error correction problem may arise in the future.

**Acknowledgements**

This work was supported financially by the Australian Research Council Centres of Excellence scheme number CE170100012. M. L. was supported by the Australian Research Council Future Fellowship (FT180100055).

## 25 – Quantum Cryptography

Eleni Diamanti and Luis Trigo Vidarte
CNRS and Sorbonne University

**Status**

Quantum cryptography encompasses primitives that use the properties of quantum mechanics to offer new cryptographic services or provide security advantages with respect to their classical counterparts. It is part of the broader field of quantum communication, which includes all technologies and applications involving the transfer of quantum states between distant locations. Quantum communication implementations typically use photons as carriers of information and optical fiber or free space (including satellite) links as communication channels. This makes integrated photonics a key player in the development of these systems, where it can facilitate scalability, reproducibility, interconnectivity and reliability, as well as a reduction in price and physical footprint [1], speeding up the adoption of these technologies at the user level in emerging global-scale quantum communication infrastructures.

Quantum key distribution (QKD) [2] is perhaps the most well-known application of quantum cryptography and commercial products are already available. It allows the exchange of a private secret key between two trusted parties over an untrusted quantum channel provided that the parties have access to an authenticated classical channel and true random number sources. The main advantage of QKD with respect to classical alternatives like post-quantum cryptography is that the security does not rely on the computational resources of the attacker, but on the faithfulness of the implementation with respect to the model used in the security proof of the protocol. QKD offers information-theoretic, long-term security, making it an application of great interest in many sectors (medicine, finance, government, defence...) where these properties are essential.

Quantum random number generation (QRNG) is very commonly associated with cryptography, since it uses entropy sources based on the laws of quantum mechanics to generate true random values, which is required for cryptography, including QKD, but is also useful for other purposes. Commercial QRNG products are already available, some of them based on photonic integrated circuits. Other interesting quantum cryptographic primitives include secret sharing, digital signatures, commitment schemes, secure multiparty computation (oblivious transfer, coin flipping) and position-based cryptography. These protocols can involve different assumptions in terms of trust between the users, levels of security and particular requirements at the system level, and offer a wide range of advanced functionalities useful for applications in quantum communication networks. Here, we focus mostly on the challenges on the way to the use of integrated photonics in QKD since these cover the main requirements for the other quantum cryptographic protocols as well.

**Current and Future Challenges**

Quantum communication primitives and QKD in particular become interesting when the quantum states can be distributed at relatively high rates and over long distances. On the quest for high-performance QKD systems, a great number of protocols have been successfully implemented [2], mostly in laboratory or field experiments, and more rarely as commercial products, but many improvements need to be made before considering them a mature market (see Fig. 1 for a general forecast). A major family of protocols are discrete-variable (DV) ones, where typically coherent states generated by lasers are attenuated and then manipulated in some degree of freedom (polarization, phase, time) before being detected by single-photon detectors. In continuous variable (CV) QKD systems, the coherent states are modulated in phase-space and detected by coherent detectors, where the quantum states are mixed with a local oscillator (generated by a laser at the receiver in most modern implementations). While in such systems the receiver needs to be trusted, measurement device independent (MDI) and twin field (TF) QKD allow for a third untrusted party to perform the measurements. Fully device independent (DI) QKD uses Bell nonlocality to remove trust assumptions from all devices but imposes stringent constraints that cannot be satisfied by currently available technology.

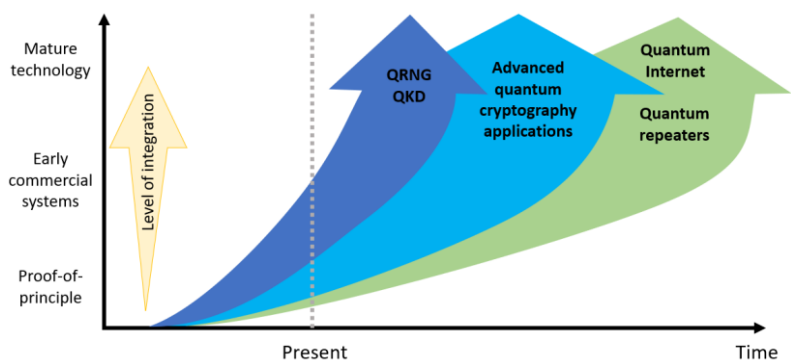

**Figure 1.** Forecast of evolution of quantum cryptography and quantum communication technologies.

System demonstrations involving photonic integrated components have been performed for DV [3], CV [4] and MDI QKD [5,6], based mostly on the Si platform but InP has also been used in some cases. These are essentially proof-of-principle experiments, and a major challenge for these systems is to reach the performance of QKD implementations based on bulk components. To increase the level of integration, it will be necessary to address important challenges that remain on the single-photon detection side and on the operation with multiple polarizations and time delays. Furthermore, in general, a particular photonic integration platform may not be optimal for all of the required components in the system. Advancing towards heterogeneous integration [7] is therefore a crucial challenge, and will also be of great importance for ultimately integrating also the electronic functionalities of the system needed for signal control and data processing. Functionalities linked to network architectures (synchronization, multiplexing, routing) should also be considered.

Since classical signal amplification techniques are detrimental for quantum signals, the range achieved by QKD and more generally by quantum communication systems is ultimately limited by the inherent exponential loss suffered by light in optical fibers. For QKD, this limits the distance over which secret key generation can be performed to a few hundred kilometers, with a strong trade-off with respect to the achievable rate. Ideally, quantum repeaters, which involve quantum memories, can be used for extending the distance. Research efforts are underway toward on-chip quantum memories [8] but several challenges will need to be addressed for such components to be sufficiently mature for use within advanced quantum cryptographic systems. Such systems also rely, for instance, on entangled-photon generation and frequency conversion, where photonic integration has considerably advanced. While developments in quantum repeaters are in process, trusted nodes can be used to extend the range, but for intercontinental distances a more interesting approach is to use satellite communications, where free-space attenuation is quadratic with the distance. Although free-space QKD using Si integrated components has shown promising results [9], to prepare the ground for satellite-based demonstrations it will be necessary to advance towards space qualification of such systems, including their photonic integrated components.

**Advances in Science and Technology to Meet Challenges**
Telecom C-band and O-band are the most common wavelengths of choice in optical fiber communications. Free-space communications can use a broader range, but the current trend is to use telecom wavelengths for compatibility. This creates a lot of synergies with classical communications, in particular for CV-QKD which shares many elements with coherent transceivers, for which implementation agreements from the Optical Internetworking Forum (e.g., TROSA) offer an industry standard. Current devices allow bandwidths of several GHz in the size of a coin suitable for high-rate classical communications, but desirable improvements for quantum systems include higher detection efficiency and narrower laser linewidths.

Efficient detection of single photons at telecom wavelengths is more problematic than coherent detection [10] and it typically requires cooling, which reduces the advantage of integration. An interesting solution in this case is to use integrated transmitters and bulk receivers, which is practical for MDI-QKD protocols as was already shown [5, 6]. Preparation of the quantum states in phase-space is straightforward (using phase, amplitude and IQ modulators) [4], but polarization and time operations can also be performed using custom systems [3, 4].

The interest in significant advances in heterogeneous (or hybrid) integration spans all quantum technologies. Of particular interest for quantum cryptography is wafer-scale silicon-on-insulator (SOI), typically interesting for telecommunication wavelengths, Si photonics matched with polymer wave guiding offering advantages in terms of manufacturability, III-V (InP and GaAs) platforms for integrating active sources, and substrates like silicon carbide (SiC), which may be interesting for quantum repeaters. It is clear, however, that for these technologies to become relevant for advanced quantum cryptographic system implementations, important developments at the component level are necessary.

**Concluding Remarks**
Quantum cryptography is a diverse field with a broad range of potential applications. Some applications like QRNG and QKD are already targeting large markets, while other protocols could greatly benefit a niche public before becoming mainstream. The nature of the protocols gives a lot of predominance to the system level of the implementation, making it interesting to provide flexible systems that can operate multiple protocols or configurations with the same hardware. The potential benefits of integrated photonics are extensive as it enables more functionalities to become available as building blocks for different technologies allowing for the design of reliable custom subsystems that can be part of quantum cryptographic systems. Improvements on heterogeneous integration of multiple platforms and packaging (including particular cases such as space qualification) will continue to expand the presence of integrated photonic devices in quantum communication systems, driving their scalable deployment in global network infrastructures.

**Acknowledgements**
The authors acknowledge funding from the European Union's Horizon 2020 Research and Innovation Programme under Grant Agreement No. 820466 (CiViQ).

## 26 – Machine Learning for Photonics
Ryan M. Camacho, Brigham Young University

**Status**

Machine learning algorithms can automatically learn patterns within a dataset, even in the absence of a physical model. In contrast to most early targets for machine learning, Maxwell equations are well-known with robust numerical techniques successfully developed and commercialized. Machine learning has become valuable in photonics primarily owing to its reduced computational cost. Traditional full-wave techniques, such as finite-difference time domain (FDTD), finite element method (FEM), and others, solve Maxwell's equations directly, but become costly with larger systems. This problem is compounded for *design* problems, in which a geometry is sought to enable a desired photonic transfer function. In such cases, many forward simulations are required along with an optimization strategy to find the optimal solution[1]. Machine learning tools greatly increase the speed of forward design, and thereby also enable rapid inverse design when coupled with techniques such as adjoint methods.

Among the most popular class of machine learning algorithms are discriminative artificial neural networks, which specify nonlinear mappings between input and output variables using matrix multiplication, neuron-based processing, and nonlinear activation functions[2]. Compared with traditional electromagnetic solvers, discriminative neural networks can evaluate forward problems orders of magnitude faster[3]. Generative neural networks using latent random variable inputs have also become popular, producing one-to-many photonic mappings that enable rapid simulation of a distribution of device layouts. In addition, several other machine learning techniques have been explored and used successfully to model photonic devices and systems.[4] An ever-increasing range of photonic devices and systems have now be successfully modelled using machine learning, ranging from integrated photonic devices[3] to ultrafast photonic systems[5] and many more.

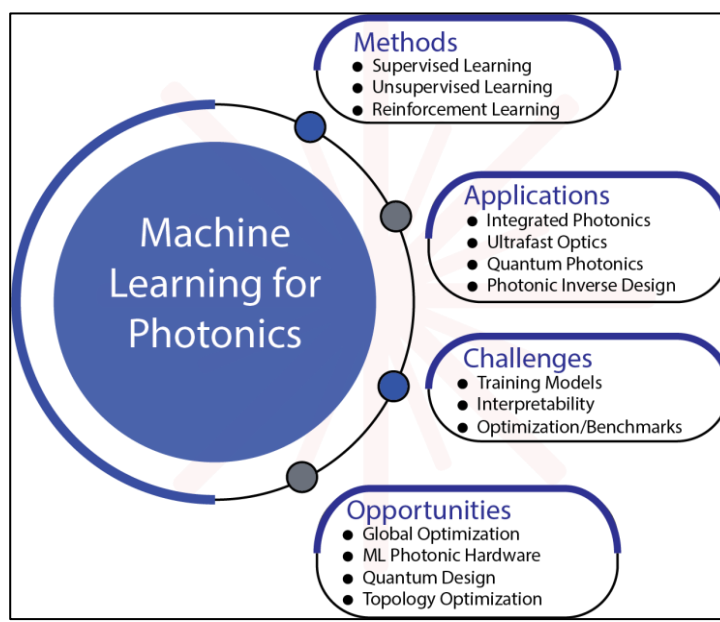

A more recent application of machine learning is for quantum photonic systems, where techniques are being developed to solve Schrodinger's equation rather than Maxwell's equations[6]. An exponential number of terms describing a quantum state can be encoded in a neural network with a polynomial number of qubits, thereby enabling efficient quantum photonic tasks such as quantum state tomography[7].

**Figure 1** Overview of Machine Learning for Photonics

**Current and Future Challenges**

There are several significant challenges and opportunities for machine learning in classical and quantum photonics. Among the most formidable is improved model training, which currently requires large datasets and significant up-front computational cost. While traditional direct solvers can be used to generate datasets, future efforts would be greatly benefitted by community collaboration to generate data sets and develop transfer learning techniques amenable to a wide class of photonic simulations. Community efforts to construct better benchmarking problems to assess new models is also needed.

Another related challenge is the residual uncertainty in the solutions provided by machine learning tools, and the relative difficulty of interpreting the results. Even the most carefully trained models are known to be at risk of diverging solutions within data points. Integrating the known structure of the underlying physical models (such as Schrodinger and Maxwell equations) with machine learning training algorithms, loss functions, and the resulting models may be the key to increasing the interpretability of the models, mitigating errors, and increasing training speed. Faster conventional solvers will also be helpful for model generation and verification. There continues to be rapid advances in forward solutions in the form of compact models and abstractions, many of which incorporate machine learning as subroutines[8].

Future efforts will also focus on incorporating more degrees of freedom and system complexity in machine learning models. Currently, most models include outputs such as reflection, transmission, scattering, absorption etc., but opportunities exist for incorporating additional properties to fully model the optical response, such phase, angular momentum, trajectories, nonlinearities, topology, and field distributions. Furthermore, opportunities exist of incorporating global optimizations that include elements of fabrication and experimental testing variables, and find hidden features and adaptations as the model is exposed to new data.

In the field of machine learning for quantum photonics, there is a need to improve optimization techniques and strategies. For example, only recently have viable techniques been introduced to find the quantum equivalent of the classical gradients used in most model optimizations[9], and further generalization as still sought. Another exciting research direction is the emerging ability to use machine learning models to suggest new and interesting quantum experiments[10].

Finally, enormous potential exists in developing improved photonic hardware to train and implement machine learning models. A major bottleneck using traditional hardware is the interconnect speed among neurons, which would be greatly enhanced using photonics, with a significant resulting societal impact. Quantum photonic implementations of neural networks may also greatly improve the ability to the enormous multi-dimensional datasets generated by quantum computers.

**Concluding Remarks**
In summary, machine learning is become a vital tool for both classical and quantum photonic systems. It is anticipated that the field will continue to grow and have increasing impact.

**Acknowledgements**
This work is supported by the National Science Foundation under grant no. 1941583.